\documentclass[reprint,twocolumn,superscriptaddress,showpacs,floatfix]{revtex4-1}

\usepackage{amsmath}
\usepackage{mathrsfs}
\usepackage{txfonts}
\usepackage{amssymb}
\usepackage{graphicx}
\usepackage{hyperref}
\usepackage{ulem}
\usepackage{overpic}
\usepackage{psfrag}
\usepackage{tabularx}
\usepackage{multirow}
\usepackage{array}
\usepackage{placeins}
\newcommand{\PreserveBackslash}[1]{\let\temp=\\#1\let\\=\temp}
\newcommand{\ket} [1] {| #1 \rangle}

\makeatletter

\begin{document}

\title{Emergent Goldstone flat bands and spontaneous symmetry breaking\\ with type-B Goldstone modes}
\author{Huan-Qiang Zhou}
\affiliation{Centre for Modern Physics, Chongqing University, Chongqing 400044, The People's Republic of China}

\author{Jesse J. Osborne}
\affiliation{School of Mathematics and Physics, The University of Queensland, St. Lucia, QLD 4072, Australia}

\author{Qian-Qian Shi}
\affiliation{Centre for Modern Physics, Chongqing University, Chongqing 400044, The People's Republic of China}

\author{Ian P. McCulloch}
\affiliation{Department of Physics, National Tsing Hua University, Hsinchu 30013, Taiwan}
\affiliation{Frontier Center for Theory and Computation, National Tsing Hua University, Hsinchu 30013, Taiwan}

\begin{abstract}
For a quantum many-body spin system undergoing spontaneous symmetry breaking with type-B Goldstone modes, a high degree of degeneracy arises in the ground state manifold. Generically, if this degeneracy is polynomial in system size, then it does not depend on the type of boundary conditions used.  However, if there exists an emergent (local) symmetry operation tailored to a specific degenerate ground state, then we show that the degeneracies are exponential in  system size and are
different under periodic boundary conditions (PBCs) and open boundary conditions (OBCs). We further show that the exponential ground state degeneracies in turn imply the emergence of Goldstone flat bands --  single-mode excitations generated by a multi-site operator and its images under the repeated action of the translation operation under PBCs or the cyclic permutation symmetry operation under OBCs.
Conversely, we also show that the presence of emergent Goldstone flat bands implies that there exists an emergent (local) symmetry operation tailored to a specific degenerate ground state. In addition, we propose an extrinsic characterization of emergent Goldstone flat bands, revealing a connection to quantum many-body scars, which violate the eigenstate thermalization hypothesis.  We illustrate this by presenting examples from the staggered ${\rm SU}(4)$ spin-1 ferromagnetic biquadratic model and the staggered ${\rm SU}(4)$ ferromagnetic spin-orbital model.
We also perform extensive numerical simulations for the more general ${\rm SO}(3)$ spin-1 bilinear-biquadratic and ${\rm SO(4)}$ ferromagnetic  spin-orbital models, containing the two aforementioned models as the endpoints in the ferromagnetic regimes respectively, and confirm the emergence of Goldstone flat bands,  as we approach these  endpoints from deep inside the ferromagnetic regimes.
\end{abstract}
\maketitle

\section{Introduction}

Spontaneous symmetry breaking (SSB) is a fundamental notion in  diverse branches of physics, including particle physics, condensed matter physics, and cosmology~\cite{andersonbook,SSBbook}.
In particular, if the broken symmetry group is continuous,  then a gapless Goldstone mode (GM) emerges~\cite{goldstone1,goldstone2,goldstone3}, where the number of GMs is equal to the number of broken symmetry generators $N_{BG}$ for a relativistic system undergoing SSB. However, for a non-relativistic system, not all SSB patterns for a continuous symmetry group fall into the same category, as was reflected in a debate between Anderson and Peierls~\cite{anderson,peierls1,peierls2}.
As a result of the long term pursuit of a proper classification of GMs~\cite{nielsen,nambu,schafer, miransky, nicolis1,nicolis2, brauner-watanabe, watanabe1,watanabe2, NG1,NG2},
it is necessary to introduce two distinct types of GMs: type-A and type-B~\cite{watanabe1,watanabe2}.  Accordingly, for an SSB pattern from the symmetry group $G$ to the residual symmetry group $H$,
a distinction between type-A and type-B GMs has to be made to understand the low-energy physics of quantum many-body systems undergoing SSB, since many of them have been revealed to exhibit SSB with type-B GMs~\cite{FMGM,hqzhou,goldensu3,spinorbitalsu4,dimertrimer,TypeBtasaki}.

Notably, quantum many-body systems undergoing SSB with type-B GMs are described by so-called frustration-free Hamiltonians~\cite{tasakibook}, as is the case in all of the 
known examples~\cite{FMGM,hqzhou,goldensu3,spinorbitalsu4,dimertrimer,TypeBtasaki}.  Hence, highly degenerate ground states arising from SSB with type-B GMs are exactly solvable. Indeed,  the orthonormal basis states spanning the ground state subspace are generated from the repeated action of the lowering operator(s) on the highest weight state and admit an exact Schmidt decomposition, which in turn implies that these orthonormal basis states exhibit  self-similarities in real space~\cite{FMGM,hqzhou}. In fact, it has been found that  the entanglement entropy scales logarithmically with the block size  in the thermodynamic limit, with the prefactor being half the number of type-B GMs, as far as the orthonormal basis states are concerned. A paradigmatic example for SSB with type-B GMs is the quantum spin-$1/2$ ${\rm SU}(2)$ ferromagnetic Heisenberg model; the entanglement entropy for this model has been investigated in Refs.~\onlinecite{popkov1,popkov2,doyon1,doyon2,ding}. Here, the ground state degeneracies under both periodic boundary conditions (PBCs) and open boundary conditions (OBCs) are the same, and are polynomial in system size in any spatial  dimensionality. However, this model alone is not sufficient to understand all the crucial features relevant to SSB with type-B GMs: this is reflected in the fact that there are quite a few  quantum many-body systems undergoing SSB with type-B GMs that have ground state degeneracies exponential in system size under both OBCs and PBCs. In contrast to the above case of polynomial degeneracy, the ground state degeneracies here depend on the type of boundary conditions adopted. Two intriguing examples are the staggered spin-1 ${\rm SU}(3)$ ferromagnetic biquadratic model~\cite{goldensu3} and the staggered ${\rm SU}(4)$ ferromagnetic spin-orbital model~\cite{spinorbitalsu4}, which are the two simplest physical realizations of the Temperley-Lieb algebra~\cite{tla,baxterbook,martin} in this context. Hence, they are exactly solvable in the Yang-Baxter sense~\cite{baxterbook,faddeev}.

A natural question arises as to whether or not there is any mechanism for explaining the origin of the exponential ground state degeneracies under PBCs and OBCs. This is particularly interesting if the mechanism enables us to establish a criterion by which one may judge  whether or not the ground state degeneracies are  exponential in system size, but different under PBCs and OBCs. The existence of such a criterion  will certainly deepen our understanding of quantum many-body spin systems undergoing SSB with type-B GMs. Physically,  any further ramifications to this criterion would be relevant to the complete classification of quantum states of matter, given that such quantum many-body spin systems exhibit exotic quantum states of matter and novel types of quantum phase transitions~\cite{goldensu3,spinorbitalsu4,dimertrimer}.

In this work, we aim to address this question from two different perspectives - one is intrinsic and the other extrinsic in nature. For our purpose, we restrict ourselves to a quantum many-body spin system undergoing SSB from $G$ to $H$ with type-B GMs, where $G$, a semisimple Lie group modulo a discrete symmetry subgroup, always contains  ${\rm SU}(2)$ in the spin space as a subgroup. We argue that if there exists an emergent (local) symmetry operation tailored to a specific degenerate ground state, then the ground state degeneracies under both PBCs and OBCs are exponential in system size. The exponential ground state degeneracies  imply the emergence of Goldstone flat bands, indicating trivial dynamics underlying the low-lying excitations. Here, an emergent Goldstone flat band is defined as a single-mode excitation  that is generated by a multi-site operator and its images under the repeated action of the translation operation or the cyclic permutation symmetry operation, if PBCs or OBCs are 
adopted, when they act on the highest weight state.  Moreover, we also show the converse relation, that the presence of  emergent Goldstone flat bands in turn implies that there exists an emergent (local) symmetry operation tailored to a specific degenerate ground state. In other words, the three properties, namely, (1) the existence of emergent (local) symmetry operations tailored to degenerate ground states, (2) the exponential ground state degeneracies in  
system size, and (3) the emergence of  Goldstone flat bands, are  equivalent  in the context of SSB with type-B GMs. This constitutes an intrinsic characterization of the emergence of Goldstone flat bands that {\it solely} depends on the features of a quantum many-body spin system undergoing SSB with type-B GMs.
We present some illustrative examples focusing on the staggered  ${\rm SU}(3)$ spin-1 ferromagnetic biquadratic model and the staggered ${\rm SU}(4)$ ferromagnetic spin-orbital model.

In addition to this intrinsic characterization, we also present an extrinsic characterization of emergent Goldstone flat bands, when  a specific quantum many-body spin system undergoing SSB with type-B GMs is treated as a special point in the parameter space of a more general quantum many-body spin system. This makes it possible to reveal a connection to quantum many-body scars~\cite{scar0,scar1,scar2,scar3}, which violate the eigenstate thermalization hypothesis (ETH)~\cite{eth1,eth2,eth3,eth4,eth5,eth6}. Meanwhile, the disordered variants of the two models under investigation exhibit Hilbert space fragmentation~\cite{tlhsf1}.
In particular, the aforementioned staggered  ${\rm SU}(3)$ spin-1 ferromagnetic biquadratic model and staggered ${\rm SU}(4)$ ferromagnetic spin-orbital model can be treated as special points that act as the endpoints in the ferromagnetic regimes for the ${\rm SO}(3)$ spin-1 bilinear-biquadratic model and the ${\rm SO(4)}$ spin-orbital model, respectively.  As it turns out, low-lying multi-magnon excitations in the ferromagnetic regimes for the two latter generic models become flat as we approach the endpoints embedding the two former models from deep inside the ferromagnetic regimes.

This extrinsic characterization reveals a drastic difference between the ${\rm SO}(3)$ spin-1 bilinear-biquadratic model and the ${\rm SO(4)}$ ferromagnetic  spin-orbital model.
Normally, quantum many-body scars only occupy a small portion of the Hilbert space (i.e.\ polynomial in system size), in contrast to the strong ergodicity breaking arising from quantum complete integrability~\cite{integrability} and many-body localization~\cite{localization1,localization2}. While this is the case for  the ${\rm SO}(3)$ spin-1 bilinear-biquadratic model, the  ${\rm SO}(4)$ spin-orbital model appears to be an exception, in the sense that at a generic non-integrable point in this model, the exactly solvable excited states occupy a large (i.e.\ exponential) portion of the Hilbert space. This is due to the fact that the model accommodates two integrable sectors, one in the spin sector and the other in the pseudo-spin sector. We remark that these two integrable sectors are identical to the  ${\rm SU(2)}$ spin-$1/2$ Heisenberg model, which itself is a paradigmatic example of quantum complete integrability in the Yang-Baxter sense~\cite{baxterbook,faddeev,sklyanin} under both PBCs and OBCs. In fact, it is readily seen that all the conserved currents for the ${\rm SU(2)}$ spin-$1/2$  Heisenberg model are also the conserved currents in  the  two integrable sectors  for the ${\rm SO}(4)$  spin-orbital model, irrespective of the boundary conditions adopted.

We perform an extensive numerical investigation of the spectral functions of the ${\rm SO}(3)$ spin-1 bilinear-biquadratic model and the ${\rm SO}(4)$  bilinear-biquadratic model, as we approach the endpoints embedding the staggered  ${\rm SU}(3)$ spin-1 ferromagnetic biquadratic model and the staggered ${\rm SU}(4)$ ferromagnetic spin-orbital model from deep inside the ferromagnetic regimes, respectively. This  makes it possible to visualize how the low-lying multi-magnon excitations flatten out, thus confirming our theoretical analysis.

The paper is organized as follows. In Section~\ref{twoexamples}, we introduce the two aforementioned  models undergoing SSB with type-B GMs, which exhibit exponential ground state degeneracies in system size that are different under both PBCs and OBCs. In Section~\ref{grd}, we briefly recall the ground state degeneracies under both PBCs and OBCs, and their connection to an emergent (local) symmetry operation tailored to a degenerate ground state.
In Section~\ref{threeitems}, we demonstrate that the three properties, as already mentioned above, are  equivalent  in the context of SSB with type-B GMs. In Section~\ref{extrinsic}, we turn to an extrinsic characterization of emergent Goldstone flat bands. We also emphasize the relevance to quantum many-body scars~\cite{scar0,scar1,scar2,scar3} and  Hilbert space fragmentation~\cite{tlhsf1}. In Section~\ref{spectralfunction}, we perform extensive numerical simulations  for the ${\rm SO}(3)$ spin-1 bilinear-biquadratic model  and the ${\rm SO(4)}$ ferromagnetic  spin-orbital model, confirming the emergence of Goldstone flat bands in the staggered ${\rm SU}(3)$ spin-1 ferromagnetic biquadratic model  and the staggered ${\rm SU}(4)$ ferromagnetic spin-orbital model. The final Section~\ref{summary} is devoted to concluding remarks.

\section{Quantum many-body spin systems undergoing SSB with type-B GMs: two examples}~\label{twoexamples}

We focus in detail on two specific examples of quantum many-body spin systems undergoing SSB with type-B GMs, which exhibit the exponential ground state degeneracies in system size under both PBCs and OBCs: the staggered ${\rm SU(3)}$ spin-1 ferromagnetic biquadratic model~\cite{barber1,barber2,barber3} and the staggered ${\rm SU(4)}$ ferromagnetic spin-orbital model~\cite{So4}. Nevertheless, there are many other models which display this behaviour~\cite{dtmodel,dimertrimer}.
Before proceeding, we stress that SSB with type-B GMs occurs in any spatial dimensions, in contrast to SSB with type-A GMs that is forbidden in one spatial dimension as a result of the Mermin-Wagner-Coleman theorem~\cite{mwc1,mwc2}. Since all the key ingredients underlying SSB with type-B GMs do not depend on the dimensionality, one may restrict oneself to investigate one-dimensional quantum many-body systems undergoing SSB with type-B GMs. In principle,  our discussion may be extended to any quantum many-body system on a lattice in two and higher spatial dimensions~\cite{2dtypeb}.

The first model we investigate is the staggered ${\rm SU(3)}$ spin-1 ferromagnetic biquadratic model, described by the Hamiltonian
\begin{equation}
	\mathscr{H}=\sum_{j}{\left(\textbf{S}_j \cdot \textbf{S}_{j+1}\right)^2}. \label{hambq}
\end{equation}
Here $\textbf{S}_j=(S^x_j,S^y_j,S^z_j)$ is the vector of the spin-1 operators at lattice site $j$.
The sum over $j$ is taken from 1 to $L$ for PBCs, and from 1 to $L-1$ for OBCs.
Unless otherwise stated, we assume that the system size $L$ is even.
The model (\ref{hambq}) possesses a staggered ${\rm SU(3)}$ symmetry, which has a total of 8 
generators. For our purpose, it is convenient to introduce the two Cartan generators  $H_1$ and $H_2$, with the raising operators $E_1$, $E_2$ and the lowering operators $F_1$, $F_2$, forming two ${\rm SU(2)}$ subgroups. In addition, there is one more ${\rm SU(2)}$ subgroup generated by $H_3 = H_1 - H_2$, with raising operator $E_3$ and lowering operator $F_3$. They satisfy the commutation relations $[E_a,E_b] = [F_a,F_b] = 0$, $[H_a,E_a] = 2E_a$, $[H_a,F_a] = -2F_a$, together with $[H_1,F_2] = -F_2$ and $[H_2,F_1] = -F_1$. All of these generators can be expressed in terms of the spin-1 operators $S^x_j,S^y_j$, and $S^z_j$ at lattice site $j$ (for the details, we refer to Ref.~\onlinecite{goldensu3}, but also see Appendix A for a brief summary). The SSB pattern is from ${\rm SU(3)} $ to ${\rm U(1)}\times {\rm U(1)}$, with two type-B GMs: $N_B=2$.

The second model we investigate is the staggered ${\rm SU(4)}$ ferromagnetic spin-orbital model, described by the Hamiltonian
\begin{equation}
	\mathscr{H}=\sum_{j}\left(\textbf{S}_j \cdot \textbf{S}_{j+1} - \frac{1}{4}\right) \left(\textbf{T}_j \cdot \textbf{T}_{j+1} -\frac{1}{4}\right).
	\label{hamist}
\end{equation}
Here  $\textbf{S}_j=(S^x_j,S^y_j,S^z_j)$ and $\textbf{T}_j=(T^x_j,T^y_j,T^z_j)$ are the vectors of the spin-$1/2$  and pseudo-spin-$1/2$ operators at lattice site $j$, respectively. The sum over $j$ is taken from $1$ to $L$ for PBCs, and from $1$ to $L-1$ for OBCs. The symmetry group is staggered ${\rm SU(4)}$, which has a total of 15 generators. It is convenient to introduce the Cartan generators  $H_1$, $H_2$ and $H_3$,  with the raising operators $E_1$, $E_2$ and $E_3$ and the lowering operators $F_1$, $F_2$ and $F_3$, forming three ${\rm SU(2)}$ subgroups, in addition to six more generators, denoted by $E_4$, $F_4$,  $E_5$, $F_5$,  $E_6$ and $F_6$.
The detailed expressions for the Cartan generators  $H_1$, $H_2$ and $H_3$ and the others in terms of
the spin-$1/2$ operators $S^x_j,S^y_j,S^z_j$ and the pseudo-spin-$1/2$ operators $T^x_j,T^y_j,T^z_j$ at lattice site $j$ may be found in Ref.~\onlinecite{spinorbitalsu4} (see also Appendix A for a brief summary).
The SSB pattern\cite{spinorbitalsu4} is from ${\rm SU(4)} $ to ${\rm U(1)}\times {\rm U(1)}\times {\rm U(1)}$ via ${\rm U(1)}\times {\rm U(1)}\times {\rm SU(2)}$, with three type-B GMs: $N_B=3$.

The commonalities of the above two models lie in the fact that both the  model  (\ref{hambq}) and the  model  (\ref{hamist}) are exactly solvable by means of the Bethe ansatz~\cite{barber1,barber2,barber3}, since they constitute (up to an additive constant)  representations of the Temperley-Lieb algebra~\cite{tla,baxterbook,martin}, and follow from a solution to the quantum Yang-Baxter equation~\cite{baxterbook,faddeev}. Moreover, both of the models share another common feature, namely that they are both frustration-free~\cite{tasakibook}: there is a ground state of the Hamiltonian $\mathscr{H}=\sum_j \mathscr{H}_{j j+1}$ which is a simultaneous ground state of each and every $\mathscr{H}_{j j+1}$, with $\mathscr{H}_{j j+1}$ being the local Hamiltonian operators acting only on the two nearest-neighbor sites $j$ and $j+1$.  Note that  the constraints imposed on such a ground state under PBCs are stricter than those under OBCs, given $j=1,2,\ldots,L$ under PBCs and $j=1,2,\ldots,L-1$ under OBCs.
For the staggered ${\rm SU(3)}$ spin-1 biquadratic model  (\ref{hambq}), this specific ground state may be chosen as $\ket{\psi_0} = \ket{\otimes_{\eta=1}^{L} \{+\}_{\;\eta}}$, which is an eigenstate of $S^z = \sum_j S^z_j$, with the eigenvalue being $L$. Here, $\vert +_{j}\rangle$,
$\vert 0_{j}\rangle$ and $\vert -_{j}\rangle$ represent the eigenstates of $S^z_j$ at lattice site $j$.
For the staggered ${\rm SU(4)}$ ferromagnetic spin-orbital model (\ref{hamist}), this specific ground state may be chosen as $\ket{\psi_0} \equiv \ket{\otimes_{\eta=1}^{L} \{\uparrow_{s}\uparrow_{t}\}_{\eta}}$, which is an eigenstate of $S^z$ and $T^z$, both taking eigenvalues to be $L/2$. Here and hereafter, $\vert \uparrow_{sj} \rangle$ and  $\vert \downarrow_{sj} \rangle$ are used to represent the eigenstates of $S^z_j$ with eigenvalues $1/2$ and $-1/2$, respectively;
and similarly, $\vert \uparrow_{tj} \rangle$ and  $\vert \downarrow_{tj} \rangle$  are used to represent the eigenstates of $T^z_j$  with eigenvalues $1/2$ and $-1/2$, respectively. Hence, $\ket{\psi_0}$ acts as the highest weight state for the symmetry group in each of the models under investigation.

However, there is a marked difference between the two models under investigation, which concerns the uniform subgroups of the symmetry groups.  In fact, the staggered ${\rm SU(3)}$ symmetry group contains  the uniform ${\rm SU(2)}$ group as a subgroup, and the staggered ${\rm SU(4)}$ symmetry group contains the uniform ${\rm SU(2)} \times {\rm SU(2)}$ group as a subgroup.  As a convention, the uniform ${\rm SU(2)}$ group is generated by the spin operators $S^{x},S^{y}$ and $S^{z}$, defined as $S^{x,y,z}= \sum_j S_j^{x,y,z}$, and the uniform ${\rm SU(2)} \times {\rm SU(2)}$ group is generated by two copies of the ${\rm SU(2)}$ generators, denoted as $S^{x},S^{y}$ and $S^{z}$, and $T^{x},T^{y}$ and $T^{z}$, respectively, defined similarly as above.
 Hence, one is always able to label degenerate ground states in terms of the eigenvalue of the Cartan generator $S^{z}$ of the subgroup ${\rm SU(2)}$ or in terms of the eigenvalues of the Cartan generators  $S^{z}$ and $T^{z}$ of the subgroup ${\rm SU(2)} \times {\rm SU(2)}$. Here we note that ${\rm SU(2)}$, homomorphic to ${\rm SO(3)}$, is simple, but ${\rm SU(2)} \times {\rm SU(2)}$, isomorphic to ${\rm SO(4)}$, is semi-simple. As we shall show in Section~\ref{extrinsic}, this difference leads to an intriguing consequence that 
the ${\rm SO(4)}$ spin-orbital model (cf. Eq.(\ref{hamist1}) in Section~\ref{extrinsic}), which accommodates
the staggered ${\rm SU(4)}$ ferromagnetic spin-orbital model  (\ref{hamist}) as a special integrable point, is generically non-integrable, but admits extensively many conserved currents, if one is only restricted to a specific spin or pseudo-spin sector. By ``extensively many" we mean infinitely many in the thermodynamic limit.  This is in sharp contrast to the ${\rm SO(3)}$ spin-1 bilinear-biquadratic model (cf. Eq.(\ref{hamso3}) in Section~\ref{extrinsic}), which in turn accommodates the staggered ${\rm SU(3)}$ spin-1 ferromagnetic biquadratic model as a special integrable point, but it does not admit any integrable sectors away from the staggered ${\rm SU(3)}$ point, if the antipodal point of the Takhtajan-Babujian model~\cite{takhtajan,babujian} is excluded (cf. Section~\ref{extrinsic}).

In general, the staggered  ${\rm SU(2s+1)}$ spin-$s$ ferromagnetic model may be re-interpreted as the staggered  ${\rm SU(2s+1)}$ biquadratic model, with the symmetry group being the staggered ${\rm SU(2s+1)}$ group~\cite{afflecksun1}, if it is expressed
in terms of the ${\rm SO(2s+1)}$ generators~\cite{son1}.  The explicit form of the Hamiltonian corresponds to $\theta =\pi/2$ in the ${\rm SO(2s+1)}$ bilinear-biquadratic model (\ref{hamso}), which we shall discuss in Section~\ref{extrinsic}. We remark that the staggered  ${\rm SU(2s+1)}$ spin-$s$ ferromagnetic model constitutes a physical realization of the Temperley-Lieb algebra, as shown in Refs.~\onlinecite{barber2,barber3}.  Specifically, the staggered  ${\rm SU(3)}$ ferromagnetic biquadratic model (\ref{hambq}) is a special case, with spin $s=1$, and the staggered  ${\rm SU(4)}$ ferromagnetic spin-orbital  model (\ref{hamist}) is unitarily equivalent to the  staggered  ${\rm SU(4)}$ spin-$3/2$ ferromagnetic model.
In this sense,  the two models under investigation here are the two simplest physical realizations of the Temperley-Lieb algebra in this context.

\section{Ground state degeneracies and emergent (local) symmetry operations tailored to degenerate ground states}~\label{grd}

For a quantum many-body spin system undergoing SSB from $G$ to $H$ with type-B GMs, the number of type-B GMs is the rank of the broken symmetry group  $G$, which is a semi-simple Lie group~\cite{FMGM,hqzhou}. Generally, the ground state degeneracies do not depend on what types of  boundary conditions adopted. Actually, the ground state subspace constitutes an irreducible representation of the symmetry group $G$, whose dimension is polynomial in $L$. Specifically,  the ground state degeneracies are linear and quadratic in $L$ for $G={\rm SU(2)}$ and  $G= {\rm SU(3)}$, respectively, in the cases of the one-dimensional ${\rm SU}(2)$ spin-$1/2$ ferromagnetic Heisenberg model and the ${\rm SU}(3)$ spin-$1$ ferromagnetic model~\cite{FMGM,hqzhou}.  However, this is not always the case.
There exist quite a few  quantum many-body systems undergoing SSB with type-B GMs, where the ground state degeneracies are exponential in $L$~\cite{goldensu3,spinorbitalsu4,dimertrimer}. In these cases, the ground state degeneracies depend on what types of boundary conditions are adopted. We denote the ground state degeneracies as ${\rm dim }(\Omega_L^{\rm PBC})$ under PBCs and 	${\rm dim }(\Omega_L^{\rm OBC})$ under OBCs. As a result, the residual entropy  per lattice site is non-zero, where the residual entropy is defined as the natural logarithm of the ground state degeneracy.

For the staggered ${\rm SU(3)}$ spin-1 ferromagnetic biquadratic model and the staggered ${\rm SU(4)}$ ferromagnetic spin-orbital model,
the recursive relations for  ${\rm dim }(\Omega_L^{\rm PBC})$ and ${\rm dim }(\Omega_L^{\rm OBC})$ have already been established~\cite{goldensu3,spinorbitalsu4}. Namely,
we have
\begin{align}
	{\rm dim }(\Omega_L^{\rm PBC})&=R^{-2L}+R^{2L}, \quad  L\geq3, \cr
	{\rm dim }(\Omega_L^{\rm OBC})&=\frac {R^{-2L-2}-R^{2L+2}}{R^{-2}-R^{2}}, \quad L\geq2 . \label{gsdsu3su4}
\end{align}
Here, $R$ is the inverse golden ratio $R=(\sqrt{5}-1)/2$ for the staggered ${\rm SU}(3)$ spin-1 ferromagnetic biquadratic model~\cite{goldensu3,read,spins,katsura-td,moudgalya}, and $R=\sqrt{2}(\! \sqrt{3}-\!1)/2$ for the staggered ${\rm SU}(4)$ ferromagnetic spin-orbital model~\cite{spinorbitalsu4}. As a consequence, the ground state degeneracies behave asymptotically as the golden spiral for  the former and  a logarithmic spiral for the latter, both of which are familiar self-similar geometric objects. Hence, the non-zero residual entropy per lattice site take the form $S_{\!r}=-2 \ln R$. 

Now we recall a notion introduced in Ref.~\onlinecite{dimertrimer} - an emergent (local) symmetry operation tailored to a specific degenerate ground state. It appears as the key ingredient in the following {\it lemma}. Suppose there exists a (local) unitary operation $g$ that does not commute with the Hamiltonian $\mathscr{H}$, i.e., $[\mathscr{H},g] \neq 0$, but satisfies $V|\Psi_0\rangle=0$ with $V=[\mathscr{H},g]$, where $|\Psi_0\rangle$ denotes a ground state, with $E_0$ being the ground state energy. Here $g$ is treated as an operation that defines a unitary operator acting on the ground state subspace, so we have exploited the same notation to denote both of them. Note that  $g$ is not necessary to be local. Given $\mathscr{H}|\Psi_0\rangle=E_0|\Psi_0\rangle$ and $|\langle \Psi_0|g|\Psi_0\rangle|\neq 1$, we have $\mathscr{H}g|\Psi_0\rangle=E_0g|\Psi_0\rangle$, where $|\Psi_0\rangle$ is assumed to be normalized. That is, $g|\Psi_0\rangle$ is a degenerate ground state. Here and hereafter, $g$  is referred to as an emergent (local) symmetry operation tailored to a specific degenerate ground state $|\Psi_0\rangle$.
If a unitary operation is an emergent symmetry operation in the above sense for any degenerate ground state, then it is said to be an emergent symmetry operation (tailored to the whole ground state subspace). Here, we emphasize that the converse of this lemma is also true. In other words, if there is a (local) unitary operation $g$ such that $g|\Psi_0\rangle$, satisfying $|\langle \Psi_0|g|\Psi_0\rangle|\neq 1$, is a ground state degenerate with $|\Psi_0\rangle$, then $g$ is an emergent (local) symmetry operator tailored to a degenerate ground state  $|\Psi_0\rangle$, in the sense that
the commutator $V=[\mathscr{H},g]$  annihilates $|\Psi_0\rangle$: $V|\Psi_0\rangle=0$. Hence, $g$ is an emergent
local symmetry operation tailored to a specific degenerate ground state $|\Psi_0\rangle$. In fact, as one can see from the above argument,  $|\Psi_0\rangle$ is not necessarily  a ground state, although we shall only consider the situation where it is.

As it turns out, this lemma is consistent with an equivalent restatement of Elitzur's theorem~\cite{elitzur} in a form restricted to a discrete local gauge symmetry, which states that a local gauge symmetry is not allowed to be spontaneously broken. Logically, this amounts to stating that if a local discrete unitary operation leads to a degenerate ground state, then it is not a local \textit{gauge} symmetry operation. In this sense, it is referred to as an \textit{emergent} (local) symmetry operation tailored to a specific ground state. One may thus expect that an emergent (local) symmetry operation tailored to a specific degenerate ground state should be present in some form and play a significant role in our investigation of quantum many-body systems exhibiting exotic quantum states of matter.

We remark that the model Hamiltonians (\ref{hambq}) and (\ref{hamist}) commute with  the (one-site) translation operation $\tau$ under PBCs. Meanwhile, the generator $\sigma$ of the cyclic permutation group $Z_L$, defined as $\sigma \equiv P_{12}P_{23} \ldots P_{L-1L}$, is an emergent permutation symmetry operation (tailored to the whole ground state subspace under OBCs), where $P_{jj+1}$ is a cyclic permutation operation acting on the two adjacent sites $j$ and $j+1$.
Actually, the action of the translation operation on a given ground state under PBCs is identical to that of
the generator $\sigma$ under OBCs, as long as this ground state remains identical as the boundary conditions vary from PBCs to OBCs. Indeed, this is guaranteed for any degenerate ground state under PBCs, since they must be a degenerate ground state under OBCs for any system undergoing SSB with type-B GMs, as it is readily seen from the fact that the constraints imposed on a ground state under OBCs are less restrictive than those under PBCs for a frustration-free Hamiltonian.  As discussed in Section~\ref{twoexamples}, this is the case for  the model Hamiltonians (\ref{hambq}) and (\ref{hamist}).

In addition to the discrete $Z_2$ SSB of the time-reversal symmetry, the discrete group $Z_L$ generated by the (one-site) translation operation $\tau$ under PBCs is \textit{partially} spontaneously broken in the models (\ref{hambq}) and (\ref{hamist}), in the sense that this SSB only happens for some, but not all degenerate ground states. There is a parallel for the (emergent) cyclic permutation symmetry operation $\sigma$ under OBCs, because all the degenerate ground states under PBCs are also degenerate ground states  under OBCs. Actually, it is this characteristic feature that imposes extra constraints on the coset spaces for the staggered ${\rm SU(3)}$ spin-1 biquadratic model  (\ref{hambq}) and the staggered ${\rm SU(4)}$ ferromagnetic spin-orbital model  (\ref{hamist}). The coset spaces, denoted as $CP^2_{\pm}$ and  $CP^3_{\pm}$, are in fact variants of the complex projective spaces $CP^2$ and
$CP^3$, respectively, which are relevant to an extension of the spin 
coherent states from  the uniform ${\rm SU(2)}$ group to
the uniform ${\rm SU(2s+1)}$ group~\cite{gilmore}, as described in Ref.~\onlinecite{hqzhou}. Here, the subscript $\pm$ indicates extra constraints on  degenerate ground states defined as linear combinations of such overcomplete basis states on a subset of the coset spaces as a result of the staggered nature of the symmetry groups. The precise meaning of  $CP^2_{\pm}$ and  $CP^3_{\pm}$ is explained in Appendix B. Moreover, there is an exchange symmetry operation between the spin operators $\textbf{S}_j$ and the pseudo-spin operators $\textbf{T}_j$ for  the staggered ${\rm SU(4)}$ ferromagnetic spin-orbital model. As we shall see, this exchange symmetry is also partially spontaneously broken, since some but not all degenerate ground states are symmetric under this  operation.

To proceed, we make a few remarks on terminology and notation. The highest weight state is denoted by $\ket{\psi_0}_q$, with the period $q$ being either one or two, as the symmetry group $G$ may be staggered.  For our purpose, we always choose the period $q$ to be one, so we simply denote the highest weight state as $\ket{\psi_0}$ throughout this work.
In addition to the highest weight state  $\ket{\psi_0}$,  it is useful to define a generalized highest weight state.  Indeed, such a generalized highest weight state is not unique, and normally depends on what types of boundary conditions are adopted. We denote a generalized highest weight state by
$|\psi_m\rangle$, where $m$ represents its level. Here $m$ is determined from the eigenvalue $L - m$ of the Cartan generator $S^z$ of the uniform  ${\rm SU(2)}$ group, which appears as a subgroup of the symmetry group $G= {\rm SU(3)}$, where $m= m_0+2m_{-1}$, with $m_0$ and $m_{-1}$ being the numbers of 
the lattice sites in the local states $|0\rangle$ and $|-1\rangle$, respectively, or from
the eigenvalues $L/2 - m_s$ and $L/2 - m_t$ of the Cartan generators $S^z$ and $T^z$ of the uniform  ${\rm SU(2)}\times {\rm SU(2)}$ group, which appears
as a subgroup of the symmetry group  ${\rm SU(4)}$, with $m=m_s+m_t$. Throughout this work, we are mainly interested in the eigenstates of $S^z$ of the uniform  ${\rm SU(2)}$ group, which are also degenerate ground states, with $m_{-1}$ being zero. Hence,  levels are up to $L/2$ for
the staggered ${\rm SU(3)}$ spin-1 biquadratic model  (\ref{hambq}).  As for the staggered ${\rm SU(4)}$ ferromagnetic spin-orbital model  (\ref{hamist}), levels are up to $L/2$ for simultaneous eigenstates of the Cartan generators $S^z$ and $T^z$ of the uniform  ${\rm SU(2)}\times {\rm SU(2)}$ group in the spin or pseudo-spin sector, or up to $L$  for simultaneous eigenstates of the Cartan generators $S^z$ and $T^z$ in the spin and pseudo-spin sectors, if they are also degenerate ground states. This stems from the fact that the time-reversal symmetry operation  maps the highest weight state and generalized highest weight states to the lowest weight state and generalized lowest weight states and vice versa. That is, both the lowest weight state and generalized lowest weight states are defined as the time-reversed partners of the highest weight state and generalized highest weight states.

Formally, a generalized highest weight state $|\psi_m\rangle$ at the $m$-th level is defined recursively as $|\psi_m\rangle \notin V_0\oplus V_1\oplus...\oplus V_{m-1}$, $E_i|\psi_m\rangle=0\mod(V_0\oplus V_1\oplus... V_{m-1})$, with $i = 1,2,3$ for the staggered ${\rm SU(3)}$ symmetry group~\cite{goldensu3}, and $i = 1,\ldots,6$ for the staggered ${\rm SU(4)}$ symmetry group~\cite{spinorbitalsu4}.
Here, $V_0$ denotes the subspace spanned by the degenerate ground states generated from the highest weight state, whereas $V_\mu$ denotes the subspace spanned by the degenerate ground states generated from a generalized highest state at the $\mu$-th level, where  $\mu=1,2, \ldots, m$.
This means that, even though $|\psi_m\rangle$ are linearly independent to the states in the subspace  $V_0\oplus V_1\oplus...\oplus V_{m-1}$, all the states generated from $|\psi_m\rangle$, with $E_1^{M_1}E_2^{M_2}|\psi_m\rangle$ ($M_1\geq0$, $M_2\geq 0$ and $M_1+M_2\geq1$) for the staggered ${\rm SU(3)}$ symmetry group or $E_1^{M_1}E_2^{M_2}E_3^{M_3}|\psi_m\rangle$ ($M_1\geq0$, $M_2\geq 0$, $M_3\geq 0$ and $M_1+M_2+M_3\geq1$) for the staggered ${\rm SU(4)}$ symmetry group, are linearly dependent to the states in the subspace $V_0\oplus V_1\oplus...\oplus V_{m-1}$. In particular, the highest weight state $\ket{\psi_0}$ can be viewed as a special case of generalized highest weight states at the $0$-th level.  As a convention, we require a generalized highest weight state to be fully factorized.

Here, we stress that there exist  generalized highest weight states that are periodic,  which we denote as $|\psi_m\rangle_p$, with $p$ being the period, which is not less than two. Actually,  such  periodic generalized highest weight states play a crucial role in unraveling the logarithmic scaling behavior of the entanglement entropy with block size for the staggered ${\rm SU(3)}$ spin-1 ferromagnetic biquadratic model and the staggered ${\rm SU(4)}$ ferromagnetic spin-orbital model~\cite{goldensu3,spinorbitalsu4}. This period $p$ can be regarded as the size of the emergent unit cell for an exact matrix product state (MPS) representation of these degenerate ground states~\cite{exactmps}.  As we shall see,  a generic generalized highest weight state is not periodic, and the total number of  generalized highest weight states is exponential in $L$. In contrast, the total number of periodic generalized highest weight states with all possible  periods is only polynomial in $L$. This polynomial versus exponential dichotomy, as a finite analogy to the dichotomy between countable and uncountable infinities, appears to be common when one deals with quantum many-body systems undergoing SSB with type-B GMs~\cite{hqzhou}. 

Throughout this work, as we shall mainly focus on nonperiodic generalized  highest weight states, we shall drop off the subscript  $p$ from $|\psi_m\rangle_p$  to denote a generalized  highest weight state. However, there are other degenerate ground states that are not fully factorized, but they are very useful in our construction. To distinguish them from generalized highest weight states,  we shall use  $|\psi_m\rangle$ with $m$ superscripts to denote a generalized highest weight state and use $|\psi_m\rangle$ with less than $m$ superscripts to denote such a degenerate ground state that is not fully factorized. If there are two or even more degenerate ground states that have the same subscripts and superscripts, then one needs to introduce an extra subscript to distinguish them, as happens when 
one discusses emergent local symmetry operations tailored to degenerate ground states at the third level (for a specific example, we refer to Appendix C for the staggered ${\rm SU}(3)$ ferromagnetic biquadratic model ).

\section{Emergent (local) symmetry operations,  exponential ground state degeneracies and emergent Goldstone flat bands}~\label{threeitems}

Here we demonstrate that the three properties, namely, (1) the existence of emergent (local) symmetry operations tailored to degenerate ground states, (2) the exponential ground state degeneracies in $L$, but different under PBCs and OBCs,  and (3) the emergence of Goldstone flat bands, are equivalent in the context of SSB with type-B GMs.
As a result,  an emergent (local) symmetry operation tailored to a specific degenerate ground state offers an intrinsic characterization of emergent Goldstone flat bands, with the exponential ground state degeneracies in $L$ being a salient feature, though they are different under PBCs and OBCs. 

This characterization offers a mechanism for explaining the origin of exponential ground state degeneracies in $L$ under both PBC and OBCs.

\subsection{From  emergent (local) symmetry operations to  exponential ground state degeneracies}~\label{emergenttoexponential}

For a quantum many-body spin system undergoing SSB from $G$ to $H$ with type-B GMs, if there is an emergent local symmetry operation tailored to a specific degenerate ground state, then the  ground state degeneracies  under OBCs and PBCs are exponential in $L$.
Here we have assumed that the entire symmetry group is $G$, which is semi-simple, modulo a discrete  symmetry subgroup.
To this end, we focus on the two specific examples of the staggered  ${\rm SU}(3)$ spin-1 ferromagnetic biquadratic model and the staggered ${\rm SU}(4)$ ferromagnetic spin-orbital model, which contain the uniform ${\rm SU(2)}$ group and the uniform ${\rm SU(2)} \times {\rm SU(2)}$ group as subgroups, respectively. However, our argument may be extended to other quantum many-body spin  systems undergoing SSB with type-B GMs. One of the candidates is the ${\rm SU}(2)$  spin-1 model with competing dimer and trimer interactions, with the symmetry group $G$ being the uniform ${\rm SU(2)}$ group or the uniform ${\rm SU(3)}$ group~\cite{dtmodel}, depending on the values of the coupling parameter. As shown in Ref.~\onlinecite{dimertrimer}, both of them admit  emergent (local) symmetry operations tailored to  degenerate ground states.

\subsubsection{The staggered ${\rm SU}(3)$ ferromagnetic biquadratic model}

For the staggered ${\rm SU}(3)$ spin-1 ferromagnetic biquadratic model (\ref{hambq}), it is proper to start from the highest weight state  $\ket{\psi_0} = \ket{\otimes_{\eta=1}^{L} \{+\}_{\;\eta}}$, which is characterized by means of an eigenvalue $L$ of $S^z$.
Let $|\Psi_0\rangle$ be the degenerate ground state defined by $S^-\ket{\psi_0} $, which is an eigenstate of $S^z$ with an eigenvalue $L-1$. Now let us consider a set of local unitary operations $g_j$,  defined as $\exp (i \pi \Sigma_j)$, with $\Sigma_j =S_j^z$ ($j=1$, \ldots, $L$). Since $[g_j, \mathscr{H}] \neq 0$ and $[g_j, \mathscr{H}]|\Psi_0\rangle = 0$, following the above mathematical lemma from Section~\ref{grd}, we are led to a set of degenerate ground states  $g_j|\Psi_0\rangle$,
which take the form: $\sum _{i=1}^L (1-2\delta_{i\;j}) S_i^-\ket{\psi_0} $, up to a multiplicative constant.
Note that each of $g_j$'s generates an emergent discrete symmetry group $Z_2$ tailored to $|\Psi_0\rangle$, because $g_j^2\vert\Psi_0\rangle= |\Psi_0\rangle$.
We are thus led to a sequence of degenerate ground states:  $\tau^j S_1^-\ket{\psi_0} $  ($j=1,2,\ldots,L$)  under PBCs and $\sigma^j S_1^-\ket{\psi_0} $ ($j=1,2,\ldots,L$) under OBCs. It is readily seen that they are formally identical to each other under both PBCs and OBCs. Here, $\tau^j$ and $\sigma^j$ denote the $j$-th power of $\tau$ and $\sigma$, respectively.
Actually, they constitute  generalized highest weight states $\vert\psi_1^j\rangle \equiv S_j^- \ket{\psi_0} $  ($j=1,2,\ldots,L$) at the first level. Physically, these generalized highest weight states  are relevant to low-lying one-magnon excitations, in the sense that they constitute a set of basis states, invariant under the translation operation under PBCs or the cyclic permutation group under OBCs, to yield  one-magnon excitations.

Now we turn to a generalized highest weight state with an eigenvalue $L-2$ of $S^z$. If $\vert\Psi_0\rangle$ is chosen to be $S^-S_1^- \ket{\psi_0} $, then we consider a local unitary operation $g = \exp (i \pi \Sigma)$, with  $\Sigma=S_1^z+S_2^z+S_L^z$ under PBCs and   $\Sigma=S_1^z+S_2^z$ under OBCs.
As follows from the lemma, we are led to a degenerate ground state  $g\vert\Psi_0\rangle$, which takes the form:
$\left(-(S_1^-)^2-S_1^-S_2^-+\sum_{j=3}^{L-1}S_1^-S_j^--S_1^-S_L^- \right)\ket{\psi_0} $ under PBCs and
$\left( (S_1^-)^2+S_1^-S_2^--\sum_{j=3}^{L-1}S_1^-S_j^- \right)\ket{\psi_0} $ under OBCs, up to a multiplicative constant. Again, $g$ generates an emergent discrete  symmetry group $Z_2$ tailored to $\vert\Psi_0\rangle$, because $g^2|\Psi_0\rangle= \vert\Psi_0\rangle$.
We are thus  led to degenerate ground states: $|\psi_2^{1}\rangle \equiv \left( (S_1^-)^2+S_1^-S_2^-+S_1^-S_L^- \right) \ket{\psi_0} 
$ and $\sum_{j=3}^{L-1}S_1^-S_j^-\ket{\psi_0} $ under PBCs and
$|\psi_2^{1}\rangle \equiv \left( (S_1^-)^2+S_1^-S_2^- \right) \ket{\psi_0} $ and $\sum_{j=3}^{L}S_1^-S_j^-\ket{\psi_0} $ under OBCs. Afterwards,
if $\vert\Psi_0\rangle$ is chosen to be $\sum_{j=3}^{L-1}S_1^-S_j^-\ket{\psi_0} $ under PBCs, then we consider
a set of local unitary operations $g_j$, defined as $\exp (i \pi \Sigma_j)$, with $\Sigma_j = S_j^z$ ($j=3$, \ldots, $L-1$). Therefore, we are led to a set of degenerate ground states  $g_j|\Psi_0\rangle$,
which take the form $\sum _{i=3}^{L-1} (1-2\delta_{i\;j}) S_1^-S_i^-\ket{\psi_0} $, up to a multiplicative constant.
Again, each of $g_j$'s generates an emergent discrete symmetry group $Z_2$ tailored to $\vert\Psi_0\rangle$, because $g_j^2|\Psi_0\rangle= |\Psi_0\rangle$.
Hence, $S_1^-S_j^-\ket{\psi_0} $ ($ j=3,\ldots,L-1$) are a sequence of degenerate ground states under PBCs.
Similarly, if $|\Psi_0\rangle$ is chosen to be $\sum_{j=3}^{L}S_1^-S_j^-\ket{\psi_0} $ under OBCs, then we consider
a set of local unitary operations $g_j$, defined as $\exp (i \pi \Sigma_j)$, with $\Sigma_j =S_j^z$ ($j=3$, \ldots, $L$). We are thus led to a set of degenerate ground states  $g_j\vert\Psi_0\rangle$,
which take the form: $\sum _{i=3}^{L} (1-2\delta_{i\;j})S_1^-S_i^-\ket{\psi_0}$, up to a multiplicative constant.
Again, each of $g_j$'s generates an emergent discrete symmetry group $Z_2$ tailored to $\vert\Psi_0\rangle$, because $g_j^2|\Psi_0\rangle= |\Psi_0\rangle$.
Hence,  $S_1^-S_j^-\ket{\psi_0}$ ($ j=3,\ldots,L$) are a sequence of degenerate ground states under OBCs.
Indeed, they constitute generalized highest weight states $\vert\psi_2^{j_1j_2}\rangle \equiv S_{j_1}^-S_{j_2}^- \ket{\psi_0}$ at the second level, where  $j_1=1,2,\ldots,L$ and $j_2$ is not less than $j_1+2$, but less than $L$ for PBCs, and $j_1=1,2$, \ldots, $L$ and $j_2$ is not less than $j_1+2$  for OBCs, respectively.  Physically, these generalized highest weight states are relevant to low-lying  two-magnon excitations, in the sense that 
they constitute a set of basis states, invariant under
the translation operation under PBCs or the cyclic permutation symmetry
operation under OBCs, to yield two-magnon excitations.

At the $m$-th level, we are led to a sequence of  generalized highest weight states $|\psi_m^{j_1,j_2,\ldots,j_m}\rangle \equiv S_{j_1}^-S_{j_2}^- \ldots S_{j_m}^- \ket{\psi_0}$,  where $j_{\alpha+1}$ is not less than $j_{\alpha}+2$ ($\alpha =1,2,\ldots,m$), with $j_m$  not less than $i+2$, but less than $L$ for PBCs, and  $j_m$  not less than $i+2$ for OBCs.  Physically, these  generalized highest weight states are relevant to low-lying $m$-magnon excitations, in the sense that  they constitute a set of basis states, invariant under
the translation operation under PBCs or the cyclic permutation symmetry
operation under OBCs, to yield  $m$-magnon excitations.
In addition, there are  also some other degenerate ground states, which are an extension of  $|\psi_2^{1}\rangle \equiv \left( (S_1^-)^2+S_1^-S_2^-+S_1^-S_L^- \right) \ket{\psi_0}$ under PBCs
and $|\psi_2^{1}\rangle \equiv \left( (S_1^-)^2+S_1^-S_2^- \right) \ket{\psi_0}$ under OBCs at the second level to the $m$-th level. We remark that  the degenerate ground states  in this category  also proliferate as $L$ increases, in the sense that the total number is exponential in $L$.

Actually, we are able to repeat this procedure until the $L/2$-th level is reached, i.e., $m=0,1,2,\ldots, L/2$, if we are mainly interested in the construction of  generalized highest weight states $|\psi_m^{j_1,j_2,\ldots,j_m}\rangle $ (for an explanation of the maximum level $L/2$, cf. Section~\ref{grd}). As an illustrative example, we present the construction of  generalized highest weight states $|\psi_m^{j_1,j_2,j_3}\rangle$ at the third level in Appendix C.
Here we remark that an explicit form of the commutator $V$ between the Hamiltonian and a local unitary operation at the $m$-th level may be, in principle, constructed. Indeed, their explicit forms  at lower levels have been constructed in Ref.~\onlinecite{dimertrimer}.

In fact, $|\psi_m^{j_1,j_2,\ldots,j_m}\rangle $, as a generalized highest weight state at the $m$-th level, is a fully factorized ground state. Hence,
the number of  generalized highest weight states at the $m$-th level, with $m$ being around $L/4$, must be exponential in $L$. More precisely, the number of  generalized highest weight states as fully factorized ground states at the $m$-th level is $ C_{L-m+1}^m -m+1$ if PBCs are adopted and $ C_{L-m+1}^m$ if OBCs are adopted, where $C_{L-m+1}^m$ denote the binomial coefficients.
This is consistent with the counting of the total number of fully factorized ground states in terms of the transfer matrix technique, as described in Ref.~\onlinecite{dtmodel}. As a result, the total number of fully factorized ground states is exponential in $L$.  On the other hand, the time-reversal symmetry is spontaneously broken, which always accompanies the SSB pattern from the staggered ${\rm SU(3)} $ group to ${\rm U(1)}\times {\rm U(1)}$ group. Hence, the total number of generalized highest weight states at all levels are exponential in $L$.
Once all the generalized highest weight states are identified,  we are able to construct highly degenerate ground states from the repeated action of the two lowering operators, e.g., $F_1$ and $F_2$, on a generalized highest weight state at each level (cf. Appendix A for the explicit expressions of the generators of the staggered ${\rm SU}(3)$ symmetry group in terms of the spin-1 operators).
We are thus led to conclude that the ground state  degeneracies are exponential in $L$. However,  the ground state degeneracies under  PBCs and OBCs are different under PBCs and OBCs, with the difference being attributed to the dependence of generalized highest weight states on the boundary conditions adopted. In particular, the ground state  degeneracy under OBCs is always greater than the ground state  degeneracy under PBCs, since all the  generalized highest weight states under PBCs are still  generalized highest weight states under OBCs, but the converse is not true.

Alternatively, one may reproduce the ground state  degeneracies  under  PBCs and OBCs, as presented in Eq.(\ref{gsdsu3su4}), for the  staggered ${\rm SU}(3)$ ferromagnetic biquadratic model (\ref{hambq}) from the requirement that they asymptotically behave as a self-similar geometric object, as a consequence of the presence of an intrinsic abstract fractal underlying the ground state subspace~\cite{hqzhou}. Since the ground state degeneracies  under PBCs and OBCs are  regarded as a function of the system size $L$,
it is plausible to restrict to  a self-similar curve in the two-dimensional setting. Hence, a logarithmic spiral must be a natural candidate for such a self-similar curve.  This requirement, together with the  Binet formula~\cite{binet} that express an integer in terms of an irrational number,  lead to the conclusion that the golden spiral, with the inverse golden ratio $R=(\sqrt{5}-1)/2$, is the choice if the system size $L$ is large enough.

\subsubsection{The staggered ${\rm SU}(4)$ ferromagnetic spin-orbital model}~\label{su4emergentlocal}

For the staggered ${\rm SU}(4)$ ferromagnetic spin-orbital model (\ref{hamist}), the highest weight state is  $\ket{\psi_0} \equiv \ket{\otimes_{\eta=1}^{L} \{\uparrow_{s}\uparrow_{t}\}_{\eta}}$, which is labeled by eigenvalues $L/2$  and $L/2$  of $S^z$ and $T^z$.
If one chooses a degenerate ground state $|\Psi_0\rangle$ to be $S^-\ket{\psi_0}$, an eigenstate of $S^z$ and $T^z$ with eigenvalues being $L/2-1$ and $L/2$, respectively, then we may consider a set of local unitary operations $g_j$, defined as $\exp (i \pi \Sigma_j)$, with $\Sigma_j =S_j^z$ ($j=1$, \ldots, $L$), irrespective of the boundary conditions adopted. Following the above mathematical lemma, we are led to a set of degenerate ground states  $g_j|\Psi_0\rangle$,
which take the form: $\sum _{i=1}^L (1-2\delta_{i\;j}) 
S_i^-\ket{\psi_0}$, up to a multiplicative constant.
We remark that each of $g_j$'s generates an emergent discrete symmetry group $Z_2$ tailored to $\vert\Psi_0\rangle$, because $g_j^2 \vert\Psi_0\rangle= -\vert \Psi_0\rangle$, since a quantum state is determined up to an overall phase factor.
We are thus led to a sequence of degenerate ground states: $S^-_j \ket{\psi_0}$  ($j=1,2,\ldots,L$) , irrespective of the boundary conditions. Hence, they constitute  generalized highest weight states $|\psi_1^j\rangle \equiv S_j^- \vert\psi_0\rangle$  ($j=1,2,\ldots,L$) in the spin sector at the first level. Similarly, one may construct  generalized highest weight states $|\psi_1^j\rangle \equiv T_j^- \vert\psi_0\rangle$  ($j=1,2,\ldots,L$)  in the pseudo-spin sector  at the first level, as a result of the exchange symmetry between the spin operators $\textbf{S}_j$ and the pseudo-spin operators $\textbf{T}_j$. It is readily seen that this exchange symmetry is partially spontaneously broken, since both the highest weight state and the lowest weight state remain to be unbroken. Physically, these states  are relevant to low-lying one-magnon excitations, 
in the sense that they constitute a set of basis states, invariant under
the translation operation under PBCs or the cyclic permutation symmetry
operation under OBCs, to yield one-magnon excitations in the spin or pseudo-spin sector.

Now we turn to a generalized highest weight state, labeled by eigenvalues $L/2-2$ and $L/2$ of  $S^z$ and $T^z$, respectively.
 If $\vert\Psi_0\rangle$ is chosen to be $ S^- S_1^-\ket{\psi_0}$, then we consider a local unitary operation $g$, defined as  $\Sigma=S_1^z+S_2^z+S_L^z$ under PBCs and $\exp (i \pi \Sigma)$, with  $\Sigma=S_1^z+S_2^z$ under OBCs.
As follows from the lemma, we are led to a degenerate ground state  $g\vert\Psi_0\rangle$, which takes the form:
$(-S_1^-S_2^-+\sum_{j=3}^{L-1}S_1^-S_j^--S_1^-S_L^-)\ket{\psi_0}$ under PBCs and
$(-S_1^-S_2^-+\sum_{j=3}^{L}S_1^-S_j^-)\ket{\psi_0}$ under OBCs, up to a multiplicative constant. Again, $g$ generates an emergent discrete  symmetry group $Z_2$ tailored to $\vert\Psi_0\rangle$, because $g^2|\Psi_0\rangle= -\vert\Psi_0\rangle$, since a quantum state is determined up to an overall phase factor.
We are thus led to degenerate ground states: $(S_1^-S_2^-+S_1^-S_L^-)\ket{\psi_0}$ and $\sum_{j=3}^{L-1}S_1^-S_j^-\ket{\psi_0}$ under PBCs and
$-S_1^-S_2^-\ket{\psi_0}$ and
$\sum _{j=3}^{L} S_1^-S_j^-\ket{\psi_0}$ under OBCs.  If we choose  $|\Psi_0\rangle$ to be $(S_1^-S_2^-+S_1^-S_L^-)\ket{\psi_0}$ under PBCs, then a local unitary operation $g$, defined as  $\exp (i \pi \Sigma)$, where $\Sigma =S_2^z$ or $\Sigma =S_L^z$, leads us to conclude that
both $-S_1^-S_2^-\ket{\psi_0}$ and
$S_1^-S_L^-\ket{\psi_0}$ are degenerate ground states under PBCs.
Afterwards,
if $|\Psi_0\rangle$ is chosen to be $\sum _{j=3}^{L-1}S_1^-S_j^-\ket{\psi_0}$ under PBCs and  $\sum _{j=3}^{L} S_1^-S_j^-\ket{\psi_0}$ under OBCs, then we consider a set of local unitary operations $g_j$, defined as $\exp (i \pi \Sigma_j)$, where $\Sigma_j =S_j^z$,  with $j=3, \ldots, L-1$ under PBCs and  $j=3, \ldots, L$\; under OBCs.
As follows from the lemma, we are led to a set of degenerate ground states  $g_j|\Psi_0\rangle$,
which take the form: $\sum _{j=3}^{L-1} (1-2\delta_{i\;j}) S_1^-S_j^-\ket{\psi_0}$ under PBCs and
$\sum _{j=3}^{L} (1-2\delta_{i\;j}) S_1^-S_j^-\ket{\psi_0}$ under OBCs, up to a multiplicative constant.
Again, each of $g_j$'s generates an emergent discrete symmetry group $Z_2$ tailored to $|\Psi_0\rangle$, because $g_j^2|\Psi_0\rangle= -|\Psi_0\rangle$, since a quantum state is determined up to an overall phase factor.
We are thus led to degenerate ground states: $ S_1^-S_j^-\ket{\psi_0}$, with $j=2,\ldots,L$ under both PBCs and OBCs.
The same construction also works for  a generalized highest weight state, labeled by eigenvalues $L/2$ and $L/2-2$ of $S^z$ and $T^z$, respectively, as follows from the exchange symmetry between the spin operators $\textbf{S}_j$ and the pseudo-spin operators $\textbf{T}_j$.
We are thus led to degenerate ground states, which constitute generalized highest weight states at the second level: $|\psi_2^{j_1j_2}\rangle \equiv S_{j_1}^-S_{j_2}^- \ket{\psi_0}$ and $|\psi_2^{j_1 j_2}\rangle \equiv T_{j_1}^-T_{j_2}^- \ket{\psi_0}$, where $j_1=1,2$,\ldots, $L$ and $j_2$ is not less than $j_1+1$, irrespective of the boundary conditions adopted. Physically, these  generalized highest weight states  are relevant to low-lying two-magnon excitations, in the sense that they constitute a set of basis states, invariant under
the translation operation under PBCs or the cyclic permutation symmetry
operation under OBCs, to yield  two-magnon excitations in the spin or pseudo-spin sector.

At the $m$-th level, we are led to a sequence of  generalized highest weight states $|\psi_m^{j_1,j_2,\ldots,j_m}\rangle \equiv S_{j_1}^-S_{j_2}^- \ldots S_{j_m}^-\ket{\psi_0}$ and $|\psi_m^{j_1,j_2,\ldots,j_m}\rangle \equiv T_{j_1}^-T_{j_2}^- \ldots T_{j_m}^-\ket{\psi_0}$,  where $j_{\alpha+1}$ is not less than $j_{\alpha}+1$ ($\alpha =1,2,\ldots,m$), under both PBCs and OBCs. The number of such fully factorized ground states, which act as generalized highest weight states at level $m$, is $C^m_L$, where $C^m_L$ denote the binomial coefficients, irrespective of the boundary conditions. Note that we may only consider levels up to $L/2$, due to the time-reversal symmetry operation that maps the highest weight state and generalized highest weight states to the lowest weight states and generalized lowest weight states and vice versa (for a detailed explanation of the maximum level $L/2$ in this case, cf. Section~\ref{grd}). Hence, the total number of fully factorized ground states is $2^L$. Indeed, this result implies that the entire sector with the eigenvalue of $S^z$ or $T^z$ being $L/2$ is contained in the ground state subspace, which simply stems from the fact that the tensor product of any state in the spin sector with the fully polarized state $\ket{\otimes_{\eta=1}^{L} \{\uparrow_{t}\}_{\eta}}$ in the pseudo-spin sector or  the tensor product of any state in the spin sector with the fully polarized state $\ket{\otimes_{\eta=1}^{L} \{\uparrow_{s}\}_{\eta}}$ in the spin sector yields a degenerate ground state. Note that other sectors generated from the repeated action of the lowering operator $S^-$ or $T^-$ of ${\rm SU}(2)\times{\rm SU}(2)$  on  $\ket{\otimes_{\eta=1}^{L} \{\uparrow_{s}\}_{\eta}}$ or $\ket{\otimes_{\eta=1}^{L} \{\uparrow_{t}\}_{\eta}}$ are also contained in the ground state subspace. In Section~\ref{extrinsic}, we shall take advantage of this fact to establish the existence of integrable sectors in the ${\rm SO}(4)$ spin-orbital model.  Physically, these  generalized highest weight states are relevant to low-lying $m$-magnon excitations, in the sense that they constitute a set of basis states, invariant under
the translation operation under PBCs or the cyclic permutation symmetry
operation under OBCs, to yield  $m$-magnon excitations  in the spin or pseudo-spin sector.

One may also construct generalized highest weight states that involve both the spin and pseudo-spin sectors, labeled by the eigenvalues $L/2-1$ and $L/2-1$ of  $S^z$ and $T^z$, respectively.
If $|\Psi_0\rangle$ is chosen to be $T^- S_1^-\ket{\psi_0}$,
then we consider a local unitary operation $g$, defined as $\exp (i \pi \Sigma)$, with $\Sigma=T_1^z+T_2^z+T_L^z$ under PBCs or $\Sigma=T_1^z+T_2^z$ under OBCs.
As follows from the lemma, we are led to a degenerate ground state  $g\vert\psi_0\rangle$, which takes the form: $ |\psi_2^{1}\rangle \equiv (S_1^-T_1^-+S_1^-T_2^--\sum_{i=3}^LS_1^-T_i^-)\ket{\psi_0}$ under OBCs and
 $|\psi_2^{1}\rangle \equiv (S_1^-T_1^-+S_1^-T_2^--\sum_{i=3}^{L-1}S_1^-T_i^-+S_1^-T_L^-)\ket{\psi_0}$ under PBCs, up to a multiplicative constant. Again, $g$ generates an emergent discrete  symmetry group $Z_2$ tailored to $\vert\Psi_0\rangle$, because $g^2|\Psi_0\rangle= -\vert\Psi_0\rangle$ under PBCs or $g^2|\Psi_0\rangle= \vert\Psi_0\rangle$ under OBCs, since a quantum state is determined up to an overall phase factor.
We are thus led to degenerate ground states: $(S_1^-T_1^-+S_1^-T_2^-+S_1^-T_L^-)\ket{\psi_0}$ and
$ \sum_{i=3}^{L-1}S_1^-T_i^-\ket{\psi_0}$ under PBCs and
$ (S_1^-T_1^-+S_1^-T_2^-)\ket{\psi_0}$ and $\sum _{i=3}^{L} S_1^-T_i^-\ket{\psi_0}$ under OBCs.
Afterwards,  if $\vert\Psi_0\rangle$ is chosen to be $ \sum _{i=3}^{L-1}  S_1^-T_i^-\ket{\psi_0}$  under PBCs, then we consider
a set of local unitary operations $g_j$, defined as $\exp (i \pi \Sigma_j)$, with $\Sigma_j = T_j^z$ ($j=3$, \ldots, $L-1$). Therefore, we are led to a set of degenerate ground states  $g_j|\Psi_0\rangle$,
which take the form $\sum _{i=3}^{L-1} (1-2\delta_{i\;j}) S_1^-T_i^-\ket{\psi_0}$, up to a multiplicative constant.
Again, each of $g_j$'s generates an emergent discrete symmetry group $Z_2$ tailored to $\vert\Psi_0\rangle$, because $g_j^2|\Psi_0\rangle= -|\Psi_0\rangle$,  since a quantum state is determined up to an overall phase factor. We are thus led to degenerate ground states: $S_1^-T_i^-\ket{\psi_0}$ ($ i=3,4,\ldots,L-1$) under PBCs. Meanwhile,
if $|\Psi_0\rangle$ is chosen to be $\sum _{i=3}^{L} S_1^-T_i^-\ket{\psi_0}$ under OBCs, then we consider
a set of local unitary operations $g_j$, defined as $\exp (i \pi \Sigma_j)$, with $\Sigma_j =T_j^z$ ($j=3$, \ldots, $L$). We are therefore led to a set of degenerate ground states  $g_j\vert\Psi_0\rangle$,
which take the form: $\sum _{i=3}^{L} (1-2\delta_{i\;j}) S_1^-T_i^-\ket{\psi_0}$, up to a multiplicative constant.
Again, each of $g_j$'s generates an emergent discrete symmetry group $Z_2$ tailored to $\vert\Psi_0\rangle$, because $g_j^2|\Psi_0\rangle= -|\Psi_0\rangle$, since a quantum state is determined up to an overall phase factor.
We are thus led to degenerate ground states:
$S_1^-T_i^-\ket{\psi_0}$ ($ i=3,4,\ldots,L$) under OBCs.
Similarly, one may swap the roles of $S^-_j$ and $T^-_j$ in the above discussion. Indeed, they constitute generalized highest weight states $|\psi_2^{ij}\rangle \equiv S_{i}^-T_{j}^- \ket{\psi_0}$ at the second level, where  $i=1,2,\ldots,L$ and $j$ is not less than $i+2$, but less than $L$ under PBCs, and $i=1,2$, \ldots, $L$ and $j$ is not less than $i+2$ under OBCs, respectively. The number of generalized highest weight states in this form at the second level is $L(L-3)$ under PBCs and $(L-1)(L-2)$ under OBCs.
This explains why the ground state degeneracies are different under OBCs and PBCs.
Physically, these generalized highest weight states are relevant to low-lying two-magnon excitations, in the sense that 
they constitute a set of basis states, invariant under
the translation operation under PBCs or the cyclic permutation symmetry
operation under OBCs, to yield  two-magnon excitations, with one in the spin sector and the other in the pseudo-spin sector, respectively.

At the $m$-th level, we are led to a sequence of  generalized highest weight states $|\psi_m^{i_1,i_2,\ldots,i_{m_s};j_1,j_2,\ldots,j_{m_t}}\rangle \equiv S_{i_1}^-S_{i_2}^- \ldots S_{i_{m_s}}^-  T_{j_1}^-T_{j_2}^- \ldots T_{j_{m_t}}^-\ket{\psi_0}$, where $m= m_s+m_t$,  $i_1 <i_2<\ldots< i_{m_s}$, $j_1 <j_2<\ldots< j_{m_t}$,  and $i_{\alpha}$ is not less than $j_{\beta}+2$, but not more than $j_{\beta+1}-2$ and vice versa ($\alpha =1,2,\ldots,m_s$ and $\beta =1,2,\ldots,m_t$). The number of generalized highest weight states $|\psi_m^{i_1,i_2,\ldots,i_{m_s};j_1,j_2,\ldots,j_{m_t}}\rangle$ may be counted and it is exponential if $m$ is around $L/2$. In particular, if one restricts  to $1 < i_1 <i_2<\ldots< i_{m_s} < L/2$ and $L/2 < j_1 <j_2<\ldots< j_{m_t} < L$, then the total number of  generalized highest weight states $|\psi_m^{i_1,i_2,\ldots,i_{m_s};j_1,j_2,\ldots,j_{m_t}}\rangle$ in this form is $2^{L-2}$.
Physically, these  generalized highest weight states $|\psi_m^{i_1,i_2,\ldots,i_{m_s};j_1,j_2,\ldots,j_{m_t}}\rangle$ are relevant to low-lying $m$-magnon excitations, in the sense that they constitute a set of basis states, invariant under
the translation operation under PBCs or the cyclic permutation symmetry
operation under OBCs, to yield  $m$-magnon excitations, with $m_s$-magnon excitations in the spin sector and $m_t$-magnon excitations in the pseudo-spin sector, respectively.
In addition, there are also other degenerate ground states, which are an extension of $|\psi_2^{1}\rangle \equiv (S_1^-T_1^-+S_1^-T_2^-+S_1^-T_L^-)\ket{\psi_0}$ under PBC and
$|\psi_2^{1}\rangle \equiv (S_1^-T_1^-+S_1^-T_2^-)\ket{\psi_0}$ under OBC at the second level to the $m$-th level.   The degenerate ground states  in this category  also proliferate as $L$ increases, in the sense that the total number is exponential in $L$.

In principle, we are able to repeat this procedure until the $L$-th level is reached, thus exhausting all the possible generalized highest weight states at level $m$ ($m=0,1,2,\ldots, L$) (for a detailed explanation of the maximum level $L$ in this case, cf. Section~\ref{grd}). Once all the generalized highest weight states are identified,  we are able to construct highly degenerate ground states from the repeated action of the three lowering operators, e.g., $F_1$, $F_2$ and $F_3$, on a generalized highest weight state at each level (cf. Appendix A for the explicit expressions of the generators of the staggered ${\rm SU}(4)$ symmetry group in terms of the spin-$/2$  and pseudo-spin-$/2$ operators).
In fact, generalized highest weight states $|\psi_m^{j_1,j_2,\ldots,j_m}\rangle $ and $|\psi_m^{i_1,i_2,\ldots,i_{m_s};j_1,j_2,\ldots,j_{m_t}}\rangle$ at the $m$-th level are fully factorized ground states. One may count the total number of fully factorized ground states in terms of the transfer matrix technique, as described in Ref.~\onlinecite{dtmodel}. As it turns out, the total number of fully factorized ground states is exponential in $L$. Note that the time-reversal symmetry  and the exchange symmetry between the spin operators $\textbf{S}_j$ and the pseudo-spin operators $\textbf{T}_j$ are (partially) broken, which always accompany the SSB pattern from the staggered ${\rm SU(4)} $ to ${\rm U(1)}\times{\rm U(1)}\times {\rm U(1)}$ via ${\rm U(1)}\times{\rm U(1)}\times {\rm SU(2)}$. Combining with this observation, we see that the total number of (linearly independent) generalized highest weight states are exponential in $L$.  Along the same reasoning as that for  the  staggered ${\rm SU}(3)$ ferromagnetic biquadratic model, we are thus led to conclude that the ground state  degeneracies under both PBCs and OBCs are exponential in $L$ under both PBCs and OBCs, but they depend on the boundary conditions adopted.

Alternatively, one may reproduce the ground state  degeneracies  under  PBCs and OBCs, as presented in Eq.(\ref{gsdsu3su4}), for the  staggered ${\rm SU}(4)$ ferromagnetic spin-orbital model (\ref{hamist}) from the requirement that they must be asymptotically behave as a self-similar geometric object. This requirement, together with the  Binet formula~\cite{binet} that express an integer in terms of an irrational number, lead us to conclude that the logarithmic spiral, with $R=\sqrt{2}(\! \sqrt{3}-\!1)/2$,  is the choice if the system size $L$ is large enough.  Note that different values of the number of type-B GMs, as a result of different SSB patterns, only affect the base of a self-similar logarithmic spiral.

\subsection{From exponential ground state degeneracies to emergent Goldstone flat bands}

For a quantum many-body spin system undergoing SSB from $G$ to $H$ with type-B GMs, if the  ground state degeneracies are exponential in $L$, but different under PBCs and OBCs,
then Goldstone flat bands emerge, indicating trivial dynamics underlying the low-lying excitations.
Note that if the symmetry group is semi-simple, modulo a discrete  symmetry subgroup, then there is {\it only} one highest weight state, from which a sequence of degenerate ground states may be generated through the repeated action of the lowering operator(s). These degenerate ground states constitute an orthonormal basis, which span an irreducible representation space of the symmetry group, with the dimension being polynomial in $L$, irrespective of the boundary conditions.
Hence, if the  ground state degeneracies under PBCs and OBCs are exponential in $L$, then there must be generalized highest weight states at distinct levels, labeled by $m$, which in turn are relevant to low-lying $m$-magnon excitations.
As a consequence,   generalized highest weight states at different levels, combining with the  translation operation $\tau$ under PBCs or the cyclic permutation symmetry
operation $\sigma$ under OBCs,   give rise to emergent Goldstone flat bands. 

Given the ground state  degeneracies  under  PBCs and OBCs are different, it is convenient to focus on those degenerate ground states that are left intact as the boundary conditions vary from PBCs to OBCs. As already stressed in Section~~\ref{grd}, all degenerate ground states under PBCs are also degenerate ground states under OBCs,  due to the frustration-free nature of the  model Hamiltonians for any quantum many-body spin systems undergoing SSB with type-B GMs~\cite{FMGM,hqzhou}. Hence, we may take advantage of this fact to restrict ourselves to degenerate ground states under PBCs . In fact, those states that become degenerate ground states {\it only} under OBCs could be dealt with as a sideline, if necessary.
Specifically, we shall take $|\psi_m^{j_1,\ldots,j_{m}}\rangle$ for the staggered ${\rm SU}(3)$ ferromagnetic biquadratic model and $|\psi_m^{j_1,\ldots,j_{m}}\rangle$ in the spin or pseudo-spin sector and  $|\psi_m^{i_1,i_2,\ldots,i_{m_s};j_1,j_2,\ldots,j_{m_t}}\rangle$ in the spin and pseudo-spin sectors for the staggered ${\rm SU}(4)$ ferromagnetic spin-orbital model to be identical under PBCs and OBCs, when we discuss simultaneous eigenstates of the respective Hamiltonian and the  translation operation under PBCs or the cyclic permutation symmetry operation under OBCs.

\subsubsection{The staggered ${\rm SU}(3)$ ferromagnetic biquadratic model}

For the staggered ${\rm SU}(3)$ ferromagnetic biquadratic model (\ref{hambq}), the  staggered ${\rm SU}(3)$ symmetry group contains the uniform ${\rm SU}(2)$ group, generated by  the spin operators $S^{x},S^{y}$ and $S^{z}$, as a subgroup. If one uses the eigenvalue of  the Cartan generator $S^z$ of the uniform ${\rm SU}(2)$ group to label highly degenerate ground states, then it is necessary to introduce generalized highest weight states, given the ground state degeneracies under PBCs and OBCs are exponential in $L$. Mathematically, this is due to the fact that the staggered ${\rm SU}(3)$  symmetry group is simple, and it shares the same highest weight state $\ket{\psi_0}$  as the subgroup ${\rm SU}(2)$, so the number of degenerate ground states generated from the repeated action of  the lowering operators, e.g., $F_1$ and $F_2$, on the highest weight state $\ket{\psi_0}$ must be quadratic in $L$. Indeed,  degenerate ground states generated from the  highest weight state $\ket{\psi_0}$ span an irreducible representation of the staggered ${\rm SU}(3)$ symmetry group, with the dimension of the representation space being quadratic in $L$. Hence, it is the presence of generalized highest weight states at different levels that are responsible for the exponential ground state degeneracies under both PBCs and OBCs. Here, we restrict ourselves to levels up to $L/2$, partially as a result of the time-reversal symmetry (for an explanation of the maximum level
$L/2$, cf. Section~\ref{grd}). As such, the number of  generalized highest weight states at all the different levels is exponential in $L$. This in turn implies that there must be exponentially many generalized highest weight states at certain levels. Indeed, as already mentioned in Subsection~\ref{emergenttoexponential}, the number of  generalized highest weight states $|\psi_m^{j_1,\ldots,j_{m}}\rangle$, which act as factorized ground states, at the $m$-th level, is
$ C_{L-m+1}^m -m+1$ under PBCs and $ C_{L-m+1}^m$ under OBCs.

Taking advantage of the  translation operation $\tau$ under PBCs or the cyclic permutation symmetry operation  $\sigma$ under OBCs,
we may define  $|\phi_{m;j_1,\ldots,j_{m}}(k)\rangle=\sum_{l=0}^{L-1}\exp(ik l) \tau ^l|\psi_m^{j_1,\ldots,j_{m}}\rangle$ or
$|\phi_{m;j_1,\ldots,j_{m}}(k)\rangle=\sum_{l=0}^{L-1}\exp(ik l) \sigma ^l|\psi_m^{j_1,\ldots,j_{m}}\rangle$
 such that $\tau  |\phi_{m;j_1,\ldots,j_{m}}(k)\rangle=\exp(ik) |\phi_{m;j_1,\ldots,j_{m}}(k)\rangle$ or
  $\sigma  |\phi_{m;j_1,\ldots,j_{m}}(k)\rangle=\exp(ik) |\phi_{m;j_1,\ldots,j_{m}}(k)\rangle$.
Here, $k$ is required to satisfy $\exp(ikL)=1$. We thus have $k=2\pi \delta/L$ ($\delta=0,1,\ldots,L-1$). In other words,  $|\phi_{m;j_1,\ldots,j_{m}}(k)\rangle$  ($m=0,1,2,\ldots, L/2$), usually {\it only} a constituent of low-lying $m$-magnon excitations, accidentally become a sequence of degenerate ground states, since they are simultaneous eigenstates of  the Hamiltonian (\ref{hambq}) and  $\tau$ under PBCs or the Hamiltonian (\ref{hambq}) and $\sigma$ under OBCs. Here, by ``accidentally" we mean
no symmetry is available to account for such a degeneracy, given that the symmetry group, as a simple Lie group, only yield  degeneracy that is polynomial in $L$. This in turn implies that  $E(k)=E_0$, where $E_0 = L$ under PBCs and $E_0 = L-1$ under OBCs.  We are thus led to the conclusion that the excitation energy is zero: $\omega (k) = 0$ for any $k \in (0, 2\pi)$ in the thermodynamic limit, valid for any low-lying  excitations.

It is convenient to express low-lying excitations in terms of the spin operators explicitly at different levels. For this purpose, we focus on generalized highest  weight states under PBCs, since all of them are  also  generalized highest  weight states under OBCs.  We start from the low-lying excitations at the first level. Given $|\psi_1^j\rangle \equiv S_j^- \vert\psi_0\rangle$  ($j=1,2,\ldots,L$), the one-magnon states may be written as $\sum _j e^{ikj} \vert\psi_1^j\rangle$.   At the second level,  given $|\psi_2^{j_1j_2}\rangle \equiv S_{j_1}^-S_{j_2}^- \ket{\psi_0}$ ($j_1=1,2,\ldots,L$ and $j_2$ is not less than $j_1+2$), it is readily seen that $\sum _j e^{ikj} |\psi_2^{j j+\chi} \rangle$  for  any fixed $\chi$, which is not less than $2$,  describe  spin-2 low-lying excitations relevant to two-magnon excitations. In addition,  since $|\psi_2^{1}\rangle \equiv \left( (S_1^-)^2+S_1^-S_2^-+S_1^-S_L^- \right) \ket{\psi_0}$ is also a  degenerate ground state under PBCs, though it is not factorized, we define  $|\psi_2^{l}\rangle \equiv \tau ^l |\psi_2^{1}\rangle$ ($l=1, \ldots,L$) and introduce a state $\sum _l e^{ikl} |\psi_2^{l}\rangle$, which is constructed to be degenerate with the ferromagnetic ground state - the highest weight state  
$\ket{\psi_0} = \ket{\otimes_{\eta=1}^{L} \{+\}_{\;\eta}}$.  From this we are led to conclude that
$\sum _j  (e^{ikj} \left(S_j^-)^2 +  2e^{ik(j+1/2)} \cos (k/2) S_j^- S_{j+1}^- \right)\ket{\psi_0}$  for any values of $k$ describe a spin-2 low-lying excitations relevant to two-magnon excitations.

The emergent Goldstone bands are therefore flat, with a trivial dispersion relation, in the sense that they are actually dispersionless low-lying excitations relevant to $m$-magnons, but accidentally become degenerate ground states. Note that the total number of emergent Goldstone flat bands, as constructed from generalized highest weight states  $|\phi_{m;j_1,\ldots,j_{m}}(k)\rangle$,  is exponential in $L$. Here, we remark that $|\phi_{m;j_1,\ldots,j_{m}}(k)\rangle$ is not periodic, as already mentioned in Section~\ref{grd}. In contrast, the total number of emergent Goldstone flat bands, as constructed from generalized highest weight states  with all possible distinct periods, is polynomial in $L$. We remark that the entanglement entropy for degenerate ground states thus constructed obeys the sub-volume 
law~\cite{goldensu3}.

\subsubsection{The staggered ${\rm SU}(4)$ ferromagnetic spin-orbital model}

For the staggered ${\rm SU}(4)$ ferromagnetic spin-orbital model (\ref{hamist}), the staggered ${\rm SU}(4)$ symmetry group contains the uniform ${\rm SU}(2)\times{\rm SU}(2)$ group, generated by  the spin operators $S^{x},S^{y}$ and $S^{z}$ and the pseudo-spin operators $T^{x},T^{y}$ and $T^{z}$, as a subgroup.  If one uses the eigenvalues of  $S^z$ and $T^z$  to label highly degenerate ground states, then it is necessary to
introduce generalized highest weight states, given the ground state degeneracies under PBCs and OBCs are exponential in 
$L$. Mathematically, this is due to the fact that the staggered ${\rm SU}(4)$ symmetry group  is simple, and it shares the same highest weight state $\ket{\psi_0}$ as the subgroup ${\rm SU}(2)\times {\rm SU}(2)$, so the number of degenerate ground states generated from the repeated action of the lowering operators, e.g., $F_1$, $F_2$ and $F_3$,  on the  highest weight state $\ket{\psi_0}$ must be cubic in $L$. Hence, it is the presence of generalized highest weight states at different levels that are responsible for the exponential ground state degeneracies under both PBCs and OBCs. Here, we  restrict ourselves to 
levels up to $L$, as a result of the time-reversal symmetry and the exchange symmetry between the spin operators $\textbf{S}_j$ and the pseudo-spin operators $\textbf{T}_j$. As such, the number of  generalized highest weight states at all the different levels is exponential in $L$.
This in turn implies that there must be exponentially many generalized highest weight states at certain levels in the spin or pseudo-spin sector as well as the spin and pseudo-spin sectors.  
Indeed, as already mentioned in Subsection~\ref{emergenttoexponential}, the number of  generalized highest weight states $|\psi_m^{j_1,\ldots,j_{m}}\rangle$ in the spin or pseudo-spin sector at the $m$-th level is $ C_{L}^m$, irrespective of the boundary conditions. Meanwhile, the total number of generalized highest weight states $|\psi_m^{i_1,i_2,\ldots,i_{m_s};j_1,j_2,\ldots,j_{m_t}}\rangle$ in  this form is $2^{L-2}$,
if one restricts  to  $1 < i_1 <i_2<\ldots< i_{m_s} < L/2$ and $L/2 < j_1 <j_2<\ldots< j_{m_t} < L$.

Taking advantage of the translation operation $\tau$ under PBCs or the cyclic permutation symmetry operation  $\sigma$ under OBCs,
we may define  $|\phi_{m;j_1,\ldots,j_{m}}(k)\rangle=\sum_{l=0}^{L-1}\exp(ik l) \tau ^l|\psi_m^{j_1,\ldots,j_{m}}\rangle$ or
$|\phi_{m;j_1,\ldots,j_{m}}(k)\rangle=\sum_{l=0}^{L-1}\exp(ikl) \sigma ^l|\psi_m^{j_1,\ldots,j_{m}}\rangle$
such that $\tau  |\phi_{m;j_1,\ldots,j_{m}}(k)\rangle=\exp(ik) |\phi_{m;j_1,\ldots,j_{m}}(k)\rangle$ or
$\sigma  |\phi_{m;j_1,\ldots,j_{m}}(k)\rangle=\exp(ik) |\phi_{m;j_1,\ldots,j_{m}}(k)\rangle$ in the spin or pseudo-spin sector.
Meanwhile, we may define $|\phi_{m;{i_1,i_2,\ldots,i_{m_s};j_1,j_2,\ldots,j_{m_t}}}(k)\rangle=\sum_{l=0}^{L-1}\exp(ik l) \tau ^l |\psi_m^{i_1,i_2,\ldots,i_{m_s};j_1,j_2,\ldots,j_{m_t}}\rangle$ under PBCs or  $|\phi_{m;{i_1,i_2,\ldots,i_{m_s};j_1,j_2,\ldots,j_{m_t}}}(k)\rangle=\sum_{l=0}^{L-1}\exp(ik l) \sigma ^l |\psi_m^{i_1,i_2,\ldots,i_{m_s};j_1,j_2,\ldots,j_{m_t}}\rangle$ under OBCs in the spin and pseudo-spin sectors.
Here, $k$ is required to satisfy $\exp(ikL)=1$. We thus have $k=2\pi \delta/L$ ($\delta=0,1,\ldots,L-1$). In other words,  $|\phi_{m;j_1,\ldots,j_{m}}(k)\rangle$ and  $|\phi_{m;{i_1,i_2,\ldots,i_{m_s};j_1,j_2,\ldots,j_{m_t}}}(k)\rangle$ are a sequence of degenerate ground states, which are simultaneous eigenstates of  the Hamiltonian (\ref{hambq}) and  $\tau$ under PBCs or the Hamiltonian (\ref{hambq}) and $\sigma$ under OBCs.
This in turn implies that  $E(k)=E_0$, where $E_0=0$, irrespective of the boundary conditions.  Mathematically, the excitation energy $\omega (k) = 0$ for any $k \in (0, 2\pi)$ in the thermodynamic limit, valid for any low-lying  excitations in the spin or pseudo-spin 
sector and in the spin and pseudo-spin 
sectors, respectively.

It is convenient to express low-lying excitations in the spin or pseudo-spin sector and  in the spin and pseudo-spin sectors in terms of the spin and pseudo-spin  operators explicitly at different levels. For this purpose, we focus on generalized highest  weight states under PBCs, since all of them are  also  generalized highest  weight states under OBCs.  We start from the low-lying excitations in the spin sector at the first level. Given $\vert\psi_1^j\rangle \equiv S_j^- \ket{\psi_0}$ ($j=1,2,\ldots,L$), the low-lying states  in the spin sector may be written as $\sum _j e^{ikj} \vert\psi_1^j\rangle$. Meanwhile, the low-lying states in the pseudo-spin sector follow from the exchange symmetry between the spin operators $\textbf{S}_j$ and the pseudo-spin operators $\textbf{T}_j$.
At the second level,  given $|\psi_2^{j_1j_2}\rangle \equiv S_{j_1}^-S_{j_2}^- \ket{\psi_0}$ ($j_1=1,2,\ldots,L$ and $j_2$ is not less than $j_1+1$), it is readily seen that $\sum _j e^{ikj} |\psi_2^{j j+\chi} \rangle$  for  any fixed $\chi$, which is not less than $1$,  describe  spin-2 low-lying states relevant to two-magnon excitations in the spin sector. Meanwhile, pseudo-spin-2 low-lying states relevant to two-magnon excitations in the pseudo-spin sector follow from the exchange symmetry between  the spin operators $\textbf{S}_j$ and the pseudo-spin operators $\textbf{T}_j$.
Moreover, $|\psi_2^{ij}\rangle \equiv S_i^-T_j^- \ket{\psi_0}$ are relevant to such two-magnon excitations at the second level, with one in the spin sector and the other in the pseudo-spin sector, where  $i=1,2,\ldots,L$ and $j$ is not less than $i+2$, but less than $L$ under PBCs. Hence, $\sum _j e^{ikj} |\psi_2^{j j+\chi} \rangle$  for  any fixed $\chi$, which is not less than $2$,  describe
low-lying states relevant to such two-magnon excitations.
In addition,  since $|\psi_2^{1}\rangle \equiv \left( S_1^-T_1^-+S_1^-T_2^-+S_1^-T_L^- \right) \ket{\psi_0}$ is also a degenerate ground state under PBCs, one may define  $|\psi_2^{l}\rangle \equiv \tau ^l |\psi_2^{1}\rangle$ ($l=1, \ldots,L$) and introduce a state $\sum _l e^{ikl} |\psi_2^{l}\rangle$, which is constructed to be degenerate with the ferromagnetic ground state - the highest weight state 
$\ket {\psi_0} \equiv \ket{\psi_0}$.  From this we are led to conclude that
$\sum _j  \left(e^{ikj} S_j^- T_j^-+  2e^{ik(j+1/2)} \cos (k/2) S_j^- T_{j+1}^- \right)\ket{\psi_0}$  for any values of $k$ describe  low-lying states relevant to two-magnon excitations, with one in the spin sector and the other in the pseudo-spin sector.

The emergent Goldstone bands are therefore flat, with a trivial dispersion relation, in the sense that they are actually dispersionless low-lying excitations relevant to $m$-magnon excitations in the spin or pseudo-spin sector and  in the spin and pseudo-spin sectors, but accidentally become degenerate ground states. Note that the total number of emergent Goldstone flat bands, as constructed from generalized highest weight states $|\psi_m^{j_1,\ldots,j_{m}}\rangle$ in the spin or pseudo-spin sector and generalized highest weight states $|\psi_m^{i_1,i_2,\ldots,i_{m_s};j_1,j_2,\ldots,j_{m_t}}\rangle$  in the spin and pseudo-spin sectors,  is exponential in $L$. 
In contrast, the total number of emergent Goldstone flat bands, as constructed from generalized highest weight states  with all possible distinct periods is polynomial in $L$.  Note that the entanglement entropy for degenerate ground states constructed from the repeated action of the lowering operators on the highest weight state and  generalized highest weight states  with distinct periods obeys the sub-volume law~\cite{spinorbitalsu4}.

\subsection{From emergent Goldstone flat bands to emergent (local) symmetry operations}

For a quantum many-body spin system undergoing SSB from $G$ to $H$ with type-B GMs, the presence of emergent Goldstone flat bands implies that  there exists an emergent (local) symmetry operation tailored to a specific degenerate ground state. Here, the broken symmetry group $G$ is assumed to be semisimple, modulo a discrete symmetry subgroup. The  explicit construction of  emergent (local) symmetry operations tailored to degenerate ground states at different levels is sketched for the staggered  ${\rm SU}(3)$ spin-1 ferromagnetic biquadratic model and the staggered ${\rm SU}(4)$ ferromagnetic spin-orbital model. Again, this construction also works for other quantum many-body spin systems undergoing SSB with type-B GMs, as long as the symmetry group $G$ contains  a uniform simple or semi-simple subgroup, such as the uniform ${\rm SU(2)}$ group or the uniform ${\rm SU(2)} \times {\rm SU(2)}$ group. In this sense, an emergent (local)  symmetry operation
may be regarded as a (local) unitary operation dynamically generated from the low-lying excitations for quantum many-body spin systems undergoing SSB with type-B GMs, with the staggered ${\rm SU}(3)$ spin-1 ferromagnetic biquadratic model and the staggered ${\rm SU}(4)$ ferromagnetic spin-orbital model being the two simplest examples.

\subsubsection{The staggered ${\rm SU}(3)$ spin-1 ferromagnetic biquadratic model}

Physically, emergent Goldstone flat bands are actually  dispersionless low-lying excitations relevant to multi-magnon excitations, but accidentally become degenerate ground states. As already stressed, it is sufficient to restrict to levels $m=0,1,2,\ldots,L/2$. Mathematically, the low-lying excitations relevant to the $m$-magnon excitations, denoted as $|\phi_{m;j_1,\ldots,j_{m}}(k)\rangle$, should
transform as $\tau  |\phi_{m;j_1,\ldots,j_{m}}(k)\rangle=\exp(ik) |\phi_{m;j_1,\ldots,j_{m}}(k)\rangle$  under the action of the translation operation $\tau$ if PBCs are adopted  or $\sigma  |\phi_{m;j_1,\ldots,j_{m}}(k)\rangle=\exp(ik) |\phi_{m;j_1,\ldots,j_{m}}(k)\rangle$  under the action of the cyclic permutation symmetry operation $\sigma$ if OBCs are adopted. Hence, $|\phi_{m;j_1,\ldots,j_{m}}(k)\rangle$ may be expressed in terms of $|\psi_m^{j_1,\ldots,j_{m}}\rangle$: $|\phi_{m;j_1,\ldots,j_{m}}^\delta(k)\rangle=\sum_{l=0}^{L-1}\exp(ik\delta l) \tau^l |\psi_m^{j_1,\ldots,j_{m}}\rangle$
under PBCs or $|\phi_{m;j_1,\ldots,j_{m}}(k)\rangle=\sum_{l=0}^{L-1}\exp(ikl) \sigma^l |\psi_m^{j_1,\ldots,j_{m}}\rangle$ under OBCs, in order to ensure that they are dispersionless: $\omega(k)=0$. They thus accidentally become degenerate ground states: $E(k)=E_0$, with $k=2\pi \delta/L$ ($\delta=0,1,\ldots,L-1$).
This in turn implies that  $|\psi_m^{i,j_1,\ldots,j_{m-1}}\rangle$ are fully factorized ground states, which may be recognized as generalized highest weight states at level $m$.

In particular, at the first level, we are led to conclude that  $|\psi_1^j\rangle \equiv S_j^- \ket{\psi_0}$  ($j=1,2,\ldots,L$) are degenerate (fully factorized) ground states.
There exists therefore an emergent local symmetry operation $g \in Z_2$,  defined as $\exp (i \pi \Sigma_j)$, with $\Sigma_j =S_j^z$ ($j=1$, \ldots, $L$),
such that  $g|\Psi_0\rangle$ is a degenerate ground state, as long as $|\Psi_0\rangle$ is chosen to be $S^-\ket{\psi_0}$ at the first level. Obviously, we have $|\langle \Psi_0|g|\Psi_0\rangle|\neq 1$.  Meanwhile, at the second level, $|\psi_2^{j_1j_2}\rangle \equiv S_{j_1}^-S_{j_2}^- \ket{\psi_0}$, with $j_1=1$, \ldots, $L$ and $j_2$ not less than $j_1+2$, but less than $L$ under PBCs and $j_1=1,2,\ldots,L$ and $j_2$ not less than $j_1+2$ under OBCs, are degenerate factorized ground states.   We remark that, at the second level, a degenerate ground state may be generated from the  action of $S^-$ on the highest weight state twice. That is, we have $(S^-)^2\ket{\psi_0}$ as a degenerate ground state. Combining with
$|\psi_2^{j_1j_2}\rangle$, we see that $|\psi_2^{1}\rangle \equiv \left( (S_1^-)^2+S_1^-S_2^-+S_1^-S_L^- \right) \ket{\psi_0}$ and $\sum_{j=3}^{L-1}S_1^-S_j^-\ket{\psi_0}$  under PBCs and $|\psi_2^{1}\rangle \equiv  \left( (S_1^-)^2+S_1^-S_2^- \right)\ket{\psi_0}$ and $\sum _{j=3}^{L} S_1^-S_j^-\ket{\psi_0}$ under OBCs are degenerate ground states. As follows from the converse form of the lemma from Section~\ref{grd},  this amounts to stating that a local unitary operation $g$, defined as  $\exp (i \pi \Sigma)$, with $\Sigma=S_1^z+S_2^z+S_L^z$ under PBCs and  $\Sigma=S_1^z+S_2^z$ under OBCs, is an emergent local symmetry operation tailored to a degenerate ground state $\vert\Psi_0\rangle$, which is chosen to be $S^-\vert 0_1+_2 \ldots +_L\rangle$. Again, we have $|\langle \Psi_0|g|\Psi_0\rangle|\neq 1$.
This argument may be extended to any level $m$, up to level $L/2$.

As a consequence, the presence of emergent Goldstone flat bands implies that  there exists an emergent (local) symmetry operation tailored to a specific degenerate ground state.  Generically, the existence of an emergent (local) symmetry operation tailored to such a specific degenerate ground state is guaranteed, if emergent Goldstone flat bands are present, as follows from the restatement of Elitzur's theorem, given that the model Hamiltonian (\ref{hambq}) is not invariant under a local $Z_2$ gauge transformation.
We remark that the total number of  emergent Goldstone flat bands, as constructed from generalized highest weight states $|\psi_m^{j_1,j_2,\ldots,j_m}\rangle$,  is exponential in $L$. 

\subsubsection{The staggered ${\rm SU}(4)$ ferromagnetic spin-orbital model}

There are two types of emergent Goldstone flat bands, which are actually dispersionless low-lying excitations relevant to the $m$-magnon excitations solely in the spin or pseudo-spin sector ($m=0,1,2,\ldots, L/2$) or dispersionless low-lying excitations relevant to the $m$-magnon  excitations in  the spin and pseudo-spin  sectors, where $m=m_s+m_t$ ($m=0,1,2,\ldots, L$, $m_s=0,1,2,\ldots, L/2$ and $m_t=0,1,2,\ldots, L/2$), with $m_s$-magnon excitations in the spin  sector and $m_t$-magnon excitations in the pseudo-spin sector. However, they accidentally become degenerate ground states.
Here we first focus on dispersionless low-lying excitations relevant to the $m$-magnon  excitations solely in the spin or pseudo-spin sector, and then move to dispersionless low-lying excitations relevant to the $m$-magnon excitations in the spin and pseudo-spin sectors.

Mathematically, low-lying excitations  relevant to  $m$-magnon excitations in the spin or pseudo-spin sector, denoted as $|\phi_{m;j_1,\ldots,j_{m}}(k)\rangle$, should
transform as $\tau  |\phi_{m;j_1,\ldots,j_{m}}(k)\rangle=\exp(ik) |\phi_{m;j_1,\ldots,j_{m}}(k)\rangle$  under the action of the translation operation $\tau$ if PBCs are adopted  or $\sigma   |\phi_{m;j_1,\ldots,j_{m}}(k)\rangle=\exp(ik) |\phi_{m;j_1,\ldots,j_{m}}(k)\rangle$ under the action of the cyclic  permutation symmetry operation $\sigma$ if OBCs are adopted. Hence, $|\phi_{m;j_1,\ldots,j_{m}}(k)\rangle$ may be expressed in terms of $|\psi_m^{j_1,\ldots,j_{m}}\rangle$: $|\phi_{m;j_1,\ldots,j_{m}}(k)\rangle=\sum_{l=0}^{L-1}\exp(ik l) \tau^l |\psi_m^{j_1,\ldots,j_{m}}\rangle$
under PBCs or $|\phi_{m;j_1,\ldots,j_{m}}(k)\rangle=\sum_{l=0}^{L-1}\exp(ik l) \sigma^l |\psi_m^{j_1,\ldots,j_{m}}\rangle$ under OBCs, in order to ensure that they are dispersionless: $\omega(k)=0$. They thus accidentally become degenerate ground states: $E(k)=E_0$, for $k=2\pi \delta/L$ ($\delta=0,1,\ldots,L-1$).
This in turn implies that  $|\psi_m^{j_1,\ldots,j_m}\rangle$ are degenerate (fully factorized) ground states, which may be recognized as generalized highest weight states in the spin or pseudo-spin sector at level $m$.

In particular, at the first level, we are led to conclude that  $|\psi_1^j\rangle \equiv S_j^-\ket{\psi_0}$ or $|\psi_1^j\rangle \equiv T_j^- \ket{\psi_0}$ ($j=1,2,\ldots,L$) are degenerate factorized ground states, due to the exchange symmetry between the spin operators $\textbf{S}_j$ and the pseudo-spin operators $\textbf{T}_j$ that is partially broken. There exists therefore an emergent local symmetry operation $g \in Z_2$,  defined as $\exp (i \pi \Sigma_j)$, with $\Sigma_j =S_j^z$ or $T_j^z$ ($j=1$, \ldots, $L$),
such that  $g|\Psi_0\rangle$ is a degenerate ground state, as long as $|\Psi_0\rangle$ is chosen to be $S^-\ket{\psi_0}$ or $T^-\ket{\psi_0}$ at the first level. Obviously, we have $|\langle \Psi_0|g|\Psi_0\rangle|\neq 1$.
Meanwhile, at the second level, $|\psi_2^{ij}\rangle \equiv S_i^-S_j^- \vert\psi_0\rangle$ or $|\psi_2^{ij}\rangle \equiv T_i^-T_j^- \vert\psi_0\rangle$  ($i=1,2,\ldots,L$ and $j$ is not less than $i+1$) are degenerate (fully factorized) ground states, due to the exchange symmetry between the spin operators $\textbf{S}_j$ and the pseudo-spin operators $\textbf{T}_j$ that is partially broken.
Note that, at the second level, a degenerate ground state may be generated from the  action of $S^-$ on the highest weight states twice. That is, we have $(S^-)^2 \; \ket{\psi_0}$ as a degenerate ground state. Combining with
$|\psi_2^{ij}\rangle$, we see that $(S_1^-T_1^-+S_1^-T_2^-+S_1^-T_L^-)\ket{\psi_0}$ and
$ \sum _{i=3}^{L-1}  S_1^-T_i^-\ket{\psi_0}$ under PBCs and
$ (S_1^-T_1^-+S_1^-T_2^-)\ket{\psi_0}$ and $\sum _{i=3}^{L}  S_1^-T_i^-\ket{\psi_0}$ under OBCs become degenerate ground states.
As follows from the converse form of the lemma, there exists therefore an emergent local symmetry operation $g \in Z_2$,  defined as $\exp (i \pi \Sigma)$, with $\Sigma=T_1^z+T_2^z+T_L^z$ under PBCs or $\Sigma=T_1^z+T_2^z$ under OBCs, if $|\Psi_0\rangle$ is chosen to be $T^- S_1^-\ket{\psi_0}$. Again, we have $|\langle \Psi_0|g|\Psi_0\rangle|\neq 1$. This argument may be extended to any level $m$ in the spin or pseudo-spin sector, up to level $L/2$.

Mathematically,  low-lying excitations relevant to the $m$-magnon excitations in the spin and pseudo-spin sectors, denoted as $|\phi_{m;i_1,\ldots,i_{m_s};j_1,\ldots,j_{m_t}}(k)\rangle$, with $m=m_s+m_t$, should
transform as $\tau |\phi_{m;i_1,\ldots,i_{m_s};j_1,\ldots,j_{m_t}}(k)\rangle=\exp(ik) |\phi_{m;i_1,\ldots,i_{m_s};j_1,\ldots,j_{m_t}}(k)\rangle$  under the action of the  translation operation $\tau$ if PBCs are adopted  or $\sigma  |\phi_{m;i_1,\ldots,i_{m_s};j_1,\ldots,j_{m_t}}(k)\rangle=\exp(ik) |\phi_{m;i_1,\ldots,i_{m_s};j_1,\ldots,j_{m_t}}(k)\rangle$ under the action of the cyclic permutation symmetry operation $\sigma$ if OBCs are adopted. Hence, $|\phi_{m;i_1,\ldots,i_{m_1};j_1,\ldots,j_{m_2}}(k)\rangle$ may be expressed in terms of $|\psi_m^{j_1,\ldots,j_{m}}\rangle$: $|\phi_{m;i_1,\ldots,i_{m_1};j_1,\ldots,j_{m_2}}(k)\rangle=\sum_{l=0}^{L-1}\exp(ik l) \tau^l|\psi_m^{i_1,i_2,\ldots,i_{m_s};j_1,j_2,\ldots,j_{m_t}}\rangle$
under PBCs or $|\phi_{m;i_1,\ldots,i_{m_s};j_1,\ldots,j_{m_t}}(k)\rangle=\sum_{l=0}^{L-1}\exp(ik l) \sigma^l |\psi_m^{i_1,i_2,\ldots,i_{m_s};j_1,j_2,\ldots,j_{m_t}}\rangle$ under OBCs, in order to ensure that they are dispersionless: $\omega(k)=0$. They thus accidentally become degenerate ground states: $E(k)=E_0$, for  $k=2\pi \delta/L$ ($\delta=0,1,\ldots,L-1$).
This in turn implies that  $|\psi_m^{i_1,i_2,\ldots,i_{m_s};j_1,j_2,\ldots,j_{m_t}}\rangle$ are degenerate (fully factorized) ground states, which may be recognized as generalized highest weight states in the spin and pseudo-spin sectors at level $m$.

In particular, at the second level, when $m=2$ with $m_s=m_t=1$, $|\psi_2^{ij}\rangle \equiv S_i^-T_j^- \vert\psi_0\rangle$ or $|\psi_2^{ij}\rangle \equiv T_i^-S_j^- \vert\psi_0\rangle$ ($i=1,2,\ldots,L$ and $j$ is not less than $i+1$) are degenerate (fully factorized) ground states, due to the exchange symmetry between the spin operators $\textbf{S}_j$ and the pseudo-spin operators $\textbf{T}_j$ that is partially broken.
Note that, at the second level, a degenerate ground state may be generated from the combined action of $S^-$ and $T^-$ on the highest weight state. That is, we have $S^- T^-\; \ket{\psi_0}$ as a degenerate ground state. Combining with
$|\psi_2^{ij}\rangle$, we see that $(S_1^-T_1^-+S_1^-T_2^-+S_1^-T_L^-)\ket{\psi_0}$ and
$ \sum _{i=3}^{L-1} S_1^-T_i^-\ket{\psi_0}$ under PBCs and
$ (S_1^-T_1^-+S_1^-T_2^-)\ket{\psi_0}$ and $\sum _{i=3}^{L}  S_1^-T_i^-\ket{\psi_0}$ under OBCs  become degenerate ground states,.
As follows from the converse form of the lemma, there exists therefore an emergent local symmetry operation $g \in Z_2$,  defined as $\exp (i \pi \Sigma)$, with $\Sigma=S_1^z+S_2^z+S_L^z$ under PBCs or $\Sigma=S_1^z+S_2^z$ under OBCs, if $|\Psi_0\rangle$ is chosen to be $S^- T_1^-\ket{\psi_0}$.  Again, we have $|\langle \Psi_0|g|\Psi_0\rangle|\neq 1$. This argument may be extended to any level $m$ in the spin  and pseudo-spin sectors, up to level $L$.

We remark that the total number of  emergent Goldstone flat bands, as constructed from generalized highest weight states $|\psi_m^{j_1,j_2,\ldots,j_m}\rangle$ in the spin or pseudo-spin sector and $|\psi_m^{i_1,i_2,\ldots,i_{m_s};j_1,j_2,\ldots,j_{m_t}}\rangle$ in the spin and pseudo-spin sectors,  is exponential in $L$. In addition, the existence of an emergent (local) symmetry operation tailored to a specific degenerate ground state is guaranteed, if emergent Goldstone flat bands are present, as one may expect from the restatement of Elitzur's theorem.

\section{An extrinsic characterization of emergent Goldstone flat bands}~\label{extrinsic}

We now turn to an extrinsic characterization of the emergent Goldstone flat bands in the staggered spin-1 ${\rm SU}(3)$ ferromagnetic biquadratic model (\ref{hambq}) and the staggered ${\rm SU}(4)$ ferromagnetic spin-orbital model (\ref{hamist}). Physically, this amounts to elaborating on how the low-lying multi-magnon excitations in the  ${\rm SO}(3)$ spin-1 ferromagnetic Heisenberg model and the  ${\rm SO}(4)$ ferromagnetic spin-orbital  model are flattening out as the  model (\ref{hambq}) and  the  model (\ref{hamist}) are approached, respectively.

For this purpose, it is convenient to consider  the ${\rm SO}(3)$ spin-1 bilinear-biquadratic model~\cite{chubukov,fath1,fath2,ivanov,rizzi,schmid,ronny,dyw} and the ${\rm SO}(4)$  bilinear-biquadratic model, which is unitarily equivalent to the ${\rm SO}(4)$ spin-orbital model~\cite{khomskii1,khomskii2,khomskii3,khomskii4,khomskii5}. Indeed, both of them may be regarded as the two  simplest cases of the ${\rm SO(2s+1)}$ bilinear-biquadratic model~\cite{son1,son2,alet}, with $s=1$ and $s=3/2$, respectively.
The ${\rm SO(2s+1)}$ bilinear-biquadratic model is described by the Hamiltonian
\begin{equation}
	\mathscr{H}=\cos\theta\sum_j\sum_{a<b} L_j^{ab}L_{j+1}^{ab}+\sin\theta\sum_j \left(\sum_{a<b}L_j^{ab}L_{j+1}^{ab}\right)^2,
	\label{hamso}
\end{equation}
where $L_j^{ab}$ $(1\leq a<b\leq 2s+1)$ denote the generators of the symmetry group ${\rm SO}(2s+1)$ at lattice site $j$, with the total number being  $s(2s+1)$, which transform in the vectorial representation. The sum over $j$ is taken from 1 to $L$ for PBCs, and from 1 to $L-1$ for OBCs. Note that the dimension of the local Hilbert space is $2s+1$, with the coupling constants being parametrized in terms of $\theta\in [0,2\pi]$. In particular, the symmetry group of the Hamiltonian (\ref{hamso}) becomes the staggered ${\rm SU}(2s+1)$ group  and it is unitarily equivalent to the staggered ${\rm SU}(2s+1)$
spin-$s$ ferromagnetic model~\cite{barber2} at $\theta = \pi/2$.

\subsection{The ${\rm SO}(3)$ spin-1 bilinear-biquadratic model}

\begin{figure}
	\centering
	\includegraphics[angle=0,totalheight=5cm]{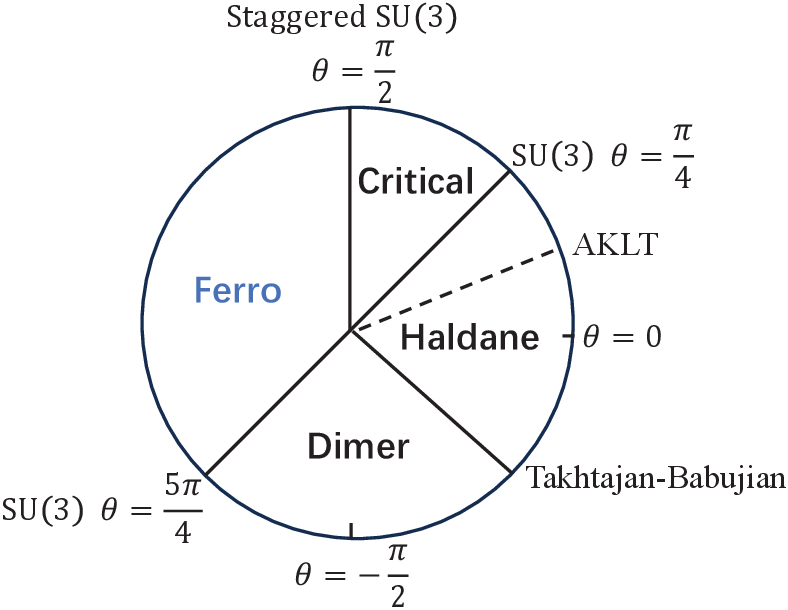} 
  	\caption{A sketch of the ground state phase diagram for the ${\rm SO}(3)$ spin-1 bilinear-biquadratic model.  It accommodates a critical phase, a Haldane phase, a dimerized phase and a ferromagnetic phase.
		Here, $\theta$
			ranges from 0 to $2\pi$, but we mainly concern the ferromagnetic regime $(\pi/2, 5\pi/4)$, with the staggered ${\rm SU(3)}$ spin-1 ferromagnetic biquadratic model, located at $\theta= \pi/2$, and the uniform ${\rm SU(3)}$ spin-1 ferromagnetic model,  located at $\theta = 5\pi/4$, as the two endpoints. We also indicate two other quantum phase transition points: one is the uniform ${\rm SU(3)}$ spin-1 antiferromagnetic model and the other is the Takhtajan-Babujian model located at $\theta = - \pi/4$, in addition to the AKLT point located at $\theta = \arctan (1/3)$ that admits four exactly solvable nearly degenerate  ground states under OBCs, though unique under PBCs.
	}\label{so3-gspd}
\end{figure}

If $s=1$, then the Hamiltonian (\ref{hamso}) is nothing but the ${\rm SO}(3)$ spin-1 bilinear-biquadratic model~\cite{chubukov,fath1,fath2,ivanov,rizzi,schmid,ronny,dyw}, described
by the Hamiltonian
\begin{equation}
	\mathscr{H}= \sum_{j}
	\left(\cos \theta \; \mathbf{S}_{j}\mathbf{S}_{j+1}
	+\sin\theta \;\left(\mathbf{S}_{j}\mathbf{S}_{j+1}\right)^2
	\right),
	\label{hamso3}
\end{equation}
where $\mathbf{S}_j=(S_{j}^{x},S_{j}^{y},S_{j}^{z})$ denotes the vector of the spin-1 operators at lattice site $j$. The sum over $j$ is taken from 1 to $L$ for PBCs, and from 1 to $L-1$ for OBCs. The ground state phase diagram for the ${\rm SO}(3)$ spin-1 bilinear-biquadratic model (\ref{hamso3}) is sketched in Fig.\ref{so3-gspd}.  It accommodates a critical phase, a Haldane phase, a dimerized phase and a ferromagnetic phase.
Here, $\theta$ ranges from 0 to $2\pi$, but we mainly concern the ferromagnetic regime $(\pi/2, 5\pi/4)$, with the staggered ${\rm SU(3)}$ spin-1 ferromagnetic biquadratic model, located at $\theta= \pi/2$, and the uniform ${\rm SU(3)}$ spin-1 ferromagnetic model~\cite{sutherland},  located at $\theta = 5\pi/4$, as the two endpoints. Two other quantum phase transition points are also indicated: one is the uniform ${\rm SU(3)}$ spin-1 antiferromagnetic model~\cite{sutherland} and the other is the Takhtajan-Babujian model located at $\theta = - \pi/4$~\cite{takhtajan,babujian}, in addition to the spin-1 Affleck-Kennedy-Lieb-Tasaki (AKLT) model located at $\theta = \arctan (1/3)$~\cite{aklt1,aklt2} that admits four exactly solvable nearly degenerate ground states under OBCs, though unique under PBCs.
In the ferromagnetic regime $(\pi, 5\pi/4)$, the SSB pattern is from ${\rm SU}(2) \approx {\rm SO}(3)$ to ${\rm U}(1)$, with the number of type-B GMs being one: $N_B=1$. Note that a completely integrable model in the ferromagnetic regime is located at $\theta = 3\pi/4$, which is the antipodal point of the Takhtajan-Babujian model~\cite{takhtajan,babujian}.
The ground state degeneracies under PBCs and OBCs are identical, equal to the dimension  $L+1$ of an irreducible representation of the symmetry group ${\rm SO}(3)$. Note that the representation space is spanned by the orthonormal basis states generated from the repeated action of the lowering operator $S^-$ on the highest weight state $\ket{\otimes_{\eta=1}^{L} \{+\}_{\;\eta}}$, irrespective of the boundary conditions.

For the ${\rm SO}(3)$ spin-1 bilinear-biquadratic model (\ref{hamso3}) in the ferromagnetic regime $(\pi/2, 5\pi/4)$,  all the eigenstates with the eigenvalues $L-1$ and $L-2$ of $S^z$
may be constructed analytically as one-magnon and two-magnon excitations~\cite{akutsu,bibikov1}.
If  the eigenvalue  of $S^z$ is  $L-1$, then we have $ \Phi_{1} (k) = \sum_{j=1}^{L} e^{ik j} S_j^- \ket{\psi_0}$, with the one-magnon excitation energy $\omega (k) = 2 |\cos \theta| \;(1-\cos k$). It is readily seen that the eigenstates $ \Phi_{1} (k)$ with different values of $k=2\pi \delta/L$ ($\delta=0,1,\ldots,L-1$) are orthogonal to each other, with different energy eigenvalues. Meanwhile, as  the staggered ${\rm SU(3)}$ spin-1 ferromagnetic biquadratic model at $\theta = \pi/2$ is approached, all  the eigenstates $ \Phi_{1} (k)$ become degenerate for any values of $k$, due to the presence of $|\cos \theta|$.
We are thus  able to conclude that the one-magnon excitations become flat, as $\theta = \pi/2$ is approached.  If  the eigenvalue  of $S^z$ is  $L-2$, then we are led to the two-magnon excitations $ \Phi_{2} (k_1,k_2) = (\sum_{j_1<j_2}^{L} a_{j_1,j_2}(k_1,k_2) S^-_{j_1} S^-_{j_2} +\sum_{j=1}^L b_j(k_1,k_2) (S^-_j)^2) \ket{\psi_0}$, with the excitation energy being $\omega (k_1,k_2) = \omega (k_1) + \omega (k_2) $, where the explicit expressions for $a_{j_1,j_2}(k_1,k_2)$ and $b_j(k_1,k_2)$ in the thermodynamic limit may be found in Refs.~\onlinecite{akutsu,bibikov1,bibikov2} (also cf. Appendix D). Hence,  the two-magnon excitations become flat, as $\theta = \pi/2$ is approached. This  is also valid for three-magnon excitations with the excitation energy $\omega (k_1,k_2,k_3) = \omega (k_1) + \omega (k_2)+ \omega (k_3)$, which have been explicitly constructed in the thermodynamic limit~\cite{bibikov2} (also cf. Appendix D). In principle, this conclusion may be extended to any $m$-magnon excitation with the excitation energy $\omega (k_1,k_2,\ldots,k_m) = \sum _\mu\omega (k_\mu)$ ($\mu =1,2,\ldots,m$). Hence, all the multi-magnon excitations become flat: $\omega (k_1,k_2,\ldots,k_m)=0$, as the staggered ${\rm SU}(3)$ spin-1 ferromagnetic biquadratic model  is approached.  In particular,  the $m$-magnon excitation states at $\theta =\pi/2$ may be expressed as a linear combination of (fully factorized ) generalized highest weight states and other degenerate ground states that are not fully factorzized at level $m$ (in the thermodynamic limit), as explicitly constructed in Appendix D for $m$ up to three.

In addition, there are a few exactly solvable excited  states $\sum _{j=1}^L (-1)^j \; S^-_j \;(S^-_{j+\chi+1} - S^-_{j+\chi- 1}) \;\ket{\psi_0}$ ($\chi>2$), $\sum _{j=1}^L \; (-1)^j \;S^-_j \;(S^-_{j+2} -2 \;S^-_j) \; \ket{\psi_0}$ and $\sum _{j=1}^L \; (-1)^j \;S^-_j \;S^-_{j+1} \; \ket{\psi_0}$ in the entire ferromagnetic regime, with their energy eigenvalues being $\cos\theta(L-4)+\sin\theta L$, $\cos\theta(L-4)+\sin\theta L$, and $\cos\theta(L-3)+\sin\theta(L+1)$ under PBCs, respectively. In particular, the first two excited states, which share the same energy eigenvalue, become linear combinations of degenerate ground states $\sum _{j=1}^L \; (-1)^j \; S^-_j \; S^-_{j+\chi+1} \;\ket{\psi_0}$ ($\chi > 2$) and $\sum _{j=1}^L \; (-1)^j \; (\;S^-_j)^2  \; \ket{\psi_0}$
at $\theta = \pi/2$, given that  $\sum _{j=1}^L \; (-1)^j \; (\;S^-_j)^2 $ is one of the eight generators of the staggered ${\rm SU}(3)$ symmetry group. Meanwhile,
$\sum _j \; (-1)^j \;S^-_j \;S^-_{j+1} \;\ket{\psi_0}$  yields the lowest energy gap above the degenerate ground states at $\theta = \pi/2$, with the gap being $1$, as confirmed numerically (in the thermodynamic limit) in Section~\ref{spectralfunction}.
These exactly solvable excited states may be interpreted as isolated quantum many-body scars in the spin-1 AKLT model~\cite{aklt1,aklt2}, which appear in quantum many-body systems with weak ergodicity breaking~\cite{scar0,scar1,scar2,scar3}. Actually, quantum many-body scars violate the ETH~\cite{eth1,eth2,eth3,eth4,eth5,eth6}, but they only occupy a small portion of the Hilbert space, in contrast to the strong ergodicity breaking arising from quantum complete integrability~\cite{integrability} and many-body localization~\cite{localization1,localization2}. In fact,  as it is readily seen from the ground state phase diagram of the  ${\rm SO}(3)$ spin-1 bilinear-biquadratic model in Fig.\ref{so3-gspd}, any quantum many-body scar in the ferromagnetic regime must be a quantum many-body scar in the Haldane phase $(-\pi/4, \pi/4)$ as well as the half part of the dimerized phase $(-\pi/2, -\pi/4)$, with the Takhtajan-Babujian model located at $\theta = - \pi/4$ in between,   since the two Hamiltonians at the two antipodal points  are identical, up to an overall  minus sign. In other words, their spectra are  simply upside down, so they share the same set of quantum many-body scars in the middle of their spectra. As a consequence, all the quantum many-body scars constructed for the spin-1 AKLT model are also quantum many-body scars at the antipodal point in the ferromagnetic regime  $(\pi/2, 5\pi/4)$.  In particular, the Arovas A and B states~\cite{arovas}, and the spin-2 magnon excitations of the AKLT model~\cite{akltscar1,akltscar2} must be  quantum many-body scars at the antipodal point in the ferromagnetic regime  $(\pi/2, 5\pi/4)$.

Indeed, this is consistent with a unified framework proposed in Ref.~\onlinecite{mark} to construct an exact tower of quantum many-body scars~\cite{schecter,fendley}  in a non-integrable quantum many-body system that does not obey the ETH, according to a mathematical lemma: for a Hamiltonian $\mathscr{H}$, if there is a subspace $W$ and an operator $Q^\dagger$ such that $Q^\dagger W \in W$ and
$ ([\mathscr{H}, Q^\dagger]- Q^\dagger) W=0$,  then $(Q^\dagger)^r |v_0\rangle$, as long as it is a non-zero vector, generate a tower of equally spaced eigenstates of $\mathscr{H}$ with the eigenvalues $E_0 + r \Omega$, where $|v_0\rangle$ is  an eigenstate of $\mathscr{H}$ and the subspace $W$  is spanned by $|v_0\rangle,  Q^\dagger|v_0\rangle, \ldots,  (Q^\dagger)^r|v_0\rangle$,  with $r$ being an integer labeling them and $\Omega$ a constant.
Note that $|v_0\rangle$ is typically chosen to be  the ground state, but the highest excited state should play the same role as well.
For the spin-1 AKLT model,  if one chooses $|v_0\rangle$ to be the ground state, then there exists a tower  of  equally spaced eigenstates $(Q^\dagger)^r|v_0\rangle$, with $Q^\dagger=\sum_j (-1)^j (S^-_j)^2$, with $r$ being up to $L/2$.  Here, we stress that this mathematical lemma targets at a ladder operator that produces an exact tower of quantum many-body scars, in contrast to  the mathematical lemma about an emergent (local) symmetry operation tailored to a specific eigenstate, which targets at an emergent (local) symmetry operation that produces degenerate eigenstates.  Actually, $Q^\dagger$ may be regarded as the ladder operator for a variant of spectrum generating algebras~\cite{gruber}, which is now tailored to a specific eigenstate. It generates a tower of exactly solvable excited states, since $\Omega$ is non-zero. In particular,  the ladder operator $Q^\dagger$ becomes one generator of the staggered  ${\rm SU}(3)$ symmetry group for the  ${\rm SU}(3)$ spin-1 ferromagnetic biquadratic model. Note that the entanglement entropy for a tower of quantum many-body scars obey the sub-volume law for the spin-1 AKLT model~\cite{akltscar2}, instead of the volume law predicted by the ETH~\cite{eth5}.

\subsection{The ${\rm SO}(4)$ spin-orbital model}

\begin{figure}
	\centering
	\includegraphics[angle=0,totalheight=5cm]{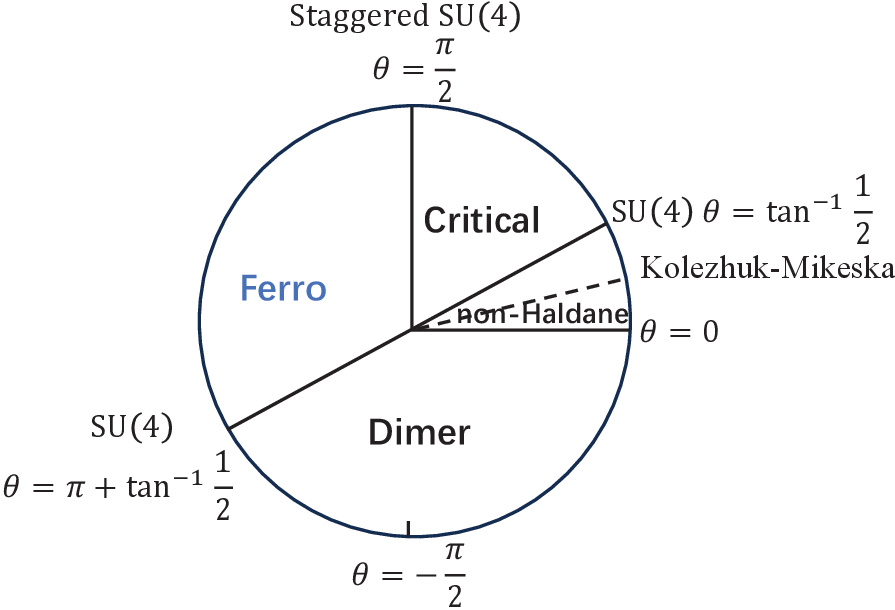}
	\caption{A sketch of the ground state phase diagram for the ${\rm SO}(4)$ bilinear-biquadratic model, unitarily equivalent to the ${\rm SO}(4)$  spin-orbital model.  It accommodates a critical phase, a non-Haldane phase, a dimerized phase and a ferromagnetic phase. Here, $\theta$ ranges from 0 to $2\pi$, but we mainly concern the ferromagnetic regime $(\pi/2, \pi + \arctan (1/2))$, with the staggered ${\rm SU(4)}$ biquadratic model located at $\theta= \pi/2$, which is unitarily equivalent to the  staggered ${\rm SU(4)}$ ferromagnetic spin-orbital model, and  the uniform ${\rm SU(4)}$ ferromagnetic  bilinear-biquadratic model,  located at $\theta = \pi + \arctan (1/2)$, which is  unitarily equivalent to the  uniform ${\rm SU(4)}$ spin-orbital model model, as the two endpoints. Moreover, two other quantum phase transition points are also indicated: one is the uniform ${\rm SU(4)}$ antiferromagnetic bilinear-biquadratic model and the other is a critical point describing two decoupled copies of the spin-$1/2$ anti-ferromagnetic Heisenberg model, in addition to the Kolezhuk-Mikeska point located at $\theta = \arctan (1/4)$ that admits exactly solvable ground states.
	}\label{so4-gspd}
\end{figure}

If $s=3/2$, then the Hamiltonian (\ref{hamso}) is unitarily equivalent to the ${\rm SO}(4)$ spin-orbital model~\cite{spinorbitalsu4,son1}.
The ${\rm SO}(4)$ spin-orbital model is described by the Hamiltonian~\cite{So4}
\begin{equation}
	\mathscr{H}=\pm\sum_{j}\left(\mathbf{S}_j\cdot \mathbf{S}_{j+1}+\zeta \right) \left(\mathbf{T}_j \cdot \mathbf{T}_{j+1}+\zeta \right),	\label{hamist1}
\end{equation}
where  $\mathbf{S}_j=(S_{j}^{x},S_{j}^{y},S_{j}^{z})$ and $\textbf{T}_j=(T_{j}^x,T_{j}^y,T_{j}^z)$ are the vectors of the spin-$1/2$ and pseudo-spin 1/2 operators at lattice site $j$. The sum over $j$ is taken from 1 to $L$ for PBCs, and from 1 to $L-1$ for OBCs.
The ground state phase diagram for the ${\rm SO}(4)$  spin-orbital model, unitarily equivalent to the ${\rm SO}(4)$ bilinear-biquadratic model, is sketched in Fig.~\ref{so4-gspd}.  It accommodates a critical phase, a non-Haldane phase, a dimerized phase and a ferromagnetic phase. Here, we mainly concern the ferromagnetic regime $(\pi/2, \pi + \arctan (1/2))$, with the  staggered ${\rm SU(4)}$ ferromagnetic spin-orbital model, unitarily equivalent to the staggered ${\rm SU(4)}$ biquadratic model located at $\theta= \pi/2$, and the  uniform ${\rm SU(4)}$ spin-orbital model model, unitarily equivalent to the uniform ${\rm SU(4)}$ ferromagnetic  bilinear-biquadratic model located at $\theta = \pi + \arctan (1/2)$, as the two endpoints.  Moreover, two other quantum phase transition points are also indicated: one is the uniform ${\rm SU(4)}$ antiferromagnetic bilinear-biquadratic model and the other is a critical point describing two decoupled copies of the spin-$1/2$ anti-ferromagnetic Heisenberg model, in addition to the Kolezhuk-Mikeska point~\cite{khomskii4} located at $\theta = \arctan (1/4)$  that admits exactly solvable ground states. Note that there is another completely integrable model located at $\theta =\pi$, which is two decoupled copies of the ${\rm SU(2)}$ spin-$1/2$ ferromagnetic Heisenberg models.
We remark that  the uniform ${\rm SU(4)}$ ferromagnetic and anti-ferromagnetic  bilinear-biquadratic models are unitarily equivalent to the uniform ${\rm SU(4)}$ spin-$3/2$ models that are exactly solvable by means of the Bethe ansatz~\cite{sutherland}.

Note that the Hamiltonian (\ref{hamist1})  at $\zeta =-1/4$, if a plus sign ``+" is taken, becomes the staggered ${\rm SU(4)}$ ferromagnetic spin-orbital model (\ref{hamist}). The model (\ref{hamist1}) shares the same ground state subspace, consisting of the ferromagnetic ground states,  in the regime $(-\infty, -1/4)$ if a plus sign ``+" is taken in the Hamiltonian  (\ref{hamist1}) and in the regime $(1/4,\infty)$ if a minus sign ``-" is taken in the Hamiltonian (\ref{hamist1}). In particular, it consists of the two decoupled ${\rm SU(2)}$ spin-$1/2$ Heisenberg ferromagnetic models at $\zeta=-\infty$ if the plus sign ``+" is taken and at $\zeta=\infty$ if the minus sign ``-" is taken, so it is frustration-free. In addition, the  Hamiltonian (\ref{hamist1})  in this ferromagnetic regime possesses  the  symmetry group ${\rm SO(4)}$ (isomorphic to ${\rm SU(2)} \times {\rm SU(2)}$), with the generators of the two copies of  ${\rm SU(2)}$ being $S^x=\sum_jS_{j}^x$, $S^y=\sum_jS_{j}^y$,  and $S^z=\sum_jS_{j}^z$, and $T^x=\sum_jT_{j}^x$, $T^y=\sum_jT_{j}^y$ and $T^z=\sum_jT_{j}^z$, respectively.
Note that there is a simple relation between $\theta$ and $\zeta$: $\zeta=(\cot\theta-1)/4$, with sign ``plus" or ``minus" in (\ref{hamist1}) determined from the sign of $\sin \theta$, given the  unitary equivalence between the ${\rm SO}(4)$ bilinear-biquadratic model and the ${\rm SO}(4)$ spin-orbital model. Hence, one may also use $\theta$ to label the ${\rm SO}(4)$ spin-orbital model.

In the ferromagnetic regime   $(\pi/2, \pi + \arctan (1/2))$ for the ${\rm SO}(4)$ spin-orbital model, the SSB pattern is from ${\rm SU}(2) \times {\rm SU}(2)$ to ${\rm U}(1) \times {\rm U}(1)$, with the number of type-B GMs being two: $N_B=2$,  because ${\rm SO}(4)$ is isomorphic to ${\rm SU}(2) \times {\rm SU}(2)$. The ground state degeneracies under PBCs and OBCs are identical to $(L/2+1)^2$,  with the ground state subspace being the tensor product of two copies of an irreducible representation of ${\rm SU}(2)$, which is spanned by the orthonormal basis states generated from the repeated action of the lowering operators $S^-$ and $T^-$ on the highest weight state $\ket{\otimes_{\eta=1}^{L} \{\uparrow_{s}\uparrow_{t}\}_{\eta}}$, which act in the spin and pseudo-spin sectors, respectively.

For the ${\rm SO}(4)$  spin-orbital model (\ref{hamist1}) in the ferromagnetic regime,  all the eigenstates with the eigenvalues $L/2-1$ and $L/2$ of $S^z$ and $T^z$ or  with the eigenvalues $L/2$ and $L/2-1$ of $S^z$ and $T^z$
may be constructed analytically as one-magnon excitations in the spin sector or the pseudo-spin sector. Indeed, it is straightforward to extend this construction to any low-lying $m$-magnon excitations in the spin sector or the pseudo-spin sector, with the eigenvalue  of $S^z$ or $T^z$ being $L/2-m$. However, instead of delving into the details of this construction,  we focus on a peculiar feature of the ${\rm SO}(4)$  spin-orbital model that is valid {\it not only} in the ferromagnetic regime {\it but also} in the entire parameter space.  Actually, although the ${\rm SO}(4)$  spin-orbital model itself is generically not completely integrable, it accommodates two completely integrable sectors, if one restricts to the spin sector with  the pseudo-spin sector being in the fully polarized state  $\ket{\otimes_{\eta=1}^{L} \{\uparrow_{t}\}_{\eta}}$ or  the pseudo-spin sector with  the spin sector being in the fully polarized state $\ket{\otimes_{\eta=1}^{L} \{\uparrow_{s}\}_{\eta}}$, due to the exchange symmetry  between the spin operators $\textbf{S}_j$ and the pseudo-spin operators $\textbf{T}_j$. Note that each of the two integrable sectors is identical to the  ${\rm SU(2)}$ spin-$1/2$  ferromagnetic Heisenberg model, up to a multiplicative factor $1/4+\zeta$, if one is restricted to the ferromagnetic regime.

Actually, the ${\rm SU(2)}$ spin-$1/2$  (ferromagnetic and antiferromagnetic)  Heisenberg model is completely integrable in the Yang-Baxter sense~\cite{baxterbook,faddeev,sklyanin} under both PBCs and OBCs. The model possesses extensively many conserved currents, as shown from the commuting transfer matrices. The commutativity of the transfer matrices in turn follows from the quantum Yang-Baxter equation under PBCs~\cite{baxterbook,faddeev} and the boundary quantum Yang-Baxter equation, i.e., the reflection equation~\cite{sklyanin} under OBCs. We stress that all these nontrivial conserved currents commute with each other. In fact, it is readily seen that all the conserved currents for the ${\rm SU(2)}$ spin-$1/2$  (both ferromagnetic and antiferromagnetic) Heisenberg model are also the conserved currents in the two integrable sectors for the ${\rm SO}(4)$ spin-orbital model (\ref{hamist1}) under both PBCs and OBCs, respectively. In particular, the model under OBCs is only a special choice of integrable boundary conditions one may construct from the boundary quantum Yang-Baxter equation~\cite{sklyanin}. Indeed, there are two sets of the conserved currents: one for $S_j^x,S_j^y$ and $S_j^z$ in the spin sector and the other for $T_j^x,T_j^y$ and $T_j^z$ in the pseudo-spin sector. Note that this is valid in the entire parameter space. Hence, the ${\rm SO}(4)$   spin-orbital model (\ref{hamist1}) at a generic point in the parameter space offers an example that is non-integrable, but still possesses extensively many conserved currents in a specific sector. Not surprisingly, one may expect that many exactly solvable states for the ${\rm SO}(4)$   spin-orbital model (\ref{hamist1}) may be constructed from the Bethe ansatz solutions for either the ${\rm SU(2)}$ spin-$1/2$  ferromagnetic Heisenberg model or  the ${\rm SU(2)}$ spin-$1/2$  antiferromagnetic Heisenberg model, depending on the locations in the parameter space. In other words, we are able to embed these exactly solvable states into the spectrum of the model  (\ref{hamist1}). The total number of these exactly solvable states is $(L+1)2^{L+1}$, in addition to other exactly solvable excited states that involve both the spin and pseudo-spin sectors, when none of them is in the fully polarized state, or in its descendant states generated from the repeated action of the lowering operator of ${\rm SU(2)}$ acting {\it only} in the spin or pseudo-spin sector. Here, $L+1$ originates from the fact that  $\ket{\otimes_{\eta=1}^{L} \{\uparrow_{s}\}_{\eta}}$ or $\ket{\otimes_{\eta=1}^{L} \{\uparrow_{t}\}_{\eta}}$ plays the role of the highest weight state in the spin or pseudo-spin sector, with the eigenvalue of $S^z$ or $T^z$ being $L/2$, one factor 2 comes from the exchange symmetry between the spin operators $\textbf{S}_j$ and the pseudo-spin operators $\textbf{T}_j$, and the remaining factor $2^L$ simply results from the dimension of the Hilbert space of the ${\rm SU(2)}$ spin-$1/2$ ferromagnetic Heisenberg model in the spin or pseudo-spin sector.  All these exactly solvable  states thus constructed in the ferromagnetic regime $(\pi/2, \pi + \arctan (1/2))$ become flat, as the staggered ${\rm SU(4)}$ ferromagnetic spin-orbital model at $\zeta =-1/4$ is approached, due to the presence of $1/4+\zeta$ as a multiplicative factor in the excitation energies, if  a plus sign ``+" is taken in the Hamiltonian  (\ref{hamist1}). In particular, all the multi-magnon excitations become flat at $\zeta=-1/4$.  This amounts to stating that all the orthonormal basis states $S_{j_1}^-S_{j_2}^-\ldots S_{j_m}^- \ket{\otimes_{\eta=1}^{L} \{\uparrow_{s}\}_{\eta}}$ ($m=0,1,2,\ldots,L/2$) in the spin  sector or $T_{j_1}^-T_{j_2}^-\ldots T_{j_m}^- \ket{\otimes_{\eta=1}^{L} \{\uparrow_{t}\}_{\eta}}$ ($m=0,1,2,\ldots,L/2$) in the pseudo-spin sector, together with their time-reversed states, are degenerate ground states  for the staggered ${\rm SU(4)}$ ferromagnetic spin-orbital model. 

Needless to say, this extrinsic characterization yields consistent results with the intrinsic characterization discussed in Section~\ref{threeitems}.  We remark that
all  these exactly solvable  states at $\theta =\pi/2$ may be expressed as linear combinations of degenerate ground states, as already constructed  for the staggered ${\rm SU(4)}$ ferromagnetic spin-orbital model in Subsection~\ref{emergenttoexponential}.  In addition, although the unique  ${\rm SU(2)}$-invariant ground state of the ${\rm SU(2)}$ spin-$1/2$ antiferromagnetic Heisenberg model is normally characterized as a conformal field theory with central charge $c=1$~\cite{cft}, it becomes the highest excited state in the spin or pseudo-spin sector, modulo the fully polarized state in the pseudo-spin or spin sector,  in the ferromagnetic regime. However, this state is smoothly embedded into the staggered ${\rm SU}(4)$ ferromagnetic spin-orbital model as a degenerate ground state, with the entanglement entropy scaling logarithmically with the block size under OBCs and PBCs~\cite{ee1,ee2,ee3,ee4,ee5}, thus obeying the sub-volume law. Moreover, the above discussion clearly shows that the exchange symmetry between the spin operators $\textbf{S}_j$ and the pseudo-spin operators $\textbf{T}_j$
is partially broken.

We stress that this is in sharp contrast to all the multi-magnon excitations in the uniform ${\rm SU(4)}$ ferromagnetic spin-orbital model at $\zeta =1/4$,  if a minus sign ``-" is taken in the Hamiltonian  (\ref{hamist1}). This is due to the fact that $1/4+\zeta$ appears as a multiplicative factor in the excitation energies. Hence,  all the exactly solvable  excited states  one may construct from the ${\rm SU(2)}$ spin-$1/2$  ferromagnetic Heisenberg model in the ferromagnetic regime $(\pi/2, \pi + \arctan (1/2))$ do not become flat, as the uniform ${\rm SU(4)}$ ferromagnetic spin-orbital model at $\zeta =1/4$ is approached.  In particular, the dispersion relations for all the multi-magnon excitations remain quadratic for the uniform ${\rm SU(4)}$ ferromagnetic spin-orbital model.

Normally, quantum many-body scars only occupy a small portion of the Hilbert space, in contrast to the strong ergodicity breaking arising from quantum complete integrability~\cite{integrability} and many-body localization~\cite{localization1,localization2}. As already mentioned above, this picture works for  the ${\rm SO}(3)$ spin-1 bilinear-biquadratic model. However, the  ${\rm SO}(4)$ spin-orbital model appears to be an exception, in the sense that  exactly solvable excited states in this model at a generic non-integrable point occupy a large portion of the Hilbert space, due to the fact that the model accommodates two integrable sectors, one in the spin sector and the other in the pseudo-spin sector. Here, by a small portion versus a large portion we mean polynomial versus exponential as a function of the system size.
Hence, the existence of extensively many conserved currents in the two integrable sectors for the ${\rm SO}(4)$   spin-orbital model (\ref{hamist1}) makes it very special, though the  ${\rm SO(4)}$ spin-orbital model is non-integrable at a generic point. This implies that exactly solvable excited states in the ${\rm SO(4)}$ spin-orbital model occupy a large portion of the Hilbert space, since both of the two sectors are exactly solvable, with the total number of exactly solvable excited states being exponential. The entanglement entropy for any state in the two integrable sectors obeys the sub-volume law or volume law~\cite{scar2}. In contrast, the entanglement entropy for a quantum many-body scar obeys the sub-volume law in the ${\rm SO}(3)$ spin-1 bilinear-biquadratic model, which only occupy a small portion of the Hilbert space, since the total number of exactly solvable excited states is at most polynomial in $L$. In other words, the ${\rm SO(4)}$ spin-orbital model is expected to behave in a way closer to completely integrable models, due to the presence of extensively many conserved currents in the two integrable sectors. Further, one may expect that level statistics follows the Poisson distribution in the two integrable sectors in the ${\rm SO}(4)$ spin-orbital model, though it should  correspond to the Gaussian Orthogonal Ensemble (GOE) in a generic non-integrable sector labeled by the  eigenvalues of $S^z$ and $T^z$, as follows from the conventional wisdom that the level statistics of an ergodic system, described by random matrix theory, corresponds to the GOE, whereas completely integrable systems instead are characterized by the Poisson statistics~\cite{bohigas,berry}.

In addition, both the staggered ${\rm SU(3)}$ spin-1 biquadratic model  (\ref{hambq}) and the staggered ${\rm SU(4)}$ ferromagnetic spin-orbital model  (\ref{hamist})  constitute specific physical realizations of the Temperley-Lieb algebra~\cite{tla,baxterbook,martin}. Although they are completely integrable as solutions to the quantum Yang-Baxter equation, their integrability is lost, once disorder is introduced in the coupling constants. However, the disordered variants of the Temperley-Lieb spin models are shown to exhibit Hilbert space fragmentation~\cite{tlhsf1} - another ergodicity breaking phenomenon  relevant to quantum many-body scars~\cite{tlhsf2,tlhsf3}, in the sense that the number of the Krylov subspaces grows exponentially, in contrast to
generic quantum many-body systems with conventional symmetries such as ${\rm Z}_2$, ${\rm U(1)}$  or ${\rm SU(2)}$. In these cases, the number of the Krylov subspaces  grows at most polynomially with the system size.
Hence, we are led to conclude that the disordered variant of the staggered ${\rm SU(4)}$ ferromagnetic spin-orbital model also provides a physically interesting example that is fragmented in an entangled basis.

Generally, one may anticipate that there exist a large class of non-integrable quantum many-body systems admitting extensively many conserved currents in a specific sector so that the number of exactly solvable excited states is exponential in system size. All of theses  exactly solvable excited states occupy a large portion of the Hilbert space, with the entanglement entropy obeying the sub-volume law or volume law~\cite{scar2}. The existence of such a class of non-integrable quantum many-body systems may be established from an inherent structure underlying a sequence of representations of the Temperley-Lieb algebra, which allow a non-integrable deformation but simultaneously keep some integrable sectors. In other words, such a non-integrable deformation admits extensively many conserved currents if one is restricted to a specific integrable sector, thus enriching our understanding of the interrelations between the ETH, quantum many-body scars, Hilbert space fragmentation and complete integrability.

\section{A numerical test: spectral functions}~\label{spectralfunction}

In order to confirm our analytical analysis, we perform
an extensive numerical investigation into the spectral functions of the $\mathrm{SO}(3)$ spin-1 bilinear-biquadratic model as a function of $\theta \in [\pi/2, 5\pi/4]$, and the $\mathrm{SO}(4)$ bilinear-biquadratic model as a function of $\theta \in [\pi/2, \pi+\arctan(1/2)]$.
We define the spectral function as $S(\omega,k) = -1/\pi \operatorname{Im} G(\omega,k)$, where $G(\omega, k)$ is the Fourier transform of the retarded Green’s function in real space $G(t,x) = -\mathrm{i} \Theta(t) \langle \psi_0|O_x^\dagger(t) O_0(0)|\psi_0 \rangle$.
Here, $\Theta(t)$ is the Heaviside step function, $O_j(t) = \mathrm{e}^{\mathrm{i}\mathscr{H}t} O_j \mathrm{e}^{-\mathrm{i}\mathscr{H}t}$ is a local operator evolved with time $t$ in the Heisenberg picture, and $\ket{\psi_0}$ is chosen to be the highest weight state $\ket{\psi_0}$ (in the thermodynamic limit), namely, $ \ket{\otimes_{\eta \in Z} \{+\}_{\;\eta}}$ for the $\mathrm{SO}(3)$ model and $\ket{\otimes_{\eta \in Z}\{\uparrow_s\uparrow_t\}_\eta}$ for the $\mathrm{SO}(4)$ model,
where $Z$ denotes the set of integers.
The presence of energy eigenstates with momentum~$k$ and excitation energy~$\omega$ that have a nonzero overlap with the operator $O_j$ applied to the highest weight state will be reflected by a nonzero value of the spectral function $S(\omega, k)$.
Thus, these spectral functions provide a useful window into the low-lying spectrum of these models.

To numerically determine the Green’s function, we first calculate the time-evolved state $\ket{\psi_j(t)} = \mathrm{e}^{-\mathrm{i}\mathscr{H}t} O_j \ket{\psi_0}$, and then evaluate $G(t,j) =  -\mathrm{i} \langle \psi_j(-t/2)|\psi_0(t/2) \rangle$.
Here, we remark that  $\ket{\psi_j(-t/2)}$ is simply the complex conjugate of $\ket{\psi_j(t/2)}$ due to time-reversal symmetry.
We find the time-evolved state~$\ket{\psi_j(t)}$ by using MPS numerics~\cite{schollwoeck2011,paeckel2019,mptoolkit}: specifically, we utilize the time-dependent variational principal (TDVP) algorithm~\cite{haegeman2016} using single-site updates with adaptive environment expansion~\cite{mcculloch2024,mcculloch2024b}.
Additionally, we adopt infinite boundary conditions (IBC)~\cite{phien2012,phien2013,milsted2013,zauner2015}, where we only evolve a finite window of sites around the local perturbation to the initial state, which we extend as the disturbance grows with time~\cite{phien2013}.
This way, we can offset the windows by $j$~sites in order to calculate the overlap $\langle \psi_j(-t/2)|\psi_0(t/2) \rangle$, and so we only need to perform the time evolution for one initial position of $O_j$.
As this strategy works directly in the thermodynamic limit $L \rightarrow \infty$, any extra complications arising from the boundary conditions are irrelevant in our numerical implementation.

\subsection{The ${\rm SO}(3)$ spin-1 bilinear-biquadratic model}

For the ${\rm SO}(3)$ spin-1 bilinear-biquadratic model, as $\theta$ varies from deep inside the ferromagnetic regime $(5\pi/4, \pi/2)$ to the staggered  ${\rm SU}(3)$ spin-1 ferromagnetic biquadratic model at $\theta =\pi/2$, the SSB pattern changes from ${\rm SU}(2)$ to ${\rm U}(1)$ to ${\rm SU(3)} $ to ${\rm U(1)}\times {\rm U(1)}$. Accordingly, the number of type-B GMs varies from $N_B=1$  to $N_B=2$.
The dispersion relation is quadratic in the ferromagnetic regime $(\pi/2, 5\pi/4)$, with the dynamical critical exponent $z=2$. However,
the dispersion relation becomes trivial for the staggered  ${\rm SU}(3)$ spin-1 ferromagnetic biquadratic model at $\theta =\pi/2$. Indeed, this is the case for one-magnon excitations, two-magnon excitations and so on.  Mathematically,  the coset space $S^2$ in the ferromagnetic regime is smoothly embedded into the coset space $CP^2_{\pm}$ for the staggered ${\rm SU}(3)$ spin-1 biquadratic model  at $\theta=\pi/2$. Hence, it is the extra complications arising from the staggered nature of the symmetry group ${\rm SU(3)}$ that lead to the emergence of Goldstone flat bands.

\begin{figure}
	\centering
	\includegraphics[angle=0,totalheight=16cm]{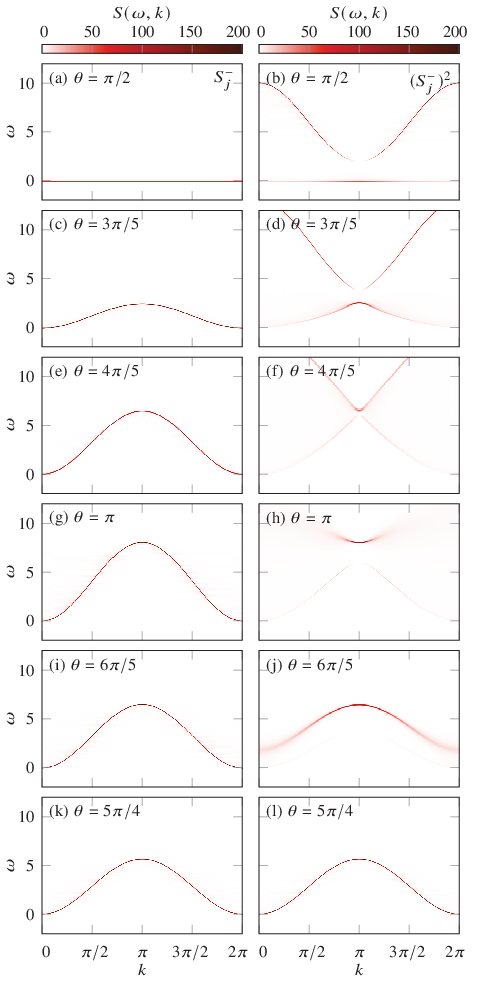}
	\caption{The spectral functions $S(\omega, k)$ generated by the two operators $S_j^-$ and  $(S_j^-)^2$ in the ${\rm SO}(3)$ spin-1 bilinear-biquadratic model.
    We calculate the spectral functions for various points in the ferromagnetic regime $\theta \in [\pi/2, 5\pi/4]$ through numerical time evolution techniques in MPS representations.
	}\label{so3}
\end{figure}

\begin{figure}
	\centering
	\includegraphics[angle=0,totalheight=16cm]{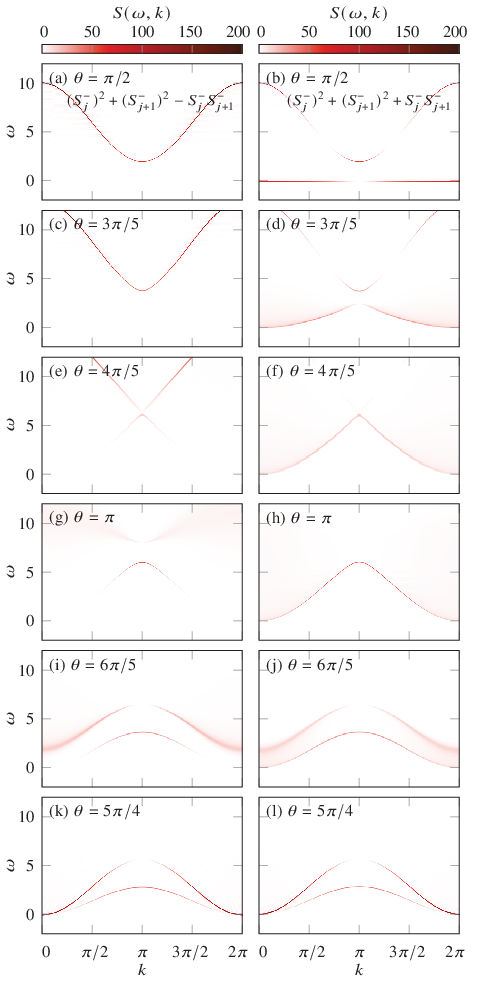}
	\caption{The spectral functions $S(\omega, k)$ generated by the two operators $(S_j^-)^2 + (S_{j+1}^-)^2 - S_j^- S_{j+1}^-$ and  $(S_j^-)^2 + (S_{j+1}^-)^2 + S_j^- S_{j+1}^-$ in the ${\rm SO}(3)$ spin-1 bilinear-biquadratic model.
	}\label{so3-2a}
\end{figure}

\begin{figure}
	\centering
	\includegraphics[angle=0,totalheight=16cm]{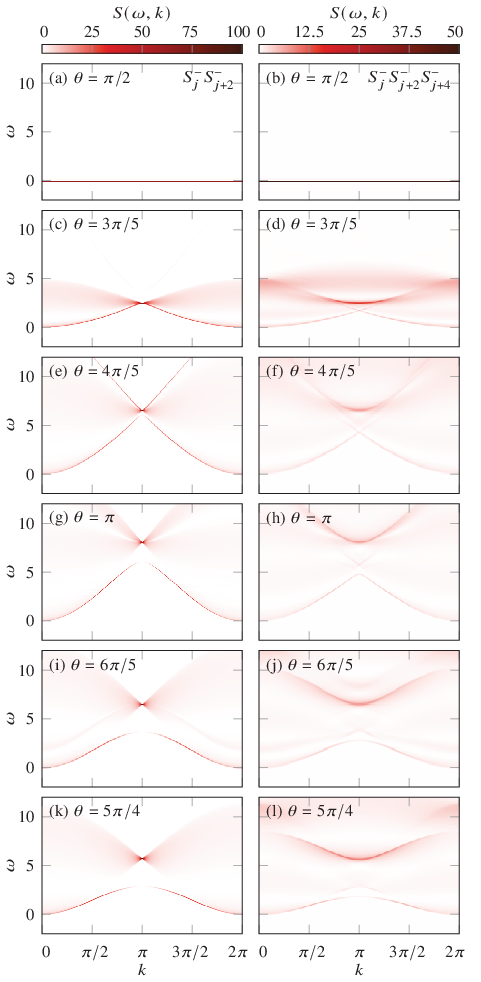}
	\caption{The spectral functions $S(\omega, k)$ generated by the two operators $S_j^- S_{j+2}^-$ and  $S_j^- S_{j+2}^- S_{j+4}^-$ in the ${\rm SO}(3)$ spin-1 bilinear-biquadratic model.
	}\label{so3-3}
\end{figure}

In Fig.~\ref{so3} and Fig.~\ref{so3-2a}, we plot four spectral functions generated by the four operators $S_j^-$,  $(S_j^-)^2$, $(S_j^-)^2 + (S_{j+1}^-)^2 - S_j^- S_{j+1}^-$ and  $(S_j^-)^2 + (S_{j+1}^-)^2 + S_j^- S_{j+1}^-$ in the ${\rm SO}(3)$ spin-1 bilinear-biquadratic model.
In Fig.~\ref{so3-3},  we plot two spectral functions  for the two operators  $S_j^- S_{j+2}^-$ and  $S_j^- S_{j+2}^- S_{j+4}^-$ in the ${\rm SO}(3)$ spin-1 bilinear-biquadratic model. 
Here, we use six values of $\theta$: $\theta = \pi/2, 3\pi/5, 4\pi/5, \pi, 6\pi/5$ and $5/4 \pi$.
While the spectral function for $S_j^-$ consistently shows a single-magnon character, reflected in its having a single spectral peak for each momentum, corresponding to the eigenstate $\sum_j \mathrm{e}^{\mathrm{i}kj} S_j^- \ket{\psi_0}$, with momentum $k$, the other spectral functions have nonzero spectral weight spread out over a range of energies, indicating the presence of multiple-magnon scattering states.
As the spectral function for each operator only shows part of the low-lying spectrum, we use a few different operators here to create a fuller picture of of the nature of this spectrum.
Furthermore, each of the spectral functions---except for the one generated by  $O_j = S_j^- S_{j+2}^- S_{j+4}^-$---only has a single peak at momentum~$k = \pi$, as the states $\sum_j (-1)^j O_j \ket{\psi_0}$ are energy eigenstates in each case.

It is readily seen that
the dispersion relations of the low-lying excited states, containing contributions from one-magnon, two-magnon and three-magnon excitations, flatten out as $\theta$ evolves from deep inside the ferromagnetic regime to the staggered  ${\rm SU}(3)$ spin-1 ferromagnetic biquadratic model, located at  $\theta =\pi/2$.
Our numerical simulations show that  the lowest energy gap above the degenerate ground states is $1$ at momentum~$k = \pi$,
indicating that this excited state must be 
$\sum _j \; (-1)^j \;S^-_j \;S^-_{j+1} \; \ket{\psi_0}$. 

In contrast, as $\theta$ varies from deep inside the ferromagnetic regime $(\pi/2,5\pi/4)$ to the uniform spin-1 ${\rm SU}(3)$ ferromagnetic model
at $\theta =5\pi/4$, the SSB pattern changes from ${\rm SU}(2)$ to ${\rm U}(1)$ to ${\rm SU(3)} $ to ${\rm SU(2)}\times {\rm U(1)}$. Accordingly, the number of type-B GMs varies from $N_B=1$  to $N_B=2$, and the coset space is changed from $CP^1$ to  $CP^2$.  We remark that $CP^1$ is diffeomorphic to $S^2$.  Mathematically,  the coset space $CP^2$ for the uniform ${\rm SU}(3)$ ferromagnetic  model accommodates, as a submanifold, the coset space $CP^1$ for the ${\rm SO}(3)$ spin-1 bilinear-biquadratic model in the ferromagnetic regime. The dispersion relation thus remains quadratic, as in the ferromagnetic regime $(\pi/2,5\pi/4)$, with the dynamical critical exponent $z=2$.

\subsection{The ${\rm SO}(4)$ spin-orbital model}

Given that the ${\rm SO}(4)$ spin-orbital model is unitarily equivalent to the ${\rm SO}(4)$ bilinear-biquadratic model, we focus on the latter formulation. As $\theta$  evolves from deep inside the ferromagnetic regime to the staggered ${\rm SU}(4)$ ferromagnetic biquadratic model,  the SSB pattern changes from ${\rm SO}(4)$ to ${\rm U}(1)\times{\rm U}(1)$ to ${\rm SU(4)} $ to ${\rm U(1)}\times {\rm U(1)}\times {\rm U(1)}$ via ${\rm U(1)}\times {\rm U(1)}\times {\rm SU(2)}$. Accordingly, the number of type-B GMs varies from $N_B=2$  to $N_B=3$.  The dispersion relation is quadratic in the ferromagnetic regime $(\pi/2, \pi +\arctan (1/2)$, with the dynamical critical exponent $z=2$. However, the dispersion relation becomes trivial for the staggered ${\rm SU}(4)$ point at $\theta =\pi/2$,  unitarily equivalent to the staggered ${\rm SU}(4)$ ferromagnetic spin-orbital model at $\zeta=-1/4$, if a plus sign ``+" is taken in the Hamiltonian  (\ref{hamist1}).
In contrast, as $\theta$  evolves from deep inside the ferromagnetic regime to the uniform ${\rm SU}(4)$ ferromagnetic  bilinear-biquadratic model at $\theta = \pi +\arctan (1/2)$, unitarily equivalent to the uniform ${\rm SU}(4)$ ferromagnetic spin-orbital model at $\zeta=1/4$, if a minus sign ``-" is taken in the Hamiltonian  (\ref{hamist1}), the dispersion relation remains  quadratic, with the dynamical critical exponent $z=2$.

\begin{figure}
	\centering
	\includegraphics[angle=0,totalheight=16cm]{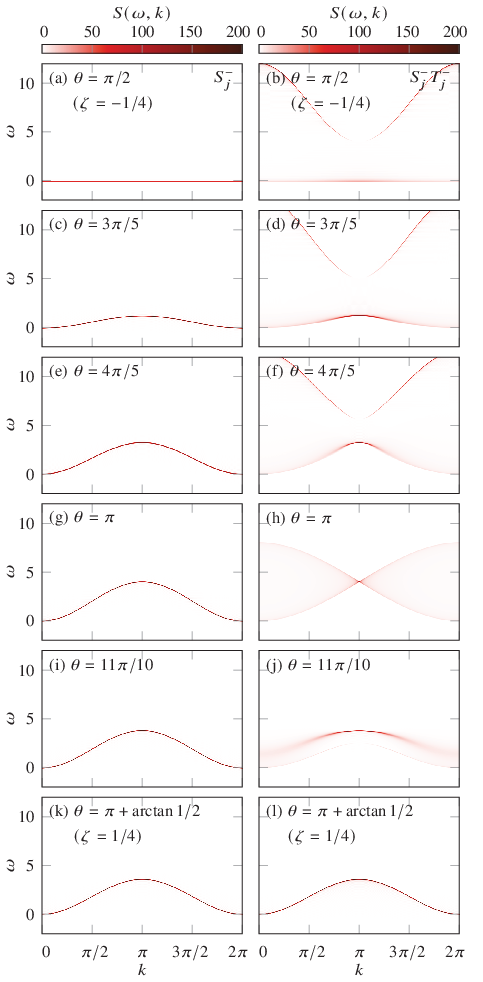}
	\caption{The spectral functions $S(\omega, k)$ generated by the two operators $S_j^-$ and  $S_j^-T_j^-$ in the ${\rm SO}(4)$ spin-orbital model.
    We calculate the spectral functions for various points in the ferromagnetic regime $\theta \in [\pi/2, \pi + \arctan (1/2)]$.
   Note that $\zeta$ is related to $\theta$ by the simple relation $\zeta=(\cot\theta-1)/4$.
	}\label{so4}
\end{figure}

\begin{figure}
	\centering
	\includegraphics[angle=0,totalheight=16cm]{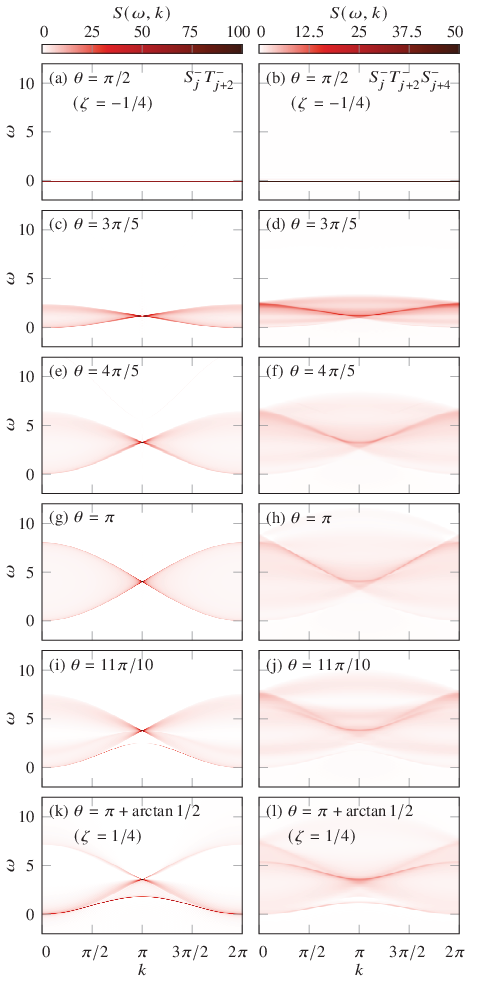}
	\caption{The spectral functions $S(\omega, k)$ generated by the two operators $S_j^-T_{j+2}^-$ and  $S_j^-T_{j+2}^-S_{j+4}^-$ in the ${\rm SO}(4)$ spin-orbital model.
	}\label{so4-2}
\end{figure}

We plot two spectral functions  generated by the two operators $S_j^-$ and   $S_j^- T_j^-$ in Fig.~\ref{so4} and two spectral functions for the two operators $S_j^-T_{j+2}^-$ and  $S_j^-T_{j+2}^-S_{j+4}^-$ in Fig.~\ref{so4-2}.
We use the six values of $\theta$, $\theta = \pi/2, 3\pi/5, 4\pi/5, \pi, 11\pi/10$ and $\pi + \arctan (1/2)$.
Similar to the $\mathrm{SO}(3)$ model considered above, the spectral function generated by $S^-_j$ only has a single peak at each momentum, while the other operators provide information on excitation continua.
In the cases of  $O_j = S^-_j T^-_j$ and $O_j = S^-_j T^-_{j+2}$, we also see only a single peak at $k = \pi$, as the states $\sum_j (-1)^j O_j\ket{\psi_0}$ generated from these two operators are energy eigenstates.

We can see that the dispersion relations, containing contributions from one-magnon, two-magnon and three-magnon excitations, flatten out as $\theta$ evolves from deep inside the ferromagnetic regime to the staggered ${\rm SU}(4)$ ferromagnetic  biquadratic model at $\theta=\pi/2$, but not as $\theta$ evolves from deep inside the ferromagnetic regime to  the uniform ${\rm SU}(4)$  ferromagnetic  bilinear-biquadratic model at   $\pi + \arctan (1/2)$, where the dispersion relations remain quadratic.

Mathematically, the coset space $CP^3$ for the uniform ${\rm SU}(4)$ ferromagnetic spin-orbital model accommodates, as a submanifold, the coset space $CP^1 \times CP^1$ for the ${\rm SO}(4)$ spin-orbital model, diffeomorphic to $S^2 \times S^2$. This explains the physics underlying the ${\rm SO}(4)$ spin-orbital model as it evolves from deep inside the ferromagnetic regime to the uniform ${\rm SU(4)}$
ferromagnetic spin-orbital model at $\theta = \pi + \arctan(1/2)$.  Moreover, there is another embedding of $S^2 \times S^2$ into the coset space $CP^3_{\pm}$ for
the  staggered ${\rm SU}(4)$ spin-orbital model (\ref{hamist}) at $\theta = \pi/2$. As already discussed in Section~\ref{extrinsic}, the multi-magnon excitations in the ferromagnetic regime become flat as  $\theta = \pi/2$ is approached. Hence, it is the staggered nature of the symmetry group ${\rm SU(4)}$ at $\theta = \pi/2$ that is responsible for the emergence of Goldstone flat bands. 

\section{Concluding remarks}~\label{summary}

We have performed a systematic investigation into the staggered  ${\rm SU}(3)$ spin-1 ferromagnetic biquadratic model and the staggered ${\rm SU}(4)$ ferromagnetic spin-orbital model, in an attempt to clarify the origin of the exponential ground state degeneracies under both PBCs and OBCs.
It has been demonstrated that, for a quantum many-body spin system undergoing SSB with type-B GMs, if there exists an emergent local symmetry operation tailored to a specific degenerate ground state, then the ground state degeneracies under both PBCs and OBCs are exponential in system size. The  exponential ground state degeneracies in turn imply the emergence of Goldstone flat bands, thus indicating trivial dynamics underlying the low-lying excitations. Moreover,  the presence of  emergent Goldstone flat bands implies that there exists an emergent (local) symmetry operation tailored to a specific degenerate ground state. In other words, the three properties, namely, (1) the exponential ground state degeneracies in system size under both PBCs and OBCs, (2) the existence of emergent (local) symmetry operations tailored to  degenerate ground states, and (3) the emergence of  Goldstone flat bands, are equivalent in the context of SSB with type-B GMs.  Although our discussion is limited to these two illustrative examples, extensions to other quantum many-body spin systems undergoing SSB with type-B GMs are straightforward, with one candidate being the ${\rm SU}(2)$  spin-1 model with competing dimer and trimer interactions~\cite{dtmodel,dimertrimer}, which deserves to be investigated on its own merits.

In addition to this intrinsic characterization, we also present an extrinsic characterization of emergent Goldstone flat bands, thus revealing a connection to quantum many-body scars and Hilbert space fragmentation. The low-lying multi-magnon excitations for the  ${\rm SO}(3)$ spin-1 ferromagnetic Heisenberg model and  the  ${\rm SO}(4)$ spin-orbital ferromagnetic model flatten out as the staggered ${\rm SU}(3)$ spin-1  ferromagnetic biquadratic model (\ref{hambq}) and the staggered ${\rm SU}(4)$ ferromagnetic spin-orbital model (\ref{hamist}) are approached, respectively. This is in contrast to the embedding of the low-lying multi-magnon excitations into the uniform ${\rm SU(3)}$ spin-1 ferromagnetic model and the uniform ${\rm SU(4)}$  spin-orbital ferromagnetic model, where the dispersion relations remain quadratic.
In particular, we have shown that multi-magnon excitations in the ferromagnetic regime for the  ${\rm SO(3)}$ spin-1 bilinear-biquadratic model become linear combinations of degenerate ground states, including generalized highest weight states, at the staggered ${\rm SU(3)}$ point.
This construction provides a means to elaborate on the precise meaning of the connection between emergent Goldstone flat bands and multi-magnon excitations, given both are relevant to (fully factorized) generalized highest weight states  and other degenerate grounds states that are not fully factorized.

Quantum many-body scars only occupy a small portion of the Hilbert space, in contrast to the strong ergodicity breaking arising from quantum complete integrability and many-body localization. In fact, this picture works for  the ${\rm SO}(3)$ spin-1 bilinear-biquadratic model.
However, the  ${\rm SO}(4)$ spin-orbital model appears to be an exception, in the sense that, at a generic point in this model, the exactly solvable excited states occupy a large portion of the Hilbert space.  Hence, the  ${\rm SO}(4)$ spin-orbital model, as it is generically non-integrable, is of theoretical importance in the investigation of the violation of the ETH. We speculate that there exist a large class of non-integrable quantum many-body systems admitting extensively many conserved currents in a specific sector so that the number of exactly solvable excited states is exponential in system size. In particular, for these exactly solvable excited states, the entanglement entropy obeys the sub-volume law or volume law. 
Furthermore, we point out that  Hilbert space fragmentation~\cite{tlhsf1}  occurs in the disordered variant of the staggered ${\rm SU(4)}$ ferromagnetic spin-orbital model.

We have carried out an extensive numerical investigation for the spectral functions of the ${\rm SO}(3)$ spin-1 bilinear-biquadratic model and the ${\rm SO}(4)$  bilinear-biquadratic model as $\theta$ evolves from deep inside the ferromagnetic regimes to the two endpoints. This lends further support to our characterization of emergent Goldstone flat bands  in the two illustrative models from both intrinsic and extrinsic viewpoints.

In closing,  we emphasize that the methodology developed here is applicable to strongly correlated itinerant fermion systems undergoing SSB with type-B GMs. As a  candidate,  we mention the ${\rm SU}(2)$ flat-band ferromagnetic Tasaki model~\cite{tasaki} and its  ${\rm SU}(2s+1)$ variant~\cite{wzhang,katsura1,katsura2}, which exhibit SSB with one and $2s$ type-B GMs~\cite{TypeBtasaki}, respectively. As already shown in Ref.~\onlinecite{TypeBtasaki}, the ground state degeneracies for the ${\rm SU}(2)$ flat-band ferromagnetic Tasaki model  are exponential in system size, if the filling is not fixed, but they are different under PBCs and OBCs. In fact,
they behave asymptotically as the golden spiral when the system size is large enough. In contrast to quantum many-body spin systems undergoing SSB with type-B GMs under investigation here, the symmetry groups of the ${\rm SU}(2)$ flat-band ferromagnetic Tasaki model and its  ${\rm SU}(2s+1)$ variant are not semi-simple, due to the presence of the ${\rm U}(1)$ symmetry group generated by the number of fermionic particles in the charge sector.
In such a situation, only a semi-simple subgroup of the symmetry group is spontaneously broken, as follows from the requirement to keep consistency with the counting rule if {\it only} type-B GMs are present, as it should be. This stems from the fact that type-A GMs are absent in one spatial dimension, as a result of the Mermin-Wagner-Coleman theorem~\cite{mwc1,mwc2}.
Hence, there exists a degenerate ground state that acts as the highest weight state in each  sector labeled by the number of fermionic particles. As a result, the ${\rm SU}(2)$ flat-band ferromagnetic Tasaki model and its  ${\rm SU}(2s+1)$ variant provide examples for a different mechanism to yield  exponential ground state degeneracies under PBCs and OBCs, which may be relevant to novel physics underlying the flat-band ferromagnetism in insulators and metals.

{\it Acknowledgment.-}  We thank Murray Batchelor and  John Fjærestad for their helpful comments and suggestions during the preparation of the manuscript.
I.P.M. acknowledges funding from the National Science and Technology Council (NSTC) Grant No. 112-2811-M-007-044 and 113-2112-M-007-MY2.

%%%%%%%%%%%%%%%%%%%%%%%%%%%%%%%%%%%%%%%%%%%%%%%%%Appendix%%%%%%%%%%%%%%%%%%%%%%%5
\onecolumngrid
\newpage
\section*{Appendices}
\twocolumngrid
\setcounter{equation}{0}
\setcounter{figure}{0}
\renewcommand{\theequation}{A\arabic{equation}}
\renewcommand{\thefigure}{A\arabic{figure}}

\subsection{ The generators of the staggered ${\rm SU}(3)$ group expressed in terms of the spin-1 operators and the generators of the staggered ${\rm SU}(4)$ group expressed in terms of the spin-$1/2$ and pseudo-spin-1/2 operators}

The staggered ${\rm SU(3)}$ symmetry group contains the two Cartan generators  $H_1$ and $H_2$, with the raising operators $E_1$, $E_2$ and the lowering operators $F_1$, $F_2$, forming two ${\rm SU(2)}$ subgroups. 
In addition, there is one more ${\rm SU(2)}$ subgroup generated by $H_3 = H_1 - H_2$, with raising operator $E_3$ and lowering operator $F_3$. They satisfy the commutation relations $[E_a,E_b] = [F_a,F_b] = 0$, $[H_a,E_a] = 2E_a$, $[H_a,F_a] = -2F_a$ for $a=1,2$, and $3$, in addition to other commutation relations, as shown in Ref.~\onlinecite{goldensu3}. All of these generators can be expressed in terms of the spin-1 operators $S^x_j,S^y_j$, and $S^z_j$. For brevity, we  introduce the local components of the generators for the staggered  ${\rm SU}(3)$ symmetry group: $H_a=\sum_jH_{a,j}$ 
 ($a=1,2$), $E_a=\sum_jE_{a,j}$ ($a=1,2,3$), $F_a=\sum_jF_{a,j}$ ($a=1,2,3$), which take the form
\begin{eqnarray*}
	H_{1,j}&=& \frac{S^z_j}{2}+(-1)^j(1-\frac{3}{2}(S^z_j)^2), \quad H_{2,j}=S^z_j, \\
	E_{1,j}&=& \frac{\sqrt{2}}{4}\left (S^x_j+iS^y_j-(-1)^j(S^z_jS^x_j+S^x_jS^z_j+iS^y_jS^z_j+iS^z_jS^y_j)\right ),\\
	E_{2,j}&=& \frac{(-1)^{j+1}}{2}\left ((S^x_j)^2-(S^y_j)^2-iS^x_jS^y_j-iS^y_jS^x_j\right ), \\
	E_{3,j}&=&\frac{\sqrt{2}}{4}\left (S^x_j+iS^y_j+(-1)^j(S^z_jS^x_j+S^x_jS^z_j+iS^y_jS^z_j+iS^z_jS^y_j)\right ),\\
	F_{1,j}&=&\frac{\sqrt{2}}{4}\left (S^x_j-iS^y_j-(-1)^j(S^z_jS^x_j+S^x_jS^z_j-iS^y_jS^z_j-iS^z_jS^y_j)\right ),\\
	F_{2,j}&=&\frac{(-1)^{j+1}}{2}\left ((S^x_j)^2-(S^y_j)^2-i(S^x_jS^y_j+S^y_jS^x_j)\right ),\\
	F_{3,j}&=& \frac{\sqrt{2}}{4}\left (S^x_j-iS^y_j+(-1)^j(S^z_jS^x_j+S^x_jS^z_j-iS^y_jS^z_j-iS^z_jS^y_j)\right ).
\end{eqnarray*}
The staggered ${\rm SU(3)}$ symmetry group accommodates the uniform ${\rm SU(2)}$ group, generated by $S^x$, $S^y$, and $S^z$, as a subgroup. 
So here  $S^x$, $S^y$, and $S^z$ are expressed in terms of the generators of the staggered ${\rm SU(3)}$ group:  
$S^x =\sqrt{2}/2(E_1+F_1+E_3+F_3) $, $S^y = \sqrt{2}i/2(-E_1-E_3+F_1+F_3) $, and $S^z= H_2$.

The staggered ${\rm SU(4)}$ symmetry group contains the  three Cartan generators  $H_1$, $H_2$ and $H_3$,  with the raising operators $E_1$, $E_2$ and $E_3$ and the lowering operators $F_1$, $F_2$ and $F_3$, forming three ${\rm SU(2)}$ subgroups. 
In addition, there are three more ${\rm SU(2)}$ subgroups generated by $H_4=H_2-H_1$, $H_5=H_3-H_2$, and $H_6=H_3-H_1$, with three raising operators $E_4$, $E_5$, and $E_6$, and three lowering operators $F_4$, $F_5$, and $F_6$, respectively.
They satisfy $[H_a,E_a]=2E_a$, $[H_a,F_a]=-2F_a$ and $[E_a,F_a]=H_a$ for $a=1,2$,\ldots, $6$, in addition to other commutation relations,
as shown in Ref.~\onlinecite{spinorbitalsu4}.  All of these generators can be expressed in terms of
the spin-$1/2$ operators $S^x_j,S^y_j$, and $S^z_j$ and the pseudo-spin-$1/2$ operators $T^x_j,T^y_j$, and $T^z_j$. 
Similarly, we introduce the local components of the generators for the staggered  ${\rm SU}(4)$ symmetry group: $H_a=\sum_jH_{a,j}$ ($a=1,2,3$), $E_a=\sum_jE_{a,j}$ ($a=1,2$, \ldots, 6), $F_a=\sum_jF_{a,j}$ ($a=1,2$, \ldots, 6), which take the form
\begin{eqnarray*}
&H_{1,j}=(\frac{1}{2}-(-1)^jS^z_j)(\frac{1}{2}+T^z_j)-(\frac{1}{2}-(-1)^jS^z_j)(\frac{1}{2}-T^z_j),\\
&H_{2,j}=(\frac{1}{2}+S^z_j)(\frac{1}{2}-(-1)^jT^z_j)-(\frac{1}{2}-S^z_j)(\frac{1}{2}-(-1)^jT^z_j),\\
&H_{3,j}=(\frac{1}{2}+S^z_j)(\frac{1}{2}+T^z_j)-(\frac{1}{2}-S^z_j)(\frac{1}{2}-T^z_j),\\
&E_{1,j}=(\frac{1}{2}-(-1)^jS^z_j)T^-_j,\quad E_{2,j}=S^-_j(1/2-(-1)^jT^z_j),\\
&E_{3,j}=-(-1)^jS^-_j T^-_j,\quad E_{4,j}=-(-1)^jS^+_jT^-_j ,\\
&E_{5,j}=(\frac{1}{2}+(-1)^jS^z_j) T^-_j,\quad E_{6,j}=S^-_j (\frac{1}{2}+(-1)^jT^z_j),	\\
&F_{1,j}=(\frac{1}{2}-(-1)^jS^z_j) T^+_j, \quad F_{2,j}=S^+_j(\frac{1}{2}-(-1)^jT^z_j),\\
&F_{3,j}=-(-1)^jS^+_j T^+_j,\quad F_{4,j}=-(-1)^jS^+_j T^-_j,\\
&F_{5,j}=(\frac{1}{2}+(-1)^jS^z_j)T^+_j,\quad F_{6,j}=S^+_j (\frac{1}{2}+(-1)^jT^z_j),
\end{eqnarray*}
where $S^{\pm}_j=S^x_j\pm iS^y_j$ and $T^{\pm}_j=T^x_j\pm iT^y_j$.

The staggered ${\rm SU(4)}$ symmetry group accommodates the uniform ${\rm SU(2)} \times {\rm SU(2)}$ group, generated by $S^x$, $S^y$, and $S^z$ and $T^x$, $T^y$, and $T^z$, as a subgroup. So here 
$S^x$, $S^y$, and $S^z$ and $T^x$, $T^y$, and $T^z$ are expressed in terms of the generators of the staggered ${\rm SU(4)}$ symmetry group:
$S^x=E_2+E_6+F_2+F_6$, $S^y =i(E_2+E_6-F_2-F_6) $, and $S^z=-H_1+H_2+H_3$, and $T^x = E_1+E_5+F_1+F_5$, $T^y =i(E_1+E_5-F_1-F_5)$, and $T^z=H_1-H_2+H_3$.  

\subsection{ $CP^2_{\pm}$ and  $CP^3_{\pm}$ as the variants of the complex projective spaces $CP^2$ and 
$CP^3$}

For the staggered ${\rm SU}(2s+1)$ spin-$s$ ferromagnetic model, the symmetry group is the staggered ${\rm SU}(2s+1)$ group. One may introduce $2s$ ${\rm SU}(2)$ subgroups, each of which is associated with one of $2s$ type-B GMs as a result of SSB.
We denote the three generators for each of the $2s$ ${\rm SU}(2)$  subgroups as $\Sigma_{\alpha}^x$, $\Sigma_{\alpha}^y$ and $\Sigma_{\alpha}^z$ ($\alpha=1,\ldots,2s$), which are defined as $\Sigma_{\alpha,j}^x=(E_{\alpha,j}+F_{\alpha,j})/2$, $\Sigma_{\alpha,j}^y=-i(E_{\alpha,j}-F_{\alpha,j})/2$ and  $\Sigma_{\alpha,j}^z=H_{\alpha,j}/2$. The generators $\Sigma_{\alpha}^x$, $\Sigma_{\alpha}^y$ and $\Sigma_{\alpha}^z$ satisfy the commutation relations: $[\Sigma_{\alpha}^{x},\Sigma_{\alpha}^{y}]=i\Sigma_{\alpha}^{z}$, $[\Sigma_{\alpha}^{y},\Sigma_{\alpha}^{z}]=i\Sigma_{\alpha}^{x}$ and $[\Sigma_{\alpha}^{z},\Sigma_{\alpha}^{x}]=i\Sigma_{\alpha}^{y}$ ($\alpha=1,\ldots,2s$).
The set of the overcomplete basis states consists of factorized (unentangled) ground states  $|\psi(\theta_1,\phi_1;\ldots;\theta_{2s},\phi_{2s})\rangle$, which are parameterized in terms of  $\theta_\alpha \in [0,\pi]$ and $\phi_\alpha \in [0,2\pi]$ ($\alpha=1,\ldots,2s$),
\begin{widetext}
\begin{equation*}
		|\psi(\theta_1,\phi_1;\ldots;\theta_{2s},\phi_{2s})\rangle=|v(\theta_1,\phi_1;\ldots;\theta_{2s},\phi_{2s})\rangle_1 \cdots |v(\theta_1,\phi_1;\ldots;\theta_{2s},\phi_{2s})\rangle_{2l-1} |v(\theta_1,\phi_1;\ldots;\theta_{2s},\phi_{2s})\rangle_{2l}\cdots |v(\theta_1,\phi_1;\ldots;\theta_{2s},\phi_{2s})\rangle_L,
\end{equation*}
with
\begin{equation*}
		|v(\theta_1,\phi_1;\ldots;\theta_{2s},\phi_{2s})\rangle_{2l-1/2l}=\exp(i\phi_{2s} \Sigma_{2s,2l-1/2l}^{z})\exp(i\theta_{2s} \Sigma_{2s,2l-1/2l}^{y})\ldots \exp(i\phi_1 \Sigma_{1,2l-1/2l}^{z})\exp(i\theta_1 \Sigma_{1,2l-1/2l}^{y})\;|s\rangle_{2l-1/2l}.
\end{equation*}
Here $|s\rangle_{2l-1/2l}$  represents the eigenstate of $S_{2l-1/2l}^z$ with the eigenvalue being $s$ at the lattice site $2l-1/2l$, where $l=1,2,\ldots,L/2$, with $L$ even. Indeed, $|\psi(\theta_1,\phi_1;\ldots;\theta_{2s},\phi_{2s})\rangle$ may be regarded as a variant of an extension of the spin coherent states~\cite{gilmore} adapted from the uniform  ${\rm SU}(2)$ group to the uniform ${\rm SU}(2s+1)$ group, which have been exploited to investigate the entanglement entropy for linear combinations on subsets of the coset space $CP^{2s}$ in  the uniform ${\rm SU}(2s+1)$ ferromagnetic model~\cite{hqzhou}. The parameter space is a $2s$-dimensional (complex) manifold, denoted as $CP^{2s}_{\pm}$. In the main text, we have referred to $CP^{2s}_{\pm}$ as the coset space for the  staggered ${\rm SU(2s+1)}$ spin-$s$ ferromagnetic model, or its unitarily equivalent spin-orbital model when $s=3/2$.

We consider $CP^{2}_{\pm}$ for the staggered ${\rm SU}(3)$  spin-1 ferromagnetic biquadratic model. The factorized components $|v(\theta_1,\phi_1;\theta_2,\phi_2)\rangle_{2l-1/2l}$ take the form
\begin{equation*}
		|v(\theta_1,\phi_1;\theta_2,\phi_2)\rangle_{2l-1/2l}=\exp(i\phi_{2} \Sigma_{2,2l-1/2l}^{z})\exp(i\theta_{2} \Sigma_{2,2l-1/2l}^{y})\exp(i\phi_1 \Sigma_{1,2l-1/2l}^{z})\exp(i\theta_1 \Sigma_{1,2l-1/2l}^{y})\;|s\rangle_{2l-1/2l}.
\end{equation*}
Specifically, at the  $(2l-1)$-th lattice site, we have
\begin{equation*}
		|v(\theta_1,\phi_1;\theta_2,\phi_2)\rangle_{2l-1}=\left (\cos(\frac{\theta_1}{2})\cos(\frac{\theta_2}{2})\exp(i\frac{\phi_1 + \phi_2}{2});
		-\sin(\frac{\theta_1}{2})\exp(-i\frac{\phi_1}{2});
		-\cos(\frac{\theta_1}{2})\sin(\frac{\theta_2}{2})\exp(i\frac{\phi_1 - \phi_2}{2}) \right )^T,
\end{equation*}
and at the $2l$-th lattice site, we have
\begin{equation*}
		|v(\theta_1,\phi_1;\theta_2,\phi_2)\rangle_{2l}=\left (\cos(\frac{\theta_2}{2})\exp(i\frac{\phi_2}{2});0;\sin(\frac{\theta_2}{2})\exp(-i\frac{\phi_2}{2}) \right )^T.
\end{equation*}
Here $T$ denotes the transpose of a vector.

As for $CP^{3}_{\pm}$ for the staggered $\rm SU(4)$ spin-$3/2$ ferromagnetic  model, or its equivalent spin-orbital model, the factorized components $	|v(\theta_1,\phi_1;\theta_2,\phi_2;\theta_3,\phi_3)\rangle_{2l-1/2l}$ take the form
\begin{align*}
		|v(\theta_1,\phi_1;\theta_2,\phi_2;\theta_3,\phi_3)\rangle_{2l-1/2l}=&\exp(i\phi_{3} \Sigma_{3,2l-1/2l}^{z})\exp(i\theta_{3} \Sigma_{3,2l-1/2l}^{y})\exp(i\phi_{2} \Sigma_{2,2l-1/2l}^{z})\nonumber \\
		&\exp(i\theta_{2} \Sigma_{2,2l-1/2l}^{y})\exp(i\phi_1 \Sigma_{1,2l-1/2l}^{z})\exp(i\theta_1 \Sigma_{1,2l-1/2l}^{y})\;|s\rangle_{2l-1/2l}.
\end{align*}
Specifically, at  the $(2l-1)$-th lattice site, we have
\begin{align*}
		|v(\theta_1,\phi_1;\theta_2,\phi_2;\theta_3,\phi_3)\rangle_{2l-1}=& \Big(
		\cos(\frac{\theta_1}{2})\cos(\frac{\theta_2}{2})\cos(\frac{\theta_3}{2})\exp(i\frac{\phi_1+ \phi_2 + \phi_3}{2});
		-\sin(\frac{\theta_1}{2})\exp(-i\frac{\phi_1}{2});\\
		&-\cos(\frac{\theta_1}{2})\sin(\frac{\theta_2}{2})\exp(i\frac{\phi_1 - \phi_2}{2});
		-\cos(\frac{\theta_1}{2})\cos(\frac{\theta_2}{2})\sin(\frac{\theta_3}{2})\exp(i\frac{\phi_1+\phi_2 - \phi_3}{2}) \Big)^T,
\end{align*}
	and at the $2l$-th lattice site, we have
	\begin{equation*}
		|v(\theta_1,\phi_1;\theta_2,\phi_2;\theta_3,\phi_3)\rangle_{2l}=\left ( \cos(\frac{\theta_3}{2})\exp(i\frac{\phi_3}{2});0;0;
		\sin(\frac{\theta_3}{2})\exp(-i\frac{\phi_3}{2}) \right )^T.
	\end{equation*}

\subsection{Emergent local symmetry operations in the staggered ${\rm SU}(3)$ ferromagnetic biquadratic model at the third level }

Here we present an illustrative example for deriving degenerate ground states, both fully factorized and not fully factorized, from emergent local symmetry operations in the staggered ${\rm SU}(3)$ ferromagnetic biquadratic model at the third level.

All these degenerate ground states, including generalized highest weight states, are eigenstates  of $S^z$, with an eigenvalue being $L-3$. If $\vert\Psi_0\rangle$ is chosen to be $S^-S_1^-S_3^- \ket{\psi_0}$.  
Then we consider a local unitary operation $g = \exp (i \pi \Sigma)$, with  $\Sigma=S_1^z+S_2^z+S_3^z+S_4^z+S_L^z$ under PBCs and   $\Sigma=S_1^z+S_2^z+S_3^z+S_4^z$ under OBCs.
As follows from the lemma, we are led to a degenerate ground state  $g\vert\Psi_0\rangle$, which takes the form: 
$\left((S_{L}^-S_{1}^-S_{3}^-+(S_{1}^-)^2S_{3}^-+S_{1}^-S_{2}^-S_{3}^-+ S_{1}^-(S_{3}^-)^2+ S_{1}^-S_{3}^-S_{4}^-) -\sum_{4< j< L} S_{j}^-S_{1}^-S_{3}^-\right)\ket{\psi_0}$ under PBCs and
$\left(((S_{1}^-)^2S_{3}^-+S_{1}^-S_{2}^-S_{3}^-+ S_{1}^-(S_{3}^-)^2+ S_{1}^-S_{3}^-S_{4}^-) -\sum_{4<j\le L} S_{j}^-S_{1}^-S_{3}^-\right)\ket{\psi_0}$ under OBCs, up to a multiplicative constant. Again, $g$ generates an emergent discrete  symmetry group $Z_2$ tailored to $\vert\Psi_0\rangle$, because $g^2|\Psi_0\rangle= \vert\Psi_0\rangle$.
We are thus led to degenerate ground states: $|\psi_3^{1}\rangle_1 \equiv \left( S_{L}^-S_{1}^-S_{3}^-+(S_{1}^-)^2S_{3}^-+S_{1}^-S_{2}^-S_{3}^-+ S_{1}^-(S_{3}^-)^2+ S_{1}^-S_{3}^-S_{4}^- \right) \ket{\psi_0}
$ and $\sum_{j=5}^{L-1}S_1^-S_3^-S_j^-\ket{\psi_0}$ under PBCs and
$|\psi_3^{1}\rangle_1 \equiv \left( (S_{1}^-)^2S_{3}^-+S_{1}^-S_{2}^-S_{3}^-+ S_{1}^-(S_{3}^-)^2+ S_{1}^-S_{3}^-S_{4}^-\right) \ket{\psi_0}$ and $\sum_{j=5}^{L}S_1^-S_3^-S_j^-\ket{\psi_0}$ under OBCs. Afterwards, if $\vert\Psi_0\rangle$ is chosen to be $\sum_{j=5}^{L-1}S_1^-S_3^-S_j^-\ket{\psi_0}$ under PBCs, then we consider a set of local unitary operations $g_j$, defined as $\exp (i \pi \Sigma_j)$, with $\Sigma_j = S_j^z$ ($j=5$, \ldots, $L-1$). Therefore, we are led to a set of degenerate ground states  $g_j|\Psi_0\rangle$,
which take the form $\sum _{i=5}^{L-1} (1-2\delta_{i\;j}) S_1^- S_3^-S_i^-\ket{\psi_0}$, up to a multiplicative constant.
Again, each of $g_j$'s generates an emergent discrete symmetry group $Z_2$ tailored to $\vert\Psi_0\rangle$, because $g_j^2|\Psi_0\rangle= |\Psi_0\rangle$.
Hence, $S_1^-S_3^-S_j^-\ket{\psi_0}$ ($ j=5,\ldots,L-1$) are a sequence of degenerate ground states under PBCs.
Similarly, if $|\Psi_0\rangle$ is chosen to be $\sum_{j=5}^{L}S_1^-S_3^-S_j^-\ket{\psi_0}$ under OBCs, then we consider a set of local unitary operations $g_j$, defined as $\exp (i \pi \Sigma_j)$, with $\Sigma_j =S_j^z$ ($j=5$, \ldots, $L$). We are thus led to a set of degenerate ground states  $g_j\vert\Psi_0\rangle$, which take the form: $\sum _{i=5}^{L} (1-2\delta_{i\;j})S_1^-S_3^-S_i^-\ket{\psi_0}\vert$, up to a multiplicative constant.
Again, each of $g_j$'s generates an emergent discrete symmetry group $Z_2$ tailored to $\vert\Psi_0\rangle$, because $g_j^2|\Psi_0\rangle= |\Psi_0\rangle$.
Hence, $S_1^-S_3^-S_j^-\ket{\psi_0}$ ($ j=5,\ldots,L$) are a sequence of degenerate ground states under OBCs.

If $\vert\Psi_0\rangle$ is now chosen to be $S^-S_1^-S_{j_1}^- \ket{\psi_0}$, with $3<j_1<L-1$,  
then one may consider a local unitary operation $g = \exp (i \pi \Sigma)$, with  $\Sigma=S_1^z+S_2^z+S_{j_1-1}^z+S_{j_1}^z+S_{j_1+1}^z+S_L^z$ under PBCs and   $\Sigma=S_1^z+S_2^z+S_{j_1-1}^z+S_{j_1}^z+S_{j_1+1}^z$ under OBCs.
As follows from the lemma, we are led to a degenerate ground state  $g\vert\Psi_0\rangle$, which takes the form: 
$\left(-(S_{L}^-S_{1}^-S_{j_1}^-+(S_{1}^-)^2S_{j_1}^-+S_{1}^-S_{2}^-S_{j_1}^-+S_{1}^-S_{j_1-1}^-S_{j_1}^-+ S_{1}^-(S_{j_1}^-)^2+ S_{1}^-S_{j_1}^-S_{j_1+1}^-) +\sum_{|j_2-j_1|>1, 3\le j_2\le L-1} S_{1}^-S_{j_1}^-S_{j_2}^-\right)\ket{\psi_0}$ under PBCs and
$\left(-((S_{1}^-)^2S_{j_1}^-+S_{1}^-S_{2}^-S_{j_1}^-+S_{1}^-S_{j_1-1}^-S_{j_1}^-+ S_{1}^-(S_{j_1}^-)^2+ S_{1}^-S_{j_1}^-S_{j_1+1}^-) +\sum_{|j_2-j_1|>1, 3\le j_2\le L}  S_{1}^-S_{j_1}^-S_{j_2}^-\right)\ket{\psi_0}$ under OBCs, up to a multiplicative constant. Again, $g$ generates an emergent discrete  symmetry group $Z_2$ tailored to $\vert\Psi_0\rangle$, because $g^2|\Psi_0\rangle= \vert\Psi_0\rangle$.
We are thus led to degenerate ground states: $\left(S_{L}^-S_{1}^-S_{j_1}^-+(S_{1}^-)^2S_{j_1}^-+S_{1}^-S_{2}^-S_{j_1}^-+S_{1}^-S_{j_1-1}^-S_{j_1}^-+ S_{1}^-(S_{j_1}^-)^2+ S_{1}^-S_{j_1}^-S_{j_1+1}^- \right) \ket{\psi_0}
$ and $\sum_{|j_2-j_1|>1, 3\le j_2\le L-1}  S_{1}^-S_{j_1}^-S_{j_2}^-\ket{\psi_0}$ under PBCs and
$\left( (S_{1}^-)^2S_{j_1}^-+S_{1}^-S_{2}^-S_{j_1}^-+S_{1}^-S_{j_1-1}^-S_{j_1}^-+ S_{1}^-(S_{j_1}^-)^2+ S_{1}^-S_{j_1}^-S_{j_1+1}^-\right) \ket{\psi_0}$ and $\sum_{|j_2-j_1|>1, 3\le j_2\le L}  S_{1}^-S_{j_1}^-S_{j_2}^-\ket{\psi_0}$ under OBCs.
If $\vert\Psi_0\rangle$ is chosen to be $\left(S_{L}^-S_{1}^-S_{j}^-+(S_{1}^-)^2S_{j}^-+S_{1}^-S_{2}^-S_{j}^-+S_{1}^-S_{j-1}^-S_{j}^-+ S_{1}^-(S_{j}^-)^2+ S_{1}^-S_{j}^-S_{j+1}^- \right)\ket{\psi_0}$ ( $3<j<L-1$) under PBCs, then we consider a local unitary operation $g$, defined as $\exp (i \pi \Sigma)$, with $\Sigma = S_1^z+S_2^z+S_L^z$. We are thus led to a set of degenerate ground states  $g|\Psi_0\rangle$,
which take the form $\left(-S_{L}^-S_{1}^-S_{j}^--(S_{1}^-)^2S_{j}^--S_{1}^-S_{2}^-S_{j}^-+S_{1}^-S_{j-1}^-S_{j}^-+ S_{1}^-(S_{j}^-)^2+ S_{1}^-S_{j}^-S_{j+1}^- \right)\ket{\psi_0}$   up to a multiplicative constant.
Again,  $g$ generates an emergent discrete symmetry group $Z_2$ tailored to $\vert\Psi_0\rangle$, because $g^2|\Psi_0\rangle= |\Psi_0\rangle$.
Hence, $|\psi_3^{1j}\rangle=S_{L}^-S_{1}^-S_{j}^-+(S_{1}^-)^2S_{j}^-+S_{1}^-S_{2}^-S_{j}^-\ket{\psi_0}$ and $|\psi_3^{j 1}\rangle=\left(S_{1}^-S_{j-1}^-S_{j}^-+ S_{1}^-(S_{j}^-)^2+ S_{1}^-S_{j}^-S_{j+1}^- \right)\ket{\psi_0}$ are two degenerate ground states under PBCs.
Similarly, if $|\Psi_0\rangle$ is chosen to be $\left((S_{1}^-)^2S_{j}^-+S_{1}^-S_{2}^-S_{j}^-+S_{1}^-S_{j-1}^-S_{j}^-+ S_{1}^-(S_{j}^-)^2+ S_{1}^-S_{j}^-S_{j+1}^- \right)\ket{\psi_0}$ with $j>3$ under OBCs, then we consider
a local unitary operation $g$, defined as $\exp (i \pi \Sigma)$, with $\Sigma= S_1^z+S_2^z$. We are thus led to a set of degenerate ground states  $g\vert\Psi_0\rangle$,
which take the form: $\left( (S_{1}^-)^2S_{j}^-+S_{1}^-S_{2}^-S_{j}^--S_{1}^-S_{j-1}^-S_{j}^-- S_{1}^-(S_{j}^-)^2-S_{1}^-S_{j}^-S_{j+1}^-\right) \ket{\psi_0}$, up to a multiplicative constant.
Again, $g$ generates an emergent discrete symmetry group $Z_2$ tailored to $\vert\Psi_0\rangle$, because $g^2|\Psi_0\rangle= |\Psi_0\rangle$.
Hence,  $|\psi_3^{1j}\rangle=\left( (S_{1}^-)^2S_{j}^-+S_{1}^-S_{2}^-S_{j}^-\right) \ket{\psi_0}$ and $|\psi_3^{j}\rangle=\left( S_{1}^-S_{j-1}^-S_{j}^-- S_{1}^-(S_{j}^-)^2-S_{1}^-S_{j}^-S_{j+1}^-\right) \ket{\psi_0}$ are two degenerate ground states under OBCs.
Afterwards, if $\vert\Psi_0\rangle$ is chosen to be $\sum_{|j_2-j_1|>1, 3\le j_2\le L-1}  S_{1}^-S_{j_1}^-S_{j_2}^-\ket{\psi_0}$ under PBCs, then we consider
a set of local unitary operations $g_j$, defined as $\exp (i \pi \Sigma_j)$, with $\Sigma_j = S_j^z$ ($|j_2-j_1|>1, 3\le j_2\le L-1$). We are thus led to a set of degenerate ground states  $g_j|\Psi_0\rangle$, which take the form $\sum_{|j_2-j_1|>1, 3\le j_2\le L-1}  (1-2\delta_{j_2\;j}) S_1^- S_{j_1}^-S_{j_2}^-\ket{\psi_0}$, up to a multiplicative constant.
Again, each of $g_j$'s generates an emergent discrete symmetry group $Z_2$ tailored to $\vert\Psi_0\rangle$, because $g_j^2|\Psi_0\rangle= |\Psi_0\rangle$.
Hence, $S_{1}^-S_{j_1}^-S_{j_2}^-\ket{\psi_0}$ ($3\le j_1\le L-1$, $3\le j_2\le L-1$, $|j_2-j_1|>1$) are a sequence of degenerate ground states under PBCs.
Similarly, if $|\Psi_0\rangle$ is chosen to be $\sum_{|j_2-j_1|>1, 3\le j_2\le L}  S_{1}^-S_{j_1}^-S_{j_2}^-\ket{\psi_0}$ under OBCs, then we consider
a set of local unitary operations $g_j$, defined as $\exp (i \pi \Sigma_j)$, with $\Sigma_j =S_j^z$ ($|j_2-j_1|>1, 3\le j_2\le L$). We are thus led to a set of degenerate ground states  $g_j\vert\Psi_0\rangle$,
which take the form: $\sum_{|j_2-j_1|>1, 3\le j_2\le L}  S_{1}^-S_{j_1}^-S_{j_2}^-\ket{\psi_0}$, up to a multiplicative constant.
Again, each of $g_j$'s generates an emergent discrete symmetry group $Z_2$ tailored to $\vert\Psi_0\rangle$, because $g_j^2|\Psi_0\rangle= |\Psi_0\rangle$.
Hence,  $S_{1}^-S_{j_1}^-S_{j_2}^-\ket{\psi_0}$ ($3\le j_1\le L$, $3\le j_2\le L$, $|j_2-j_1|>1$) are a sequence of degenerate ground states under OBCs.

As a consequence, we are led to generalized highest weight states $\vert\psi_2^{j_1j_2j_3}\rangle \equiv S_{j_1}^-S_{j_2}^- S_{j_3}^-\ket{\psi_0}$ at the third level, where  $j_1=1,2,\ldots,L$, $j_2$ is not less than $j_1+2$, and $j_3$ is not less than $j_2+2$, but less than $L$ when $j_1=1$ for PBCs, and  $j_1=1,2,\ldots,L$, $j_2$ is not less than $j_1+2$, and $j_3$ is not less than $j_2+2$ for OBCs, respectively.  Physically, these generalized highest weight states are relevant to low-lying  three-magnon excitations, in the sense that  they constitute a set of basis states, invariant under
the translation operation under PBCs or the cyclic permutation symmetry operation under OBCs, to yield three-magnon excitations.

In addition, if $\vert\Psi_0\rangle$ is chosen to be $S^-((S_1^-)^2+S_1^-S_2^-+S_1^-S_L^-)\ket{\psi_0}$ under PBCs and  $S^-((S_1^-)^2+S_1^-S_2^-)\ket{\psi_0}$ under OBCs, then we may consider a local unitary operation $g = \exp (i \pi \Sigma)$, with  $\Sigma=S_1^z+S_2^z+S_3^z+S_{L-1}^z+S_L^z$ under PBCs and  $\Sigma=S_1^z+S_2^z+S_3^z$ under OBCs. 
As follows from the lemma, we are led to a degenerate ground state  $g\vert\Psi_0\rangle$, which takes the form: 
$(S_1^-+S_2^-+S_3^-+S_{L-1}^-+S_L^-)((S_1^-)^2+S_1^-S_2^-+S_1^-S_L^-) -\sum_{3<j<L-1} S_j^-((S_1^-)^2+S_1^-S_2^-+S_1^-S_L^-)\ket{\psi_0}$ under PBCs and
$(S_1^-+S_2^-+S_3^-)((S_1^-)^2+S_1^-S_2^-) -\sum_{j>3} S_j^-((S_1^-)^2+S_1^-S_2^-)\ket{\psi_0}$ under OBCs, up to a multiplicative constant. Again, $g$ generates an emergent discrete  symmetry group $Z_2$ tailored to $\vert\Psi_0\rangle$, because $g^2|\Psi_0\rangle= \vert\Psi_0\rangle$.
We are thus  led to degenerate ground states: $|\psi_3^{1}\rangle_2 \equiv(S_1^-+S_2^-+S_3^-+S_{L-1}^-+S_L^-)((S_1^-)^2+S_1^-S_2^-+S_1^-S_L^-) \ket{\psi_0}
$ and $\sum_{3<j<L-1} S_j^-((S_1^-)^2+S_1^-S_2^-+S_1^-S_L^-)\ket{\psi_0}$ under PBCs and
$|\psi_3^{1}\rangle_2 \equiv (S_1^-+S_2^-+S_3^-)((S_1^-)^2+S_1^-S_2^-) \ket{\psi_0}$ and $\sum_{j>3} S_j^-((S_1^-)^2+S_1^-S_2^-)\ket{\psi_0}$ under OBCs. Afterwards,
if $\vert\Psi_0\rangle$ is chosen to be $\sum_{3<j<L-1} S_j^-((S_1^-)^2+S_1^-S_2^-+S_1^-S_L^-)\ket{\psi_0}$ under PBCs, then we consider
a set of local unitary operations $g_j$, defined as $\exp (i \pi \Sigma_j)$, with $\Sigma_j = S_j^z$ ($j=4$, \ldots, $L-2$). Therefore, we are led to a set of degenerate ground states  $g_j|\Psi_0\rangle$,
which take the form $\sum _{i=5}^{L-1} (1-2\delta_{i\;j}) S_j^-((S_1^-)^2+S_1^-S_2^-+S_1^-S_L^-)\ket{\psi_0}$, up to a multiplicative constant.
Again, each of $g_j$'s generates an emergent discrete symmetry group $Z_2$ tailored to $\vert\Psi_0\rangle$, because $g_j^2|\Psi_0\rangle= |\Psi_0\rangle$.
Hence, $|\psi_3^{1j}\rangle_2=S_j^-((S_1^-)^2+S_1^-S_2^-+S_1^-S_L^-)\ket{\psi_0}$ ($ j=4,\ldots,L-2$) are a sequence of degenerate ground states under PBCs.
Similarly, if $|\Psi_0\rangle$ is chosen to be $\sum_{j>3} S_j^-((S_1^-)^2+S_1^-S_2^-)\ket{\psi_0}$ under OBCs, then we consider
a set of local unitary operations $g_j$, defined as $\exp (i \pi \Sigma_j)$, with $\Sigma_j =S_j^z$ ($j=4$, \ldots, $L$). We are thus led to a set of degenerate ground states  $g_j\vert\Psi_0\rangle$,
which take the form: $\sum _{i=5}^{L} (1-2\delta_{i\;j}) S_j^-((S_1^-)^2+S_1^-S_2^-)\ket{\psi_0}$, up to a multiplicative constant.
Again, each of $g_j$'s generates an emergent discrete symmetry group $Z_2$ tailored to $\vert\Psi_0\rangle$, because $g_j^2|\Psi_0\rangle= |\Psi_0\rangle$.
Hence,  $|\psi_3^{1j}\rangle=S_j^-((S_1^-)^2+S_1^-S_2^-)\ket{\psi_0}$ ($ j=4,\ldots,L$) are a sequence of degenerate ground states under OBCs.

\end{widetext}

\subsection{ The one-magnon, two-magnon and three-magnon excitations for the ${\rm SO}(3)$  spin-1  bilinear-biquadratic model in the ferromagnetic regime}

For convenience, we summarize the explicit expressions for the one-magnon, two-magnon and three-magnon excitations in the ${\rm SO}(3)$  spin-1 ferromagnetic bilinear-biquadratic model. Although the one-magnon excitations are well understood, it is quite challenging to find  the explicit expressions for the two-magnon excitations~\cite{akutsu,bibikov1} and the three-magnon excitations ~\cite{bibikov2} even in the thermodynamic limit. Indeed, an advantage to work directly in the thermodynamic limit is that any complications arising from the boundary conditions vanish.   We shall take advantage of these explicit expressions to establish a connection between emergent Goldstone flat bands and multi-magnon excitations for the staggered ${\rm SU}(3)$  spin-1 ferromagnetic biquadratic model (\ref{hambq}), up to the third level.  In principle, this construction may be extended to multi-magnon excitations. However, the problem to construct the explicit expressions for the multi-magnon excitations becomes increasingly complicated as the number of magnons involved increases, as indicated in Ref.~\onlinecite{bibikov2} for four-magnon excitations.

As we shall show,  $m$-magnon excitation states $\Phi_{m}$ in the ferromagnetic regime $\pi/2 < \theta < 5\pi/4$,  as the staggered ${\rm SU}(3)$  spin-1 ferromagnetic biquadratic model located at $\theta = \pi/2$ is approached, become a linear combination of degenerate ground states, including  generalized highest weight states at the $m$-th level, up to $m=3$.

\subsubsection{The one-magnon excitations}

From the time-independent ${\rm Schr\ddot odinger}$ equation, it is simple to figure out  the excitation energy $\omega (k)$ for a one-magnon excitation, which is defined to be
\begin{equation*}
	 \Phi_{1} (k) = \sum_{j} e^{ik j} S_j^-\ket{\psi_0}.
\end{equation*}
Here, we recall that $\ket{\psi_0}$  is the highest weight state $\ket{\otimes_{\eta \in Z} \{+\}_{\;\eta}}$ in the thermodynamic limit, as already introduced in Section~\ref{spectralfunction}.        
As a result, we have $\omega (k) = 2|\cos \theta| \;(1-\cos k$).  This is valid in the entire ferromagnetic regime including the two endpoints: $\pi/2 \le \theta \le 5\pi/4$. In particular, 
at $\theta=\pi/2$, the model becomes the staggered $\rm SU(3)$ spin-1 ferromagnetic model. As argued in Subsection~\ref{emergenttoexponential}, all the states $\vert\psi_1^j\rangle \equiv S_j^-\ket{\psi_0}$, as a result of an emergent local symmetry operation tailored to the permutation-invariant ground state $\sum_{j}  S_j^- \ket{\psi_0}$, are degenerate ground states. In fact, they are  generalized highest weight states at the first level. As a consequence, the one-magnon excitation state $\Phi_{1} (k)$, as $\theta = \pi/2$ is approached, simply becomes a linear combination of these degenerate ground states, with the excitation energy being zero.

\begin{widetext}
\subsubsection{The two-magnon excitations}

In the thermodynamic limit, a two-magnon excitation $\Phi_{2}$ takes the form
\begin{equation*}
		\Phi_{2}  = \Big(\sum_{j_1<j_2}a_{j_1,j_2}S^-_{j_1} S^-_{j_2} +\sum_{j}b_j (S^-_j)^2 \Big) \ket{\psi_0}.
\end{equation*}

According to the time-independent ${\rm Schr\ddot odinger}$ equation, we have
\begin{eqnarray}
		|\cos(\theta)\; (4a_{j_1,j_2}-a_{j_1-1,j_2}-a_{j_1+1,j_2}-a_{j_1,j_2-1}-a_{j_1,j_2+1})=\omega a_{j_1,j_2},\;j_2-j_1>1,\label{aijk121}\\
		|\cos(\theta)|\left ((3\!-\!\tan\theta)a_{j,j+1}-a_{j-1,j+1}-a_{j,j+2}+(\tan\theta-1)(b_{j}+b_{j+1})\right )=\omega a_{j,j+1},\label{aijk12f}\\
		|\cos(\theta)|\left (2(2-\tan\theta)b_{j}-\tan\theta(b_{j-1}+b_{j+1})+(\tan\theta-1)(a_{j-1,j}+a_{j,j+1})\right )=\omega b_{j},\label{aijk12}
\end{eqnarray}
where $\omega$ denotes the excitation energy for the two-magnon excitation.  For our purpose, we assume that $a_{j_1,j_2}$, $b_j$, and $\omega$ are parameterized in terms of $k_1$ and $k_2$:
$a_{j_1,j_2} \equiv a_{j_1,j_2} (k_1,k_2)$, $b_j \equiv b_j(k_1,k_2)$, and $\omega \equiv \omega (k_1,k_2)$. Hence $\Phi_{2}$ becomes $\Phi_{2}(k_1,k_2)$.

At $\theta=5\pi/4$, the model becomes the uniform $\rm SU(3)$ ferromagnetic model. Given $\tan \theta =1$, the last terms in Eq.~(\ref{aijk12f}) and Eq.~(\ref{aijk12}) vanish. As such,  equations (\ref{aijk121}) and  (\ref{aijk12f}) for $a_{j_1,j_2}$ are independent to equation (\ref{aijk12}) 
for $b_j$.  Then it is readily seen that the equation for $b_j$ yields a solution $b_j(k_1,k_2)={\rm e}^{i(k_1+k_2)j}$, with $\omega(k_1,k_2)$ being formally identical to the excitation energy for a one-magnon excitation with $k = k_1+k_2$, i.e., $\omega(k_1,k_2)= \sqrt{2}\left (1-\cos (k_1+k_2)\right )$. Further, one may solve the two equations for $a_{j_1,j_2}(k_1,k_2)$ to yield another solution, which takes the same form as that given in Eq.(\ref{aj1j2}) below, with $A(k,\tilde k)=-1+{\rm e}^{i\tilde k}+
3{\rm e}^{ik}+{\rm e}^{i(k+2\tilde k)}-2{\rm e}^{2ik}-4{\rm e}^{i(k+\tilde k)}+3{\rm e}^{i(2k+\tilde k)}-{\rm e}^{2i(k+ \tilde k)}$ and $\omega(k_1,k_2)$ being given by $\omega(k_1,k_2)=\omega(k_1)+\omega(k_2)= \sqrt{2}(2-\cos k_1-\cos k_2)$, which is the sum of the excitation energies for two one-magnon excitations with $k = k_1$ and $k=k_2$. Physically, this stems from the fact that there are two type-B GMs at this special endpoint. Mathematically, this is due to the fact that {\it not only} $\sum_j S_j^z$ {\it but also} $\sum_j (S_j^z)^2$ commute with the model Hamiltonian, as a result of the uniform $\rm SU(3)$ symmetry group.

For the ferromagnetic regime $\pi/2 < \theta < 5\pi/4$, $a_{j_1,j_2}(k_1,k_2)$ and $b_j(k_1,k_2)$ are subject to the following Ansatz
\begin{equation}
		a_{j_1,j_2}(k_1,k_2)=A(k_1,k_2)\;{\rm e}^{i(k_1j_1+k_2j_2)}-A(k_2,k_1)\;{\rm e}^{i(k_2j_1+k_1j_2)},\quad
		b_j(k_1,k_2)=B(k_1,k_2)\;{\rm e}^{i(k_1+k_2)j}.\label{aj1j2}
\end{equation}
Following Eq.~(\ref{aijk12f}), we have $\omega(k_1,k_2)=\omega(k_1)+\omega(k_2)$, 
where $\omega(k_1)$ and $\omega(k_2)$ are  the excitation energies $\omega (k)$ for one-magnon excitations with $k=k_1$ and $k=k_2$, respectively.
Meanwhile, it is readily seen that Eq.~(\ref{aijk12}) may be reduced  to the form
\begin{eqnarray*}
		\Big(1+{\rm e}^{i(k_1+k_2)}-(1+\tan\theta){\rm e}^{ik_2}\Big) \; A(k_1,k_2)-\Big(1+{\rm e}^{i(k_1+k_2)}-(1+\tan\theta){\rm e}^{ik_1}\Big)\;
		A(k_2,k_1)+(\tan\theta-1)\Big(1+{\rm e}^{i(k_1+k_2)}\Big)\;B(k_1,k_2)=0,\nonumber\\
		(\tan\theta\!-\!1)\Big({\rm e}^{-ik_1}+{\rm e}^{ik_2}\Big)A(k_1,k_2)-(\tan\theta\!-\!1)\Big({\rm e}^{ik_1}+{\rm e}^{-ik_2}\Big)A(k_2,k_1)+2\Big(\cos{k_1}+\cos{k_2}-\tan\theta(\cos{(k_1+k_2)}+1)\Big)B(k_1,k_2)=0.
\end{eqnarray*}
They yield the  solutions as  follows
\begin{equation}
		A(k,\tilde k)={\rm e}^{i\tilde k}+{\rm e}^{i(k+2\tilde k)}-(1+\tan\theta)\;{\rm e}^{2ik}-(1+3\tan\theta)\;{\rm e}^{i(k+\tilde k)}+
		\tan\theta\;\Big(3{\rm e}^{ik}+3{\rm e}^{i(2k+\tilde k)}-1-{\rm e}^{2i(k+ \tilde k)}\Big),\label{ak1k2}
\end{equation}
	and
\begin{equation}
		B(k_1,k_2)=(1-\tan\theta)\;\Big({\rm e}^{i k_2}-{\rm e}^{ik_1}\Big)\Big(1+{\rm e}^{i(k_1+k_2)}\Big).\label{bk1k2}
\end{equation}
	Here $k=k_1$ and $\tilde k=k_2$ or  $k=k_2$ and $\tilde k=k_1$.

We are particularly interested in the endpoint $\theta=\pi/2$. At this point, the above solutions for the two-magnon excitation $ \Phi_{2} (k_1,k_2)$ become
\begin{equation}
		A(k,\tilde k)=-{\rm e}^{2ik}-3{\rm e}^{i(k+\tilde k)}+3{\rm e}^{ik}+3{\rm e}^{i(2k+\tilde k)}-1-{\rm e}^{2i(k+ \tilde k)},\label{ak1k2n}
\end{equation}
	and
\begin{equation}
		B(k_1,k_2)=-\Big({\rm e}^{i k_2}-{\rm e}^{ik_1}\Big)\Big(1+{\rm e}^{i(k_1+k_2)}\Big).\label{bk1k2n}
\end{equation}

Given that this two-magnon excitation yields zero excitation energy, one may expect that it must be a linear combination of degenerate ground states, as a result of an emergent local symmetry operation tailored to the permutation-invariant ground state $(\sum_{j}  S_j^-)^2 \ket{\psi_0}$. As argued in Subsection~\ref{emergenttoexponential}, all the states $\vert\psi_2^{j_1j_2}\rangle \equiv S_{j_1}^-S_{j_2}^-\ket{\psi_0}$ ($j_2\ge j_1+2$) act as  generalized highest weight states at the second level. In addition, all the states
$|\psi_2^{j}\rangle=(S_{j-1}^-S_{j}^-+ (S_{j}^-)^2+ S_{j}^-S_{j+1}^-) \ket{\psi_0}$, as follows from 
$|\psi_2^{1}\rangle \equiv (S_{1}^-S_{L}^-+ (S_{1}^-)^2+ S_{1}^-S_{2}^-) \ket{\psi_0}$ under the translation operation under PBCs (in the thermodynamic limit), are also degenerate ground states. It is readily seen that  the two-magnon excitation state $\Phi_{2} (k_1,k_2)$ at $\theta = \pi/2$ is simply a linear combination of these degenerate ground states, including all the generalized highest weight states at the second level. More precisely,  we have
\begin{equation*} 
	\Phi_{2} (k_1,k_2)=\sum_{j_2-j_1>1} a_{j_1,j_2}(k_1,k_2) \vert\psi_2^{j_1j_2}\rangle+ \sum_{j} b_j(k_1,k_2)|\psi_2^{j}\rangle.
\end{equation*}
This follows from the relation, which is {\it only} valid at $\theta = \pi/2$:
\begin{equation*} 
a_{j,j+1}(k_1,k_2)=b_j(k_1,k_2)+b_{j+1}(k_1,k_2).
\end{equation*}

\subsubsection{The three-magnon excitations}

In the thermodynamic limit, a three-magnon excitation $\Phi_{3}$ takes the form
\begin{equation*}
		\Phi_{3}  = \Big(\sum_{j_1<j_2<j_3} a_{j_1,j_2,j_3}S^-_{j_1} S^-_{j_2} S^-_{j_3} +\sum_{j_1<j_2} \Big(b_{j_1,j_2}^1 (S^-_{j_1})^2S^-_{j_2}+b_{j_1,j_2}^2S^-_{j_1}(S^-_{j_2})^2 \Big) \Big) \ket{\psi_0}.
\end{equation*}

According to the time-independent ${\rm Schr\ddot odinger}$ equation,  we have
\begin{align}
		&|\cos\theta|(6a_{j_1,j_2,j_3}-a_{j_1+1,j_2,j_3}-a_{j_1,j_2+1,j_3}-a_{j_1,j_2,j_3+1}-a_{j_1-1,j_2,j_3}-a_{j_1,j_2-1,j_3}-a_{j_1,j_2,j_3-1})=\omega a_{j_1,j_2,j_3},\label{eqa123}
\end{align}
if $j_2-j_1>1$ and $j_3-j_2>1$, and
\begin{eqnarray}
		&|\cos\theta|\left ((5-\tan\theta)a_{j_1-1,j_1,j_2}\!-\!a_{j_1-1,j_1+1,j_2}\!-\!a_{j_1-1,j_1,j_2+1}\!-\!a_{j_1-2,j_1,j_2}-a_{j_1-1,j_1,j_2-1}\!+\!(\tan\theta\!-\!1)(b_{j_1-1,j_2}^1\!+\!b_{j_1,j_2}^1)\right )\!=\!\omega a_{j_1-1,j_1,j_2},\label{eqabm3n1}\\
		&|\cos\theta|\left ((5-\tan\theta)a_{j_1,j_2,j_2+1}\!-\!a_{j_1,j_2,j_2+2}\!-\!a_{j_1+1,j_2,j_2+1}\!-\!a_{j_1,j_2-1,j_2+1}\!-\!a_{j_1-1,j_2,j_2+1}\!+\!(\tan\theta\!-\!1)(b_{j_1,j_2}^2\!+\!b_{j_1,j_2+1}^2)\right )\!=\!\omega a_{j_1,j_2,j_2+1},\label{eqabm3n2}\\
		&|\cos\theta|\left (2(3-\tan\theta)b_{j_1,j_2}^1-b_{j_1,j_2+1}^1-b_{j_1,j_2-1}^1-\tan\theta(b_{j_1-1,j_2}^1+
		b_{j_1+1,j_2}^1)+(\tan\theta-1)(a_{j_1,j_1+1,j_2}+a_{j_1-1,j_1,j_2})\right )=\omega b_{j_1,j_2}^1,
		\label{eqabm3n3}\\
		&|\cos\theta|\left (2(3-\tan\theta)b_{j_1,j_2}^2-b_{j_1+1,j_2}^2-b_{j_1-1,j_2}^2-\tan\theta(b_{j_1,j_2-1}^2+b_{j_1,j_2+1}^2)+(\tan\theta-1)(a_{j_1,j_2,j_2+1}+a_{j_1,j_2-1,j_2})\right )=\omega 
 b_{j_1,j_2}^2,\label{eqabm3n4}
\end{eqnarray}
if $j_2-j_1>1$,  and
	\begin{align}
		&|\cos\theta|\{2(2-\tan\theta)\;a_{j-1,j,j+1}-a_{j-1,j,j+2}-a_{j-2,j,j+1}+(\tan\theta-1)(b_{j-1,j+1}^1+b_{j,j+1}^1+b_{j-1,j}^2+b_{j-1,j+1}^2)\}=\omega\;a_{j-1,j,j+1},\label{eqab12}\\
 	&	|\cos\theta|\left ((4-\tan\theta)\;b_{j,j+1}^1-b_{j,j+2}^1-\tan\theta\; b_{j-1,j+1}^1-b_{j,j+1}^2+(\tan\theta-1)
		\;a_{j-1,j,j+1}\right )=\omega\;b_{j,j+1}^1,\label{eqb12n1}\\
		&|\cos\theta|\left ((4-\tan\theta)\;\;b_{j-1,j}^2-b_{j-2,j}^2-\tan\theta\; b_{j-1,j+1}^2-b_{j-1,j}^1+(\tan\theta-1)\;a_{j-1,j,j+1}\right )=\omega\;b_{j-1,j}^2.\label{eqb12n2}
	\end{align}
For our purpose, we assume that $a_{j_1,j_2,j_3}$, $b^1_{j_1,j_2}$ and $b^2_{j_1,j_2}$ are generically parameterized in terms of  $k_1$, $k_2$, $k_3$, $\tilde k_1$, $\tilde k_2$, and $\tilde k_3$: $a_{j_1,j_2,j_3} \equiv a_{j_1,j_2,j_3}(k_1,k_2,k_3,\tilde k_1,\tilde k_2,\tilde k_3)$, $b^1_{j_1,j_2} \equiv b^{1}_{j_1,j_2}(k_1,k_2,k_3,\tilde k_1,\tilde k_2,\tilde k_3)$, and $b^2_{j_1,j_2} \equiv b^{2}_{j_1,j_2}(k_1,k_2,k_3,\tilde k_1,\tilde k_2,\tilde k_3)$. 
Hence $\Phi_{3}$ becomes $\Phi_{3}(k_1,k_2,k_3,\tilde k_1,\tilde k_2,\tilde k_3)$.
Note that $k_1$, $k_2$, $k_3$, $\tilde k_1$, $\tilde k_2$, and $\tilde k_3$ are subject to the constraints
\begin{equation*}
		\prod_{\alpha=1}^3{\rm e}^{i\tilde k_\alpha}=\prod_{\alpha=1}^3{\rm e}^{ik_\alpha} \;\;{\rm and} \;\;\quad\sum_{\alpha=1}^3\omega(\tilde k_\alpha)=\sum_{\beta=1}^3\omega(k_\beta).
\end{equation*}
In addition, $\omega$ is parameterized in terms of  $k_1$, $k_2$ and $k_3$: $\omega \equiv \omega(k_1,k_2,k_3)$, where
$\omega(k_1,k_2,k_3)=\sum_{\alpha=1}^3\omega( k_\alpha)$. Here $\omega(k_1)$, $\omega(k_2)$ and $\omega(k_3)$ are  the excitation energies $\omega (k)$ for one-magnon excitations with $k=k_1$, $k=k_2$ and $k=k_3$, respectively. However, the two completely integrable cases at $\theta=5\pi/4$ and 
$\theta=3\pi/4$ are special, in the sense that they admit a parametrization in terms of  $k_1$, $k_2$, $k_3$: $a_{j_1,j_2,j_3} \equiv a_{j_1,j_2,j_3}(k_1,k_2,k_3)$, $b^1_{j_1,j_2} \equiv b^{1}_{j_1,j_2}(k_1,k_2,k_3)$, and $b^2_{j_1,j_2} \equiv b^{2}_{j_1,j_2}(k_1,k_2,k_3)$. 
Hence $\Phi_{3}$ becomes $\Phi_{3}(k_1,k_2,k_3)$. In particular, the model located at $\theta = 3\pi/4$ in the ferromagnetic regime is completely integrable, since the Takhtajan-Babujian model~\cite{takhtajan,babujian} is located at its antipodal point (cf. Fig.~\ref{so3-gspd}).

At $\theta=5\pi/4$, the model becomes the uniform $\rm SU(3)$ ferromagnetic model. Given $\tan \theta =1$,  Eq.~(\ref{eqa123}), Eq.~(\ref{eqabm3n1}), Eq.~(\ref{eqabm3n2}), and Eq.~(\ref{eqab12}) for   $a_{j_1,j_2,j_3}$ are independent from  Eq.~(\ref{eqabm3n3}), Eq.~(\ref{eqabm3n4}) Eq.~(\ref{eqb12n1}),  and Eq.(\ref{eqb12n2}) for $b^{1}_{j_1,j_2}$ and $b^{2}_{j_1,j_2}$. Physically, this stems from the fact that there are two type-B GMs at this special endpoint, since the two generators, i.e.,  $\sum_j S_j^z$ and $\sum_j (S_j^z)^2$, commute with the model Hamiltonian, as a result of the uniform $\rm SU(3)$ symmetry group. 
They yield a solution
\begin{eqnarray}
		&a_{j_1,j_2,j_3}(k_1,k_2,k_3)=\sum_{\alpha,\beta,\gamma=1}^3\varepsilon_{\alpha \beta \gamma}A(k_\alpha,k_\beta)A(k_\alpha,k_\gamma)A(k_\beta,k_\gamma){\rm e}^{i(k_\alpha j_1+k_\beta j_2+k_\gamma j_3)},
		\label{ab1b2n1}\\
		&b^{1}_{j_1,j_2}(k_1,k_2,k_3)=\frac{1}{2}\sum_{\alpha,\beta,\gamma=1}^3\varepsilon_{\alpha \beta \gamma}B(k_\alpha,k_\beta)A(k_\alpha,k_\gamma)A(k_\beta,k_\gamma){\rm e}^{i\left ((k_\alpha+k_\beta)j_1+k_\gamma j_2\right )},\label{ab1b2n2}\\
		&b^{2}_{j_1,j_2}(k_1,k_2,k_3)=\frac{1}{2}\sum_{\alpha,\beta,\gamma=1}^3\varepsilon_{\alpha \beta \gamma}A(k_\alpha,k_\beta)A(k_\alpha,k_\gamma)B(k_\beta,k_\gamma){\rm e}^{i[\left (k_\alpha j_1+(k_\beta+k_\gamma)j_2\right )},\label{ab1b2}
\end{eqnarray}
where $A(k,\tilde k)$ and $B(k,\tilde k)$ have been presented in Eq.(\ref{ak1k2}) and Eq.(\ref{bk1k2}), and $\varepsilon_{\alpha \beta \gamma}$ is the completely antisymmetric tensor. Note that the same expressions also yield a solution at  $\theta=3\pi/4$.

For the ferromagnetic regime $\pi/2 < \theta < 5\pi/4$,  $a_{j_1,j_2,j_3}(k_1,k_2,k_3,\tilde k_1,\tilde k_2,\tilde k_3)$, $b^{1}_{j_1,j_2}(k_1,k_2,k_3,\tilde k_1,\tilde k_2,\tilde k_3)$ and $b^{2}_{j_1,j_2}(k_1,k_2,k_3,\tilde k_1,\tilde k_2,\tilde k_3)$ are subject to the Ansatz
\begin{eqnarray}
		a_{j_1,j_2,j_3}(k_1,k_2,k_3,\tilde k_1,\tilde k_2,\tilde k_3)&=&\varphi(\tilde k_1,\tilde k_2,\tilde k_3)X_{j_1,j_2,j_3}(k_1,k_2,k_3)-
		\varphi(k_1,k_2,k_3)X_{j_1,j_2,j_3}(\tilde k_1,\tilde k_2,\tilde k_3),\label{a123}\\
		b^{u}_{j_1,j_2}(k_1,k_2,k_3,\tilde k_1,\tilde k_2,\tilde k_3)&=&\varphi(\tilde k_1,\tilde k_2,\tilde k_3)Y^{u}_{j_1,j_2}(k_1,k_2,k_3)-
		\varphi(k_1,k_2,k_3) Y^{u}_{j_1,j_2}(\tilde k_1,\tilde k_2,\tilde k_3), \; u=1,2,\label{b123}
\end{eqnarray}
where
\begin{equation*}
		\varphi(k_1,k_2,k_3)=\Big({\rm e}^{ik_1}-{\rm e}^{ik_2}\Big)\Big({\rm e}^{ik_2}-{\rm e}^{ik_3}\Big)\Big({\rm e}^{ik_3}-{\rm e}^{ik_1}\Big)
		\prod_{\alpha=1}^3(1-{\rm e}^{ik_\alpha}).
\end{equation*}
Here $X_{j_1,j_2,j_3}(k_1,k_2,k_3)$, $Y^{1}_{j_1,j_2}(k_1,k_2,k_3)$, and $Y^{2}_{j_1,j_2}(k_1,k_2,k_3)$ take the same form as Eq.~(\ref{ab1b2}). In other words, we have
\begin{eqnarray}
		&X_{j_1,j_2,j_3}(k_1,k_2,k_3)=\sum_{\alpha,\beta,\gamma=1}^3\varepsilon_{\alpha \beta \gamma}A(k_\alpha,k_\beta)A(k_\alpha,k_\gamma)A(k_\beta,k_\gamma){\rm e}^{i(k_\alpha j_1+k_\beta j_2+k_\gamma j_3)},
		\label{AB1B2n1}\\
		&Y^{1}_{j_1,j_2}(k_1,k_2,k_3)=\frac{1}{2}\sum_{\alpha,\beta,\gamma=1}^3\varepsilon_{\alpha \beta \gamma}B(k_\alpha,k_\beta)A(k_\alpha,k_\gamma)A(k_\beta,k_\gamma){\rm e}^{i\left ((k_\alpha+k_\beta)j_1+k_\gamma j_2\right )},\label{AB1B2n2}\\
		&Y^{2}_{j_1,j_2}(k_1,k_2,k_3)=\frac{1}{2}\sum_{\alpha,\beta,\gamma=1}^3\varepsilon_{\alpha \beta \gamma}A(k_\alpha,k_\beta)A(k_\alpha,k_\gamma)B(k_\beta,k_\gamma){\rm e}^{i\left (k_\alpha j_1+(k_\beta+k_\gamma)j_2\right )}.\label{AB1B2n3}
\end{eqnarray}
We remark that, at $\theta=3\pi/4$,  the model admits two three-magnon excitations, degenerate in energy, which are given by Eq.~(\ref{ab1b2n1}), Eq~(\ref{ab1b2n2}) and Eq.~(\ref{ab1b2}) and by Eq.~(\ref{AB1B2n1}), Eq.~(\ref{AB1B2n2}) and Eq.~(\ref{AB1B2n3}). Note that they are generically different from each other, unless two of $k_1$, $k_2$ and $k_3$ are identical, or one of them is zero modulo $2\pi$.

At $\theta=\pi/2$,  $a_{j_1,j_2,j_3}(k_1,k_2,k_3,\tilde{k}_1,\tilde{k}_2,\tilde{k}_3)$, $b^{1}_{j_1,j_2}(k_1,k_2,k_3,\tilde{k}_1,\tilde{k}_2,\tilde{k}_3)$ and $b^{2}_{j_1,j_2}(k_1,k_2,k_3,\tilde{k}_1,\tilde{k}_2,\tilde{k}_3)$ follow from Eq.(\ref{a123}) and Eq.(\ref{b123}), with $A(k,\tilde{k})$ and $B(k,\tilde{k})$ in Eq.~(\ref{AB1B2n1}), Eq.~(\ref{AB1B2n2}) and Eq.~(\ref{AB1B2n3}) for $X_{j_1,j_2,j_3}(k_1,k_2,k_3)$, $Y^{1}_{j_1,j_2}(k_1,k_2,k_3)$, and $Y^{2}_{j_1,j_2}(k_1,k_2,k_3)$ being given by Eq.(\ref{ak1k2n}) and Eq.(\ref{bk1k2n}). 
As argued in Appendix~C, in addition to fully factorized ground states, acting as generalized highest weight states at the third level, as a result of an emergent local symmetry operation tailored to the permutation-invariant ground state $(\sum_{j}  S_j^-)^3\ket{\psi_0}$: $\vert\psi_{3}^{j_1j_2j_3}\rangle \equiv S_{j_1}^-S_{j_2}^-S_{j_3}^- \ket{\psi_0}$ ($|j_2-j_1|>1,|j_3-j_2|>1,|j_1-j_3|>1$), there are three types of degenerate ground states that are not fully factorized at the third level (in the thermodynamic limit):
\begin{eqnarray*}
&&\vert\psi_{3}^{j}\rangle_1 \equiv \left (S_{j-1}^-S_{j}^-S_{j+2}^-+(S_{j}^-)^2S_{j+2}^-+S_{j}^-S_{j+1}^-S_{j+2}^-+ S_{j}^-(S_{j+2}^-)^2+ S_{j}^-S_{j+2}^-S_{j+3}^-\right ) \ket{\psi_0},\nonumber \\
&&\vert\psi_{3}^{j}\rangle_2 \equiv \sum_{k=j,j\pm1,j\pm2}S_{k}^-\left ((S_{j}^-)^2+S_{j-1}^-S_{j}^-+S_{j}^-S_{j+1}^- \right )\ket{\psi_0},\nonumber \\
&&\vert\psi_{3}^{j_1j_2}\rangle \equiv S_{j_2}^-\left ((S_{j_1}^-)^2+S_{j_1-1}^-S_{j_1}^-+S_{j_1}^-S_{j_1+1}^- \right )\ket{\psi_0}, \; |j_1-j_2|>2.
\end{eqnarray*}
Note that they follow from $\vert\psi_{3}^{1}\rangle_1 $, $\vert\psi_{3}^{1}\rangle_2$ and $\vert\psi_{3}^{1j_2}\rangle$ under the translation operation under PBCs (in the thermodynamic limit). 
Here we have introduced extra subscripts in $\vert\psi_{3}^{j}\rangle_1$ and $\vert\psi_{3}^{j}\rangle_2$ to label two different degenerate ground states that are not fully factorized. A notable feature is that degenerate ground states $\vert\psi_{3}^{j}\rangle_1$ and $\vert\psi_{3}^{j}\rangle_2$ are not orthogonal, but linearly independent to each other. Hence the three-magnon excitation state $\Phi_{3}(k_1,k_2,k_3,\tilde k_1,\tilde k_2,\tilde k_3)$ at $\theta=\pi/2$, with zero excitation energy, is simply a linear combination of  these degenerate ground states:
\begin{align*} 
	&\Phi_{3} (k_1,k_2,k_3,\tilde{k}_1,\tilde{k}_2,\tilde{k}_3)=\sum_{j_2-j_1>1,j_3-j_2>1} a_{j_1,j_2,j_3}(k_1,k_2,k_3,\tilde{k}_1,\tilde{k}_2,\tilde{k}_3)\vert\psi_3^{j_1j_2j_3}\rangle+\nonumber \\
	&	\sum_{|j_2-j_1|>2} c_{j_1j_2}(k_1,k_2,k_3,\tilde{k}_1,\tilde{k}_2,\tilde{k}_3)|\psi_3^{j_1j_2}\rangle+\frac{1}{3}\sum_{j} \left (b_{jj+1}^1(k_1,k_2,k_3,\tilde{k}_1,\tilde{k}_2,\tilde{k}_3)+b_{jj+1}^2(k_1,k_2,k_3,\tilde{k}_1,\tilde{k}_2,\tilde{k}_3)\right )|\psi_3^{j}\rangle_1+\nonumber \\
	&	\frac{1}{3}\sum_{j} \left (2b_{jj+1}^1(k_1,k_2,k_3,\tilde{k}_1,\tilde{k}_2,\tilde{k}_3)-b_{jj+1}^2(k_1,k_2,k_3,\tilde{k}_1,\tilde{k}_2,\tilde{k}_3)\right )\vert\psi_{3}^{j}\rangle_2.
\end{align*}
Here $c_{j_1j_2}(k_1,k_2,k_3,\tilde{k}_1,\tilde{k}_2,\tilde{k}_3)=b_{j_1j_2}^1(k_1,k_2,k_3,\tilde{k}_1,\tilde{k}_2,\tilde{k}_3)$ for  $j_2>j_1$ and $c_{j_1j_2}(k_1,k_2,k_3,\tilde{k}_1,\tilde{k}_2,\tilde{k}_3)=b_{j_1j_2}^2(k_1,k_2,k_3,\tilde{k}_1,\tilde{k}_2,\tilde{k}_3)$ for $j_2<j_1$. There are a few relations, which are {\it only} valid at $\theta = \pi/2$:
\begin{eqnarray*}
&&a_{jj+1j+2}(k_1,k_2,k_3,\tilde{k}_1,\tilde{k}_2,\tilde{k}_3)=b_{j+1j+2}^1(k_1,k_2,k_3,\tilde{k}_1,\tilde{k}_2,\tilde{k}_3)+b_{jj+2}^1(k_1,k_2,k_3,\tilde{k}_1,\tilde{k}_2,\tilde{k}_3),\nonumber \\
&&a_{jj+1j+2}(k_1,k_2,k_3,\tilde{k}_1,\tilde{k}_2,\tilde{k}_3)=b_{jj+1}^2(k_1,k_2,k_3,\tilde{k}_1,\tilde{k}_2,\tilde{k}_3)+b_{jj+2}^2(k_1,k_2,k_3,\tilde{k}_1,\tilde{k}_2,\tilde{k}_3),\nonumber \\
&&a_{j_1j_1+1j_2}(k_1,k_2,k_3,\tilde{k}_1,\tilde{k}_2,\tilde{k}_3)=b_{j_1j_2}^1(k_1,k_2,k_3,\tilde{k}_1,\tilde{k}_2,\tilde{k}_3)+b_{j_1+1j_2}^1(k_1,k_2,k_3,\tilde{k}_1,\tilde{k}_2,\tilde{k}_3),\; j_2-j_1>2,\nonumber \\
&&a_{j_1j_2j_2+1}(k_1,k_2,k_3,\tilde{k}_1,\tilde{k}_2,\tilde{k}_3)=b_{j_1j_2}^2(k_1,k_2,k_3,\tilde{k}_1,\tilde{k}_2,\tilde{k}_3)+b_{j_1j_2+1}^2(k_1,k_2,k_3,\tilde{k}_1,\tilde{k}_2,\tilde{k}_3)\; j_2-j_1>2, \nonumber \\
&&b_{jj+1}^1(k_1,k_2,k_3,\tilde{k}_1,\tilde{k}_2,\tilde{k}_3)+2b_{j+1j+2}^1(k_1,k_2,k_3,\tilde{k}_1,\tilde{k}_2,\tilde{k}_3)=2b_{jj+1}^2(k_1,k_2,k_3,\tilde{k}_1,\tilde{k}_2,\tilde{k}_3)+b_{j+1j+2}^2(k_1,k_2,k_3,\tilde{k}_1,\tilde{k}_2,\tilde{k}_3).
\end{eqnarray*}

In principle, the multi-magnon excitation states in the ferromagnetic regime  $\pi/2 < \theta < 5\pi/4$ must tend to be a linear combination of degenerate ground states  at $\theta = \pi/2$, including those  generalized highest weight states at the $m$-th level, which are fully factorized ground states. Note that the excitation energy tends to vanish, as $\theta = \pi/2$ is approached. This amounts to stating that the set of generalized highest weight states constructed at $\theta=\pi/2$, together with other degenerate ground states that are not fully factorized at the $m$-th level, constitutes a complete set of  the basis states to construct multi-magnon states, which in turn imposes a constraint on $m$-magnon excitations in the ferromagnetic regime $\pi/2 < \theta < 5\pi/4$.

\end{widetext}

\end{document}